\documentclass[11pt]{article}
\usepackage{graphics,cite,amssymb,epsfig,float}
\usepackage[usenames,dvips]{color}
\usepackage{amsmath, mathbbol}

\usepackage{multirow}
\makeatletter

\@addtoreset{equation}{section}
\makeatother

\def\lsim{\;\raise0.3ex\hbox{$<$\kern-0.75em\raise-1.1ex\hbox{$\sim$}}\;}
\def\gsim{\;\raise0.3ex\hbox{$>$\kern-0.75em\raise-1.1ex\hbox{$\sim$}}\;}

\def\ben{\begin{enumerate}}  \def\een{\end{enumerate}}
\def\bit{\begin{itemize}}    \def\eit{\end{itemize}}
\def\beq{\begin{equation}}   \def\eeq{\end{equation}}
\def\ba{\begin{array}}       \def\ea{\end{array}}
\def\bea{\begin{eqnarray}}   \def\eea{\end{eqnarray}}

\begin{document}

\begin{titlepage}
\renewcommand{\thefootnote}{\fnsymbol{footnote}}
\setcounter{footnote}{0}

\vspace*{-2cm}
\begin{flushright}
LPT Orsay 10-48 \\
CFTP 10-010 \\
PCCF RI 1003\\

\vspace*{2mm}
\end{flushright}

\begin{center}
\begin{center}
\vspace*{15mm}
{\Large\bf Interplay of LFV and slepton mass splittings at the LHC as
  a probe of the SUSY seesaw} \\
\vspace{1cm}
{\bf A. Abada$^{a}$, A. J. R. Figueiredo$^{b}$, J. C. Rom\~ao$^{b}$ and 
A. M. Teixeira$^{c}$
}

 \vspace*{.5cm} 
$^{a}$ Laboratoire de Physique Th\'eorique, CNRS -- UMR 8627, \\
Universit\'e de Paris-Sud 11, F-91405 Orsay Cedex, France

\vspace*{.2cm} 
$^{b}$ Centro de F\'{\i}sica Te\'orica de Part\'{\i}culas, 
Instituto Superior T\'ecnico, \\ Av. Rovisco Pais 1, 
1049-001 Lisboa, Portugal

\vspace*{.2cm} 
$^{c}$ Laboratoire de Physique Corpusculaire, CNRS/IN2P3 -- UMR 6533,\\ 
Campus des C\'ezeaux, 24 Av. des Landais, F-63177 Aubi\`ere Cedex, France

\end{center}

\vspace*{10mm}
\begin{abstract}

\vspace*{3mm}
We study the impact of a type-I SUSY seesaw concerning lepton flavour
violation (LFV) both at low-energies and at the LHC.  The study of the
di-lepton invariant mass distribution at the LHC allows to reconstruct
some of the masses of the different sparticles involved in a decay
chain. In particular, the combination with other observables renders feasible
the reconstruction of the masses of the intermediate sleptons involved
in $ \chi_2^0\to \tilde \ell \,\ell \to  \ell \,\ell\,\chi_1^0$ decays.
Slepton mass splittings can be either interpreted as a signal of
non-universality in the SUSY soft breaking-terms (signalling a
deviation from constrained scenarios as the cMSSM) or as being due to
the violation of lepton flavour.  In the latter case, in addition to
these high-energy processes, one expects further low-energy
manifestations of LFV such as radiative and three-body lepton decays.
Under the assumption of a type-I seesaw as the source of neutrino
masses and mixings, all these LFV observables are related.
Working in the framework of the cMSSM  
extended by three right-handed neutrino superfields, we conduct a systematic
analysis addressing the simultaneous implications of the SUSY seesaw
for both high- and low-energy lepton flavour violation.
We discuss how the confrontation of slepton mass splittings as
observed at the LHC and low-energy LFV observables may provide
important information about the underlying mechanism of LFV.

\end{abstract}
\end{center}

\vspace*{3mm}

\end{titlepage}

\renewcommand{\thefootnote}{\arabic{footnote}}
\setcounter{footnote}{0}
\setcounter{page}{2}

\vspace*{5mm}
\section{Introduction}

The experimental observation of non-vanishing neutrino masses and
mixings~\cite{neutrino}, constitutes clear evidence 
for physics beyond the Standard
Model (SM), and  as of today, little is known about the underlying
model of new physics. Since neutrinos are very weakly
interacting particles and their masses lie orders of magnitude 
below the other fermion masses of the SM, additional experimental 
input will be instrumental to shed some light on the new physics
model.

In extensions of the SM where $\nu$ oscillations (and hence massive
neutrinos) can be naturally accommodated, many other new
phenomena could in principle be expected. Among them, and given that 
neutrino oscillations indisputably signal 
lepton flavour violation (LFV) in the neutral sector, 
it is only natural to expect that charged lepton flavour will
also be violated in these extensions (for a review, see 
Ref.~\cite{Raidal:2008jk}).    
The search for manifestations of charged LFV constitutes the goal of 
several experiments~\cite{Brooks:1999pu, Aubert:2005wa, Aubert:2005ye, Bellgardt:1987du, Aubert:2003pc, Akeroyd:2004mj, Aysto:2001zs, 
Kuno:2005mm, Bona:2007qt, Kiselev:2009zz, Ritt:2006cg, PRIME,
Hayasaka:2007vc}, exclusively dedicated to look for signals of  
processes such as  rare radiative as well as three-body decays and lepton
conversion in muonic nuclei.

In parallel to these low-energy searches, the high-energy Large Hadron
Collider (LHC) has started its quest of unveiling the 
mechanism of electroweak (EW) symmetry breaking and of possibly
providing a solution to the SM hierarchy problem. 
Supersymmetry (SUSY) is a well motivated solution for the hierarchy
problem that also offers an elegant solution for the existence of 
non-baryonic dark matter 
(DM) in the Universe~\cite{Jungman:1995df, Bertone:2004pz, WMAP}.  
If the LHC indeed finds signatures of 
SUSY, it is then extremely appealing to consider
supersymmetric models that can also accommodate neutrino oscillation
phenomena. One of the most economical and elegant possibilities is 
perhaps to 
embed a seesaw mechanism~\cite{seesaw:I, seesaw:II, seesaw:III} in a
supersymmetric framework, the so-called SUSY seesaw.  
 
If a type-I seesaw~\cite{seesaw:I} is  at work and explains  the observed
neutrino masses and leptonic mixings, then the neutrino Yukawa
couplings could leave their imprint in the slepton mass matrices: 
as first shown in~\cite{Borzumati:1986qx}, starting from flavour 
diagonal soft supersymmetry breaking terms at some high energy 
unification scale, flavour violation appears at low-energies 
due to the renormalisation group (RG) evolution of the SUSY 
soft-breaking parameters~\cite{Hall:1985dx, Donoghue:1983mx}. 
Having natural values for the neutrino Yukawa couplings  
implies that the seesaw scale 
(e.g. the right-handed neutrino mass scale in the
case of a type-I seesaw) is very high, close to the Grand Unified
Theory (GUT) scale ($M_\text{GUT} \sim 10^{16}$ GeV). 
Moreover, the flavour off-diagonal structure of the neutrino
Yukawa couplings  required to comply with the observed 
large mixing in the lepton sector~\cite{Schwetz:2008er,
GonzalezGarcia:2010er}, can then induce potentially large lepton flavour
violation in the slepton sector. Low-energy manifestations of LFV 
in the framework of the SUSY seesaw include sizable branching ratios
(BR) for radiative decays as $l_i \to l_j \gamma$, three-body decays, $l_i \to 3
l_j$  and  $\mu-e$ transitions in heavy 
nuclei~\cite{Hisano:1995cp, Hisano:1995nq, Hisano:1998fj, Buchmuller:1999gd, Kuno:1999jp, Casas:2001sr, Lavignac:2001vp, Bi:2001tb, Ellis:2002fe,Deppisch:2002vz, Fukuyama:2003hn, Brignole:2004ah,  Masiero:2004js, Fukuyama:2005bh, Petcov:2005jh,  Arganda:2005ji, Deppisch:2005rv, Yaguna:2005qn, Calibbi:2006nq, Antusch:2006vw, Hirsch:2008dy, Arganda:2007jw, Arganda:2008jj}.
In the presence of CP violation, one can also have T- and P-odd 
asymmetries in LFV decays  and contributions to lepton electric dipole
moments (see for example~\cite{Okada:1999zk, Ellis:2001xt,
  Ellis:2001yza, Masina:2003wt}).  

The quest for new physics is currently being pursued along different avenues: 
high-energy colliders like the LHC are the ideal laboratory to directly  
discover the particle content of the SM extension; low-energy
experiments probe the new physics contributions (arising from new particles
and/or interactions) to several observables (muon anomalous magnetic
moment, electric dipole moments, LFV, B-physics, etc.).
A successful (or even partial) reconstruction of the underlying model of 
new physics will necessarily rely on the complementarity of the information  
derived from direct and indirect searches, which can be further 
strengthened by data from neutrino experiments, dark matter searches
and cosmological observations.  

In this work, we study the impact of a type-I SUSY seesaw concerning 
flavour violation both at low-energies and at the LHC. 
At the LHC, there are three possible signals of LFV: 
firstly, one  can have sizable widths for  LFV decay processes like
$\chi_2^0 \to \ell_i^\pm\, \ell_j^\mp\, \chi_1^0$~\cite{Arkanihamed:1996au,
Hinchliffe:2000np, Carvalho:2002jg, Hirsch:2008dy, Carquin:2008gv} ;
secondly, one  can have flavoured slepton mass splittings (MS). 
These can be identified since under certain conditions, 
one can effectively reconstruct
slepton masses via observables such as the kinematic end-point of the 
invariant mass distribution of the leptons coming from the cascade decays
$\chi_2^0 \rightarrow \tilde{\ell}^{\pm} \ell^{\mp} \rightarrow
 \ell^{\pm} \ell^{\mp}\chi_1^0$. If the slepton in the decay chain is
real, the di-lepton invariant mass spectrum has a kinematical edge that
might then  be measured  with very high precision (up to 0.1
\%)~\cite{Paige:1996nx, Hinchliffe:1996iu, Bachacou:1999zb}. 
Together with data arising from other observables, this information 
allows to reconstruct the slepton masses \cite{Paige:1996nx, Hinchliffe:1996iu,
Bachacou:1999zb,  Ball:2007zza,ATLAS}.
Finally, one  can observe multiple edges in di-lepton invariant mass
distributions $\chi_2^0 \to \chi_1^0 \ell_i^{\pm} \ell_i^{\mp}$, arising from the
exchange of a different flavour slepton $\tilde l_j$ 
(in addition to the left- and right-handed sleptons, $\tilde l_{L,R}^i$).
Slepton mass splittings can be either interpreted as a signal of
non-universality in the SUSY soft breaking-terms (hinting towards a
deviation from flavour-blind scenarios of SUSY breaking such as 
the constrained Minimal Supersymmetric Standard Model (cMSSM)) 
or as being due to the violation of lepton flavour.

The potential of LHC experiments in probing the allowed seesaw
parameters through measurements of masses and branching ratios of
supersymmetric particles has also been discussed in
Refs.~\cite{Hisano:1998wn, Blair:2002pg, Buckley:2006nv}.  Recently,
another study of slepton mass-splittings as a probe of LFV at the LHC
was performed~\cite{Buras:2009sg} for scenarios with an effective
parametrization of flavour violation. In our case, we consider the
specific framework of a type-I SUSY seesaw, where the source of
flavour violation for both the LHC and the low-energy observables is
unique - the neutrino Yukawa couplings -
implying that all these LFV observables will be correlated.  Working
in the framework of the cMSSM extended by three right-handed neutrino
superfields, and taking into account the DM constraints~\cite{WMAP},
we conduct a systematic analysis addressing the simultaneous
implications of the SUSY seesaw for both high- and low-energy LFV.

Under the assumption of a type-I SUSY seesaw, the interplay of a joint
measurement of LFV branching ratios and of the Chooz angle
$\theta_{13}$ has been shown to be a powerful tool to shed some light
on the SUSY seesaw parameters (see for example~\cite{Antusch:2006vw}).
Here we will focus on how the confrontation of slepton mass splittings
(as potentially observable at the LHC) and of low-energy LFV
observables may provide important information about the underlying
mechanism of LFV. After having identified regions in the cMSSM
parameter space, where the slepton masses could in principle be
reconstructed from the kinematical edges of di-lepton mass
distributions (i.e.  $\chi_2^0 \to \ell_i^\pm\, \ell^\mp_i \,
\chi_1^0$ can occur with a non-negligible number of events), we study
the different slepton mass splittings arising in this case from small
flavour-conserving radiative effects and $LR$ slepton mixing.  We then
discuss the effect of implementing a type-I seesaw for the slepton
mass splittings, also exploring the implications for LFV decays. We
investigate several scenarios in which the SUSY seesaw can be tested
and propose, in addition to two already existing LHC benchmark points,
other minimal Supergravity (mSUGRA) inspired benchmarks embedded in a
type-I seesaw.

As we will show in this work, if the seesaw is indeed the source of
both neutrino masses and leptonic mixings and accounts for low-energy
LFV observables within future sensitivity reach, interesting slepton
phenomena are expected to be observed at the LHC: in addition to the
mass splittings, the most striking effect will be the possible
appearance of new edges in di-lepton mass distributions.  From the
comparison of the predictions for the two sets of observables (high
and low energy) with the current experimental bounds and future
sensitivities, one can either derive information about the otherwise
unreachable seesaw parameters, or disfavour the type-I SUSY seesaw as
the unique source of LFV.

The paper is organised as follows.  In Section~\ref{sec:LHC} we
discuss how lepton masses can be reconstructed from observation at the
LHC, describing the mechanisms for production, the favoured decay
chains and the kinematical observables.  In
Section~\ref{susy:seesaw:lfv} we define the model, providing a brief
overview on the implementation of a type-I seesaw in the constrained
MSSM, as well as its implications for low-energy LFV observables. We
also comment on the possibility of generating the observed BAU from
leptogenesis, and how complying with present observation can constrain
the SUSY seesaw parameters.  In Section~\ref{lfv:lhc} we study, both
for the cMSSM and its type-I seesaw extension, the impact of LFV at
the LHC.  Our results are presented in Section~\ref{sec:results}
where, after briefly considering the cMSSM case, we study the
different high- and low-energy observables in the seesaw case. This
will also allow to draw some conclusions on the viability of a type-I
seesaw as the underlying mechanism of LFV. Further discussion is
presented in the concluding Section~\ref{sec:conclusions}.

\section{Slepton masses and invariant mass distributions at the
  LHC}\label{sec:LHC}

In this work we are interested in the study of  slepton mass
differences to probe deviations from the cMSSM, and possibly derive
some information about the underlying theory of flavour violation in
the (s)lepton sector. We briefly outline in this section how 
slepton masses can be reconstructed from observation at the LHC. We
describe the mechanisms for production, the favoured decay chains and 
finally the kinematical observables used to reconstruct  
the slepton masses and hence their mass splittings. 

We recall that the  cMSSM is defined by its superpotential, 
\begin{equation}\label{eq:W:def}
\mathcal{W}^\text{MSSM}\,=\,\hat U^c\,Y^u\,\hat Q \, \hat H_2 \,+\,
\hat D^c\,Y^l\,\hat Q \, \hat H_1 \,+\,
\hat E^c\,Y^l\,\hat L \, \hat H_1 \,+\,
\mu \,\hat H_1 \,\hat H_2 \,,
\end{equation}
and by the mSUGRA-inspired conditions imposed on the soft-breaking
SUSY Lagrangian: universal gaugino masses ($M_1=M_2=M_3=M_{1/2}$),
universal scalar masses for Higgs bosons, squarks and sleptons
($m_{H_1, H_2} =m_{\tilde Q, \tilde U, \tilde D } = m_{\tilde L,
\tilde E} =m_0$) and universal trilinear couplings ($A_{u,d,l}=A_0
Y^{u,d,l}$), the universality being imposed at some high energy scale,
which we chose to be the gauge coupling unification scale.  The model
is further defined by the ratio of the vacuum expectation value of the
Higgs fields, $\tan \beta = v_2/v_1$ and $\operatorname{sign}(\mu)$,
leading to a total of 4 continuous and one discrete parameter.

\subsection{Slepton production at the LHC}

If R-parity is preserved, SUSY particles are produced in pairs, and
decay to the lightest SUSY particle (LSP), which is stable. The usually
complex decay cascades lead to signatures involving in general  multiple
jets and/or missing transverse energy from the LSPs escaping the detector. 
Several reconstruction methods have been proposed 
(see, e.g.~\cite{Paige:1996nx, Hinchliffe:1996iu,Bachacou:1999zb} 
and references therein) allowing to extract  very precise 
combinations of masses and branching ratios from several 
experimental measurements. In particular, the analysis of endpoints in
kinematical distributions for specific final states allows to determine 
fundamental parameters of the model, especially in the case of simple
SUSY realisations as the cMSSM. In favourable cases, where one expects 
to observe a large number of events, 
and if the signal to background ratio is large, the cMSSM
parameters are likely to be measured with very good
accuracy~\cite{Ball:2007zza, ATLAS}. 

Provided the SUSY breaking scale is not too high, supersymmetric 
particles are expected to be abundantly produced
at the LHC, operating at a centre of mass (c.o.m.) energy 
$\sqrt s=$ 7 TeV - 14 TeV. 
The production of coloured SUSY sparticles
will dominantly occur from quark-antiquark annihilation
and gluon-gluon fusion, and possibly also via (strong) quark-quark
scattering and quark-gluon fusion.
QCD-singlet particles as sleptons can be directly produced via
Drell-Yan processes  ($s$-channel $Z$- or $\gamma$-exchange) 
or arise from gaugino-like neutralino decays. However, in the
first case the associated production
cross sections are in general small and detection is compromised
due to the large SM backgrounds. 
The second process is more favourable, since neutralinos
can be produced directly, or arise from cascade decays of squarks; 
if kinematically allowed, squark decays
lead to a large number of chains with intermediate slepton states
(like for instance $\tilde q_L\to q_L \,\chi_2^0\to q_L \, \tilde \ell\,
\ell \to q_L \,\ell \,\ell \, \chi_1^0$). 

At the LHC, squarks might  be pair produced $p p \to \tilde{q} \tilde{q}^*,
\, \tilde{q} \tilde{q}$ and single produced $pp \to
\tilde{q} \tilde{g}$~\cite{Beenakker:1996ch}.   Squarks can then
decay to a ${\chi}^0_i \, q\,$ pair, while the gluino 
preferably decays to $\tilde{q}_R q, \tilde{t}_1 t$. 
Direct neutralino production goes either through pure electroweak
interactions ($pp \to {\chi}^0_2 {\chi}^0_i, \,{\chi}^0_2
{\chi}^{\pm}_i$) or mixed EW-strong ($pp \to {\chi}^0_2 \tilde{q}^i_{L,R},\,
{\chi}^0_2 \tilde{g}$, with $\tilde q$ possibly decaying into
${\chi}^0_2 q$).  

Here we will distinguish between three primary production modes:
``direct'' neutralino production ($pp \to \chi_2^0 X$), squark-decay
($pp \to \tilde q_L Y$) and gluino-gluino mode ($\tilde g \tilde g$).
In the cMSSM framework, the process 
$pp \to \tilde{g} \tilde{g}$ is in general kinematically 
suppressed ($m_{\tilde{g}} > m_{\tilde{q}}$). We
also consider separately the prospects for 
at least one- and exactly two-$\chi_2^0$ production.

\subsection{Di-lepton invariant mass
  distributions}\label{subsec:di-lepton} 
As extensively discussed in the literature, in scenarios where the
${\chi}^0_2$ is sufficiently heavy to decay via a real (on-shell)
slepton, the process ${\chi}^0_2 \to
\ell^{\pm}\,{\ell^{\mp}}\,{\chi}^0_1$ is greatly enhanced while
providing a very distinctive signal~\cite{Paige:1996nx,
Hinchliffe:1996iu,Bachacou:1999zb,ATLAS}: same-flavour
opposite-charged leptons with missing energy.  Moreover, the momentum
of the leptons is expected to be easily reconstructed (accounting for
smearing effects in $\tau$'s), thus allowing to extract indirect
information on the mass spectrum of the involved sparticles.

As previously mentioned, the best approach to reconstruct the
intermediate sparticle masses in a decay chain is the construction of
invariant kinematical quantities, which are comparatively easy to
measure (even in the presence of large amounts of missing energy).  In
particular, the di-lepton invariant mass distribution presents
kinematical edges (di-particle or tri-particle), which allow to derive
information on the mass of the exchanged sparticles.

In order to reduce the SM background, several cuts have to be applied
in the reconstruction of the events.  It has been
shown~\cite{Hinchliffe:1996iu,Bachacou:1999zb} that one of those was
having two isolated leptons with large transverse momentum, $p_T > 10$
GeV. We will therefore always require hard outgoing leptons in our
analysis.
From the SUSY decay chain, 
$\tilde q_L \to \chi_2^0\, q \to \tilde \ell_{L,R} \,  \ell q \to
\chi_1^0 \ell\,  \ell\,  q$, 
one can construct several invariant quantities~\cite{Bachacou:1999zb,Allanach:2000kt}:
\begin{itemize}
\item[(i)] 3 di-particle invariant masses
\begin{align}
& m_{\ell \ell}^\text{max}\,\  =\,
 M(m_{\chi_2^0},m_{\tilde \ell_{L,R}},m_{\chi_1^0})\,,\\
& m_{\ell^\text{near}q}^\text{max}\,=\, M(m_{\tilde q_L},m_{\chi_2^0},
m_{\tilde \ell_{L,R}})\,,\\
& m_{\ell^\text{far}q}^\text{max}\,\  =\, M^\prime(m_{\tilde q_L},m_{\chi_2^0},
m_{\chi_1^0}, m_{\tilde \ell_{L,R}})\,,
\end{align}
where 
\begin{equation}
M(x,y,z)=1/y \sqrt{(x^2-y^2)(y^2-z^2)}\,, \quad \quad 
M^\prime(x,y,z,w)=1/w\sqrt{(x^2-y^2)(w^2-z^2)}\,, \nonumber
\end{equation}
and whose end-points have a common structure;
\item[(ii)] tri-particle invariant mass 
\begin{equation}
m_{\ell \ell q}^\text{max}\,=\, M(m_{\tilde q_L},m_{\chi_2^0},m_{\chi_1^0})\,.
\end{equation}
\end{itemize}

\noindent 
Here, we shall focus on di-lepton invariant mass
distributions:  
\begin{equation}\label{eq:mll}		
m_{\ell \ell} \equiv \sqrt{ \left( p_{{\ell'}} + p_{\ell} \right)^2 } = 
m^{\text{(max)}}_{\ell \ell} \cos\frac{\theta}{2}, \, \quad \quad
m^{\text{(max)}}_{\ell \ell} = \frac{1}{m_{\tilde{\ell}}} \sqrt{ 
\left( m^2_{{\chi}^0_2} - m^2_{\tilde{\ell}} \right) 
\left( m^2_{\tilde{\ell}} - m^2_{{\chi}^0_1} \right) }, \,	
\end{equation}
where $\pi - \theta$ is the angle between the two leptons 
in the slepton's rest frame.
In general terms, the ${\chi}^0_2 \to {\chi}^0_1 \,
\ell'^{\pm}\,{\ell^{\mp}}$ decay process occurs via: 
$t$- and $u$-channel with  charged slepton exchange;      
exchange of the lightest Higgs boson\footnote{In mSUGRA scenarios
the exchange of the heaviest CP-even ($H$) or of the
CP-odd ($A$) Higgs bosons are off-shell suppressed.}, 
$h$; or via a $Z$ boson.

Considering the complete decay process, i.e., 
via on-shell and off-shell intermediate states, the di-lepton
invariant mass distribution has ``true'' start- and end-points 
given by
\begin{equation}		
\overline{m}^{\text{min}}_{\ell \ell} = 
m_{\ell'} + m_{{\ell}}, \, \quad \quad 
\overline{m}^{\text{max}}_{\ell \ell} = m_{{\chi}^0_2} - m_{{\chi}^0_1} \,, 	
\end{equation}	
respectively. It can be easily verified that the $\overline{m}_{\ell
\ell}$ end-point matches the on-shell end-point for a slepton of mass
$m_{\tilde{\ell}} = \sqrt{ m_{{\chi}^0_2} m_{{\chi}^0_1} }$, in which
case no decreasing event rate is expected to be observed beyond
$\overline{m}^{\text{max}}_{\ell \ell}$.

The invariant mass distributions can also be used to extract the mass
splittings of the intermediate sleptons by looking at distinctive
two-edge distributions which are expected to emerge whenever two
different sleptons $\tilde{\ell}_{1,2}$ have sufficiently high rates
for ${\chi}^0_2 \to \tilde{\ell}_{1,2} \, \ell_1 \to{\chi}^0_1 \,
\ell_2 \, \ell_1$. In our analysis we will study the di-muon and
di-electron invariant mass distributions, looking for edges that
correspond to the exchanged selectron and smuon states, thus allowing
to reconstruct the $\tilde e_L$ and $\tilde \mu_L$ masses (and in some
cases, $\tilde e_R$ and $\tilde \mu_R$ as well).  Hard outgoing taus
can decay hadronically and can also be identified, however the
background will be much larger in this case.  Nevertheless, we also
address $\tilde \mu - \tilde \tau$ mass differences.

As we will discuss in detail in the following section, the mass
differences of sleptons of the first two generations are expected to
be extremely small.  However, if slepton universality is broken
(e.g. via diagonal, but non-universal soft-breaking slepton masses),
or if lepton flavour is violated in the (s)lepton sector, distinct
two-edge distributions can be observable provided there is sufficient
resolution to be sensitive to a certain amount of mass
splitting~\cite{Bartl:2005yy}.  The kinematical edge is expected to be
measurable at LHC with a precision up to $0.1 \%$~\cite{Ball:2007zza,
ATLAS, Paige:1996nx,Hinchliffe:1996iu,Bachacou:1999zb}.  The relative
slepton mass splittings, which are defined as
\begin{equation}\label{eq:MS:def}
\frac{\Delta m_{\tilde \ell}}{m_{\tilde \ell}} (\tilde \ell_i, \tilde
  \ell_j) \, = \, 
\frac{|m_{\tilde \ell_i}-m_{\tilde \ell_j}|}{<m_{\tilde
    \ell_{i,j}}>}\,,
\end{equation}
can then be inferred from the kinematical edges with a sensitivity of
$\mathcal{O}(0.1\%)$~\cite{Allanach:2008ib} for $\tilde e_L - \tilde
\mu_L$ and $\mathcal{O}(1\%)$ for $\tilde \mu_L - \tilde \tau_2$.
Even if already impressive, the edge splitting can be further enhanced
by considering the so-called fractional shift of the kinematical edge
in the di-lepton invariant mass distribution:	
\begin{equation}		
\frac{\Delta m_{\ell \ell}}{m_{\ell \ell}} = \frac{\Delta
  m_{\tilde{\ell}}}{m_{\tilde{\ell}}}  
\left[ \frac{ m^2_{{\chi}^0_2} m^2_{{\chi}^0_1} - m^4_{\tilde{\ell}} }{ 
\left( m^2_{{\chi}^0_2} - m^2_{\tilde{\ell}} \right) 
\left( m^2_{\tilde{\ell}} - m^2_{{\chi}^0_1} \right) } \right].
\end{equation}

\bigskip
Even though this will be discussed in greater detail in
Section~\ref{sec:results}, it is clear from the above discussion that
certain conditions must be fulfilled in order to render feasible the
study of slepton mass splittings.  Firstly, sleptons must be produced
in non-negligible amounts: this translates into having a not
excessively heavy SUSY spectrum (to allow for abundant squark and
$\chi_2^0$ production), and in a neutralino-slepton hierarchy such
that the decays of $\chi_2^0$ into real sleptons are kinematically
viable. As already noticed in~\cite{Buras:2009sg}, in the regions of
the cMSSM where the latter decays are allowed, the BR($\chi_2^0 \to
\chi_1^0 \ell \ell$) is in general enhanced when compared to the case
of virtual intermediate sleptons.  Secondly, an efficient
tagging/event selection requires ``hard'' - highly energetic -
outgoing leptons, implying the following requirement for the phase
space: $m_{{\chi}^0_2} - m_{\tilde{e}_L,\tilde{\mu}_L,\tilde{\tau}_2}
\geq 10$ GeV.  In summary, the experimental study of slepton mass
differences at the LHC will only be possible if the specific
realisation of the SUSY model meets the above requirements.

\section{Lepton flavour violation in the SUSY
  seesaw}\label{susy:seesaw:lfv} 

Extensions of the SM by heavy states such as fermionic
singlets~\cite{seesaw:I} or fermionic triplets~\cite{seesaw:III} or
scalar triplets~\cite{seesaw:II}, allow to explain the smallness of
the neutrino masses via seesaw-like mechanisms. In these realisations,
the violation of lepton flavour number can be easily accommodated
in the neutral lepton sector and parametrized by a leptonic mixing
matrix. One may also have lepton flavour violation in the charged
sector through four-fermion dimension-six effective operators (see
for example~\cite{Abada:2007ux}, where several lepton violation processes
were studied in the three different seesaw types).  In this study, we
will consider a type-I seesaw (heavy fermionic singlets with masses at
a sufficiently high scale to have large enough Yukawa couplings:
$~10^{10} \text{ GeV}- 10^{15}$ GeV) embedded in the framework of
supersymmetric theories as a source of lepton flavour violation in the
charged lepton sector.  Within the so-called SUSY seesaw, flavour
violation in the neutrino sector is transmitted to the charged leptons
via radiative effects involving the neutrino Yukawa couplings $Y^\nu$.
Even under GUT scale universality conditions, the RGE-induced flavour
violation is sufficiently large to account for sizable rates of LFV
observables such as radiative ($\ell_i \to \ell_j \gamma$) and
three-body ($\ell_i \to \ell_j \ell_j \ell_k$) decays, and $\mu-e$
conversion in nuclei. It may also account for potentially large mass
splittings for the slepton masses (in addition to the usual
$LR$-mixing) and, in the presence of complex $Y^\nu$, for CPV
observables, such as T- and P-odd asymmetries in radiative and
three-body decays as well as contributions to the lepton electric
dipole moments (EDMs). Remarkably, having a unique source of flavour
violation in the lepton sector implies that all the above mentioned
observables will be strongly related.

In this section, we briefly overview the implementation of a type-I
seesaw in the constrained MSSM, as well as its implications for
low-energy LFV observables.  We also comment on the possibility of
generating the observed baryon asymmetry of the Universe (BAU) from
leptogenesis, and how complying with present observation on the baryon
asymmetry can constrain the SUSY seesaw parameters.  The impact of LFV
for high-energy experiments, manifest in observables such as slepton
mass splittings or direct flavour violation in sparticle decays, will
be discussed in Section~\ref{lfv:lhc}.

\subsection{Type-I SUSY seesaw}
We consider an extension of the MSSM 
to which three right-handed neutrino superfields are added. 
Each supermultiplet $\hat N^c$ contains the right-handed    
neutrinos $\nu_R$ and their superpartners $\tilde \nu_R$. 
The SUSY type-I seesaw is defined by the superpotential $\mathcal W$
of the MSSM extended by two additional terms involving $\hat N^c$. 
The leptonic part of $\mathcal W$ is then given by:
\begin{equation}\label{eq:Wlepton:def}
\mathcal{W}^\text{lepton}\,=\,\hat N^c\,Y^\nu\,\hat L \, \hat H_2 \,+\,
\hat E^c\,Y^l\,\hat L \, \hat H_1 \,+\,
\frac{1}{2}\,\hat N^c\,M_N\,\hat N^c\,.
\end{equation}
The lepton Yukawa couplings $Y^{l,\nu}$ and the
Majorana mass $M_N$ are $3\times 3$ matrices in lepton flavour
space. Hereafter we will always assume, without loss of generality, 
that we are in a basis where both $Y^l$ and $M_N$ are diagonal:
\begin{equation}
Y^l\, =\,\operatorname{diag}  (Y^e, Y^\mu,Y^\tau)\,,
\quad 
M_N\, =\,\operatorname{diag}  (M_{N_1},M_{N_2},M_{N_3})\,.
\end{equation}
The slepton part of the soft-SUSY breaking Lagrangian is specified by 
\begin{align}\label{eq:Lsoftslepton:def}
\mathcal{V}_\text{soft}^\text{slepton}\,=-\mathcal{L}^\text{slepton}\,&=
m_{\tilde L}^2 \, \tilde l_L \,\tilde 
l_L^*\,+\,m_{\tilde E}^2 \tilde l_R \,\tilde l_R^*
\,+\,m^2_{\tilde \nu_R}\,\tilde \nu_R\,\tilde \nu_R^*\,+
\nonumber \\
& +\, \left(
A_l\, 
H_1\,\tilde l_L \,\tilde l_R^*
\,+\,A_\nu\, 
H_2\,\tilde \nu_L \,\tilde \nu_R^* 
\,+\, B_{\nu}\,\tilde \nu_R\,\tilde \nu_R\, +\text{H.c.}
\right)\,.
\end{align}

\noindent
Motivated by SUSY breaking schemes mediated by flavour-blind
gravitational interactions (minimal supergravity inspired), we work
within the framework of the constrained MSSM, where the soft-SUSY
breaking parameters are assumed to be universal at some high-energy
scale $M_X$, which we choose to be the gauge coupling unification
scale $M_\text{GUT} \sim 10^{16}$ GeV.  Thus, at $M_X$, the additional
parameters in $\mathcal{L}^\text{slepton}$ also obey the following
universality conditions:
\begin{align}\label{eq:cMSSM:univcond}
&
\left(m_{\tilde L}\right)^2_{ij}\,=\,
\left(m_{\tilde E}\right)^2_{ij}\,=\,
\left(m_{\widetilde {\nu}_R}\right)^2_{ij}\,=\, m_0^2\,\delta_{ij}\,,\,\,
\nonumber \\
& \left(A_l\right)_{ij}\,=\, A_0 \, \left(Y^l \right)_{ij},\,\,
\left(A_\nu \right)_{ij}\,=\, A_0 \, \left(Y^\nu \right)_{ij},
\end{align}
where $m_0$ and $A_0$ are the universal scalar soft-breaking mass and
trilinear coupling of the cMSSM, and $i,j$ denote lepton flavour
indices ($i,j=1,2,3$).  

After electroweak symmetry breaking (EWSB), the Dirac mass terms for the
charged leptons and neutrinos are 
\begin{equation}
m_l\,=\,Y^l\,\,v_1\,, \quad \quad
m_D^\nu\,=\,Y^\nu\,v_2\,,
\end{equation}
where $v_i$ are the vacuum expectation values (VEVs) of the neutral
Higgs scalars, $v_{1(2)}= \,v\,\cos (\sin) \beta$ with $v=174$
GeV. Assuming $Y^l$ diagonal in flavour space, one has $m_l=
\operatorname{diag} (m_e, m_\mu,m_\tau)$, while the masses of the
physical Majorana neutrinos are given by the eigenvalues of the
$6\times6$ neutrino mass matrix,
\begin{equation}\label{eq:seesaw:def}
M^\nu\,=\,\left(
\begin{array}{cc}
0 &{m_D^\nu}^{T} \\
m_D^\nu & M_N
\end{array} \right)\,.
\end{equation}
In the seesaw limit (i.e. $M_{N_i}\,\gg\,v$), and at lowest order in
the $(m_D^\nu/M_N)^n$ expansion, the above matrix can be
block-diagonalized, leading to the usual seesaw equation for the light
neutrino mass matrix,
\begin{equation}\label{eq:seesaw:light}
m_\nu\,=\, - {m_D^\nu}^T M_N^{-1} m_D^\nu  \,, 
\end{equation}
while the masses of the heavy eigenstates are simply given by 
$M_{N_i}$. 

The light neutrino mass matrix $m_\nu$ is diagonalized by the 
Maki-Nakagawa-Sakata unitary matrix
$U^{\text{MNS}}$~\cite{PMNS},
\begin{align}\label{eq:MNS:physicalmasses}
m_{\nu}^\text{diag}&
\,=\,{U^\text{MNS}}^T \,m_{\nu}\, U^\text{MNS} 
\,=\, \text{diag}\,(m_{\nu_1},m_{\nu_2},m_{\nu_3})\,,
\end{align}
where under the standard parametrization $U^\text{MNS}$
is given by
\begin{equation}
U^\text{MNS}=
\left( 
\begin{array}{ccc} 
c_{12} \,c_{13} & s_{12} \,c_{13} & s_{13} \, e^{-i \delta} \\ 
-s_{12}\, c_{23}\,-\,c_{12}\,s_{23}\,s_{13}\,e^{i \delta} 
& c_{12} \,c_{23}\,-\,s_{12}\,s_{23}\,s_{13}\,e^{i \delta} 
& s_{23}\,c_{13} \\ 
s_{12}\, s_{23}\,-\,c_{12}\,c_{23}\,s_{13}\,e^{i \delta} 
& -c_{12}\, s_{23}\,-\,s_{12}\,c_{23}\,s_{13}\,e^{i \delta} 
& c_{23}\,c_{13}
\end{array} \right) .\, V\,,
\label{Umns}
\end{equation}
with
\begin{equation}
 V\,=\,\text{diag}\,(e^{-i\frac{\varphi_1}{2}},e^{-i\frac{\varphi_2}{2}},1)\,,
\end{equation}
and $c_{ij} \equiv \cos \theta_{ij}$, $s_{ij} \equiv \sin \theta_{ij}$.
$\theta_{ij}$ are the leptonic mixing angles, $\delta$ is the Dirac
CPV phase and $\varphi_{1,2}$ the Majorana CPV phases. 

Current (best-fit) analyses of the low-energy neutrino data favour the
following intervals for the mixing angles~\cite{GonzalezGarcia:2010er}
\begin{align}\label{eq:mixingangles:data}
& 
\theta_{12}\,=\ (34.4\pm 1.0)^\circ , 
\quad 
\theta_{23}\,=\, (42.8\,  ^{+ 4.7}_{-2.9} )^\circ ,
\quad 
\theta_{13}\,=\, (5.6\,  ^{+ 3.0}_{-2.7} )^\circ \,(\leq 12.5^\circ),
\end{align}
while for the mass-squared differences one has
\begin{align}\label{eq:lightmasses:data}
& 
\Delta\, m^2_\text{21} \,=\,(7.6\, \pm 0.2)\,\times 10^{-5}\,\,\text{eV}^2\,,
\quad 
\Delta \, m^2_\text{31} \,=\left\{ \begin{array}{l} \,(-2.36\, \pm \
    0.11)\,\times 10^{-3}\,\,\text{eV}^2\,\\  
\,(+2.46\, \pm \ 0.12)\,\times 10^{-3}\,\,\text{eV}^2\ 
 \end{array}\right. \,,
\end{align}
where the two ranges for $\Delta \, m^2_\text{31}$ correspond to
normal and inverted neutrino spectrum.

A convenient means of parametrizing the neutrino Yukawa couplings,
while at the same time allowing to accommodate the experimental data,
is given by the Casas-Ibarra parametrization~\cite{Casas:2001sr},
which reads at the seesaw scale $M_N$
\begin{equation}\label{eq:seesaw:casas}
Y^\nu v_2=m_D^\nu \,=\, i \sqrt{M^\text{diag}_N}\, R \,
\sqrt{m^\text{diag}_\nu}\,  {U^\text{MNS}}^{\dagger}\,.
\end{equation}
In the above $R$ is a complex orthogonal $3 \times 3$ matrix that
encodes the possible mixings involving the right-handed neutrinos, in
addition to those of the low-energy sector
(i.e. $U^{\text{MNS}}$). $R$ can be parametrized in terms of three 
complex angles $\theta_i$ $(i=1,2,3)$ as
\begin{equation}\label{eq:Rcasas}
R\, =\, 
\left( 
\begin{array}{ccc} 
c_{2}\, c_{3} & -c_{1}\, s_{3}\,-\,s_1\, s_2\, c_3
& s_{1}\, s_3\,-\, c_1\, s_2\, c_3 \\ 
c_{2}\, s_{3} & c_{1}\, c_{3}\,-\,s_{1}\,s_{2}\,s_{3} 
& -s_{1}\,c_{3}\,-\,c_1\, s_2\, s_3 \\ 
s_{2}  & s_{1}\, c_{2} & c_{1}\,c_{2}
\end{array} 
\right)\,,
\end{equation}
with $c_i\equiv \cos \theta_i$, $s_i\equiv \sin\theta_i$.  Before
advancing, it is worth commenting that out of the 18 parameters
involved in the seesaw (as readily verified from either side of
Eq.~(\ref{eq:seesaw:casas})), in practice only the degrees of freedom
related to the light neutrinos (masses, leptonic mixings angles, and
potentially 2 of the 3 CPV phases) can be effectively reconstructed
from low-energy data and cosmological observations.  Unless the seesaw
scale is very low, in which case $Y^\nu$ is very small, this implies
that the dynamics of the right-handed neutrino sector is unreachable,
and may only be indirectly probed.

\subsection{Radiative LFV in the slepton sector}

In the presence of mixings in the lepton sector,  
$Y^\nu$ is clearly non-diagonal in flavour space. At the seesaw scale 
$Y^\nu$ satisfies Eq.~(\ref{eq:seesaw:casas}), and
the running from $M_X$ down to the seesaw  scale will induce flavour mixing
in the otherwise (approximately) flavour conserving SUSY breaking
terms. The low-energy parameters are obtained by solving the full set
of renormalisation group equations (RGEs), which include additional
terms and equations due to the extended neutrino and sneutrino 
sectors. In our work, the running is carried in several steps: 
the full set of equations is first run down from
$M_X$ to the seesaw scales; below the seesaw threshold, after 
the right-handed neutrinos (and sneutrinos) decouple, the new RGEs are
then run down to EW scale, where the low-energy mass
matrices and couplings are finally computed.

Due to the mixing induced by the RGE running in the slepton mass matrices, 
 at low energies, the charged slepton squared mass matrix,
$M_{\tilde{l}}^2$, can be decomposed in four blocks ($LL$, $RR$, $LR$
and $RL$) whose elements are given by (see, e.g.~\cite{Martin:1997ns}) 
\begin{align}\label{eq:slepton:mass6}
M_{LL}^{ij \, 2} & \,=  \,
m_{\tilde{L}, ij}^2 \, + \, v_1^2  \,\left( {Y^l}^{\dagger} \, Y^l 
\right)_{ij} \, + \, 
M_Z^2  \,\cos 2 \beta \, \left(-\frac{1}{2} \,+ \, \sin^2 \theta_{W}
\right)  \,  \delta_{ij} \,, \nonumber \\
M_{RR}^{ij \, 2} & \,=  \,
m_{\tilde{E}, ij}^2 \, + \, v_1^2 \, \left( {Y^l} \, {Y^l}^{\dagger}
\right)_{ij} \, -  \, 
M_Z^2 \, \cos 2 \beta  \,\sin^2 \theta_{W}  \,\delta_{ij} \,, \nonumber \\
M_{LR}^{ij \, 2} & \,=  \,
v_1  \,\left({{A_l}^{\dagger}}\right)_{ij}  \,- \,
v_2\,\mu \, {Y^l}^ \dagger_{ij} \,, 
\nonumber \\
M_{RL}^{ij \, 2} & \,=  \,\left(M_{LR}^{ji \, 2}\right)^{*} \, ,
\end{align}
where $M_Z$ is the $Z$-boson mass and $\theta_W$ the weak mixing
angle.
The low-energy sneutrino mass eigenstates are dominated by the 
$\tilde \nu_L$ components~\cite{Grossman:1997is} 
(the right-handed sneutrinos having decoupled at the seesaw scale),
and are described by the following mass matrix:
\begin{equation}\label{eq:sneutrino:mass3}
(M_{\tilde{\nu}}^2)_{ij} 
\,=\,
m_{\tilde{L}, ij}^2  + \frac{1}{2}\, M_Z^2 \,\cos 2 \beta \, 
\delta_{ij}\,.
\end{equation} 

Although in the numerical studies of Section~\ref{sec:results} a full
2-loop RGE evaluation is conducted, a useful analytical estimation of
the amount of flavour violation induced from RGE running  on the
slepton mixing matrices can be obtained using the leading logarithmic
approximation (LLog). At leading order, one has the following
radiative corrections to the soft slepton mass matrices entering in
Eqs.~(\ref{eq:slepton:mass6}, \ref{eq:sneutrino:mass3}):
\begin{align}\label{eq:slepton:RGE:LLog}
& (m_{\tilde{L}}^2)_{ij} \, =\,\left(
m_0^2 + 0.5\,M_{1/2}^2-  m_0^2 \,|y|\,({Y^l})^2_{ij}  \right)\,
\delta_{ij} + (\Delta m_{\tilde{L}}^2)_{ij} \,, \nonumber \\
& (m_{\tilde{E}}^2)_{ij} \, =\,\left(
m_0^2 + 0.15\,M_{1/2}^2-  2 \ m_0^2 \,|y|\,({Y^l})^2_{ij}    \right)\,
\delta_{ij} + (\Delta m_{\tilde{E}}^2)_{ij} \,, 
\end{align}
with
\begin{equation}\label{eq:slepton:RGE:y}
|y|\, \approx\,\frac{1}{8 \pi^2}\, \left(3+\frac{A_0^2}{m_0^2}
\right)\, \log(\frac{M_X}{m_\text{SUSY}}) 
\end{equation} 
where $m_\text{SUSY}$ represents a generic (average)
SUSY mass, 
and where the terms $\Delta m^2$ and also the
correction to the trilinear coupling, $\Delta A_l$, are only present 
for non-vanishing neutrino Yukawa couplings:
\begin{align}\label{eq:LFV:LLog}
(\Delta m_{\tilde{L}}^2)_{ij}&\,=\,
-\frac{1}{8\, \pi^2}\, (3\, m_0^2+ A_0^2)\, ({Y^{\nu}}^\dagger\, 
L\, Y^{\nu})_{ij} 
\,,\nonumber \\
(\Delta A_l)_{ij}&\,=\,
- \frac{3}{16 \,\pi^2}\, A_0\, Y^l_{ij}\, ({Y^{\nu}}^\dagger\, L\, Y^{\nu})_{ij}
\,,\nonumber \\
(\Delta m_{\tilde{E}}^2)_{ij}&\,=\,
0\,\,;\, L_{kl}\, \equiv \,\log \left( \frac{M_X}{M_{N_k}}\right) \,
\delta_{kl}\,.
\end{align}
These terms can give rise to flavour mixing in the slepton mass
matrix, originated by the running from $M_X$ to the right-handed
threshold $M_N$. The amount of flavour violation is
encoded in the matrix elements $({Y^\nu}^\dagger L Y^\nu)_{ij}$ of
Eq.~(\ref{eq:LFV:LLog}), which can be related to high- and 
low-energy neutrino parameters using Eq.~(\ref{eq:seesaw:casas}).

As can be seen from the above equations, the RGE corrections have an
impact regarding both flavour non-universality and flavour violation
in the charged slepton sector: 
(i) the charged lepton Yukawa couplings (in particular $Y^\tau$)
induce flavour non-universality, i.e. a splitting between the soft
masses of the third and the first two slepton generations (the latter
remaining approximately degenerate);
(ii) the neutrino Yukawa couplings contribute to both flavour
non-universality and flavour violating effects. Due to the underlying
seesaw mechanism, the $Y^\nu$ can be sizable (even $\mathcal{O}(1)$), 
so that the associated RGE corrections can be important.
From the previous equations it is also manifest that $LR$ mixing is
only significant for the third generation ($\tau$).

The physical masses and states are obtained by diagonalizing the
previous mass matrices, leading to
\begin{align}\label{eq:slepton:Rmatrix}
{M_{\tilde l}^2}^\text{diag} & \,=\, 
R^{\tilde l}  \,M_{\tilde l}^2  \,R^{\tilde l\,\dagger} \, = \,
\text{diag} \,(m_{\tilde l_1}^2,..,m_{\tilde l_6}^2) \,,
\nonumber \\
{M_{\tilde \nu}^2}^\text{diag}  & \,=  \,
R^{\tilde \nu}  \,M_{\tilde \nu}^2  \,R^{\tilde \nu\,\dagger} \, = \,
\text{diag}\,(m_{\tilde \nu_1}^2, \,
m_{\tilde \nu_2}^2, \,m_{\tilde \nu_3}^2)\,,
\end{align}
where $R^{\tilde l}$ and $R^{\tilde \nu}$ are unitary ($6\times 6$ and
$3\times 3$, respectively) rotation matrices.

\subsection{Low energy LFV observables}

The exact formulae for the branching ratios of the radiative and
three-body LFV lepton decays can be found in~\cite{Raidal:2008jk}, and are
incorporated in the {\sc SPheno} code~\cite{Porod:2003um} 
 used for the numerical analysis. 

Radiative decays $\ell_i \to \ell_j \gamma$ receive contributions from
sneutrino-chargino and slepton-neutralino loop. However, a simple and
illustrative expression can be obtained using the LLog approximation:
since the dominant contribution to the 
transitions stems from the RGE induced flavour violating entry
$(\Delta m_{\tilde{L}}^2)_{ij}$, one has
\begin{equation}\label{eq:BR:MIA:LL}
\frac{
\text{BR}(\ell_i \to \ell_j \,\gamma)
}{
\text{BR}(\ell_i \to \ell_j \,\nu_i\, \bar \nu_j)
}\,=\, 
\frac{\alpha^3\, \tan^2 \beta}{G_F^2\, m_\text{SUSY}^8}\,
\left|
\frac{1}{8\,\pi^2}\, \left(3\, m_0^2+ A_0^2\right)\, \left({Y^{\nu}}^\dagger\, 
L\, Y^{\nu}\right)_{ij} 
\right|^2\,,
\end{equation}
where $G_F$ is the Fermi constant, $\alpha$ the electromagnetic
coupling constant.

The full computation of the three-body decays $\ell_i \to 3 \ell_j$ 
 includes photon-, $Z$- and Higgs-penguins as well as box diagrams.
Since the dominant contribution is found to originate from the
photon-penguin diagrams as occurs in the case of the radiative
decays~\cite{Brignole:2004ah,Arganda:2005ji}, the BR for the $\ell_i
\to 3 \, \ell_j$ decay can be approximately related to that of the
radiative decay as follows:
\begin{equation}\label{eq:BR3body:BRrad}
 \text{BR}(\ell_i \to 3 \ell_j)\,=\,
 \frac{\alpha}{3\,\pi}\,
\left(\log\frac{m_{l_i}^2}{m_{l_j}^2}\,-\,\frac{11}{4}\right)\,
 \times\,
\text{ BR}(\ell_i \to \ell_j\, \gamma)\, .
\end{equation}           

From Eqs.~(\ref{eq:BR:MIA:LL}, \ref{eq:BR3body:BRrad}) it is
straightforward to derive the dependence of the observables on the
relevant SUSY parameters. The impact of the seesaw parameters
(right-handed neutrino masses, light neutrino mass hierarchy, $R$-matrix
angles and $\theta_{13}$) on the BRs has been studied
in~\cite{Antusch:2006vw}, and can be analytically understood from
explicitly writing $({Y^{\nu}}^\dagger\, L\, Y_{\nu})_{ij}$, using
Eq.~(\ref{eq:seesaw:casas}).

Equally interesting LFV observables are $\mu-e$ conversions in heavy
nuclei such as aluminium, gold or titanium (for detailed discussions
see, e.g.~{\cite{Kitano:2002mt}). In the limit of  
photon-penguin dominance, the conversion rate CR($\mu-e$) 
in nuclei and BR($\mu \to e \gamma$) are strongly correlated, 
since both observables are sensitive to the same leptonic mixing
parameters~\cite{Arganda:2007jw}. 
Typically, the SUSY seesaw predictions regarding the conversion rates
are smaller than BR($\mu \to e \gamma$) by approximately two orders of
magnitude (the actual factor depending on the mSUGRA parameters and on
the properties of the muonic nucleus)~\cite{Yaguna:2005qn}.  However,
and although significant improvements are expected regarding the
experimental sensitivity to $\mu \to e \gamma$ ($ <
10^{-13}$~\cite{Kiselev:2009zz}), the most challenging experimental
prospects arise for the CR($\mu-e$) in heavy nuclei such as titanium
or gold. The possibility of lowering the sensitivities to values as
low as $ \sim10^{-18}$ renders this observable an extremely powerful
probe of LFV in the muon-electron sector.

We summarise in Table~\ref{table:LFV:bounds} the current bounds 
on the above discussed LFV observables, as well as the future 
sensitivity of dedicated experimental facilities.\\

\begin{table}[h!]
\begin{center}
\begin{tabular}{|l|c r|c r|}
\hline
LFV process & Present bound & & Future sensitivity & \\
\hline
BR($\mu \to e \gamma$) & $1.2 \times 10^{-11}$&
\cite{PDG}&$10^{-13} $& \cite{Kiselev:2009zz} \\ 
BR($\tau \to e \gamma$) &$1.1 \times 10^{-7}$ & \cite{Aubert:2005wa}&
$ 10^{-9}$& \cite{Bona:2007qt} \\ 
BR($\tau \to \mu \gamma$) & $4.5 \times 10^{-8}$&
\cite{Hayasaka:2007vc}&$ 10^{-9}$ & \cite{Bona:2007qt} \\ 
\hline
BR($\mu \to 3 e $) &$1.0 \times 10^{-12}$ &
\cite{PDG}&
&  \\ 
BR($\tau \to 3 e $) & $3.6 \times 10^{-8}$& \cite{PDG}&$2
\times 10^{-10}$ & \cite{Bona:2007qt} \\ 
BR($\tau \to 3 \mu$) & $3.2 \times 10^{-8}$& \cite{PDG}&$2
\times 10^{-10}$ & \cite{Bona:2007qt} \\ 
\hline
CR($\mu-e$, Ti) & $4.3 \times 10^{-12}$& \cite{PDG}&
${\cal{O}}(10^{-16})$ (${\cal{O}}(10^{-18})$)&
\cite{Glenzinski:2010zz}~(\cite{Cui:2009zz}) \\ 
CR($\mu-e$, Au) & $7 \times 10^{-13}$& \cite{PDG}& 
& 
 \\
\hline 
CR($\mu-e$, Al) && & ${\cal{O}}(10^{-16})$&  \cite{Cui:2009zz}\\\hline
\end{tabular}
\end{center}
\caption{Present bounds and future sensitivities for several LFV
  observables discussed in the text.}
\label{table:LFV:bounds}
\end{table}

\subsubsection{Lepton electric dipole moments}

The bounds on the LFV BRs mostly constrain the source of mixing
(i.e.~off-diagonal elements) while the bounds on the lepton EDMs
constrain the flavour-conserving CP-violating phases. Notice that CP
violation in the lepton sector is also a consequence of the seesaw.
Both low and high-energy CPV phases will give rise to complex soft
breaking terms, potentially contributing to charged lepton EDMs. The
present upper bound on the electron (muon) EDM is $1.4\times10^{-27}$
($7.1\times10^{-19}$) e cm~\cite{PDG} while the future experiments are
expected to reach a sensitivity of $10^{-31}$ e cm for the electron
EDM~\cite{DeMille:2006wy} and $10^{-24}$ e cm for the muon
EDM~\cite{EDM-JPARC}.

\subsection{Implications of the SUSY seesaw for thermal leptogenesis} 
   
As mentioned in the Introduction, in addition to explaining the
smallness of neutrino masses, the seesaw can also provide an
interesting explanation to the observed baryon asymmetry of the
Universe. The minimal thermal leptogenesis
scenario~\cite{Fukugita:1986hr} (for a recent review,
see~\cite{Davidson:2008bu}) is based on the type-I seesaw mechanism,
consisting of the SM extended by 2 or 3 right-handed (RH) Majorana
neutrinos with hierarchical masses, which can be easily generalized to
supersymmetric extensions of the SM.  In these scenarios, the lightest
RH neutrino $N_1$, produced in the thermal bath after inflation by
inverse decays and scatterings, decays through out-of-equilibrium
processes that violate lepton number, C and CP symmetries. These
processes induce a dynamical production of a lepton asymmetry, which
can be later converted into a BAU through (B+L)-violating sphaleron
interactions. In supersymmetric scenarios, the constraint from the
reheating temperature $T_{\text{RH}}$ (arising from the so-called
gravitino problem~\cite{Ellis:1984eq}) already sets an upper bound on
the mass of the lightest RH neutrino. Assuming an optimal washout
(efficiency) and a successful BAU leads in turn to the following
interval (lower bound) on $M_{N_1}$, $M_{N_1} \simeq 10^9 \text{ GeV}-
10^{10}$ GeV~\cite{Antusch:2006gy}.

In order to ensure that CP is indeed violated (via interference
between loop and tree level decays), the neutrino Yukawa couplings
have to be complex and CP violation is encoded in the $R$ and
$U^\text{MNS}$ matrices ($Y^\nu={i\over v_2}
\sqrt{M_{N}^\text{diag}}\,R\,  \sqrt{m_\nu^\text{diag}} \
{{U^\text{MNS}}^{\dagger} }$). 
It has been recently shown that a correct formulation of the lepton
asymmetry should be done considering each flavour separately (the
number of distinguishable lepton flavours depending on the energies at
which leptogenesis occurs)~\cite{Nardi:2006fx, Abada:2006ea,
Abada:2006fw}.  Having flavours play an important r\^ole in
leptogenesis also means that both low- and high-energy CPV phases
contribute to the CP asymmetry; however, the flavoured BAU can be
accounted for exclusively with $R$ phases (for any value of
$\theta_{13},\ \delta,\ \varphi_{1},\ 
\varphi_{2}$)~\cite{Davidson:2008pf, Antusch:2006gy} - 
in other words even if the $U^\text{MNS}$ phases are measured, 
the BAU can have any value.

Of course one cannot use a successful leptogenesis requirement to
derive constraints on the CP violating sources, since leptogenesis is
not an observable (contrarily to EDMs and LFV widths).  However, one
can have an idea about the range of variation of the complex angles
$\theta_i$ of the $R$ matrix that succeed in accounting for a viable
(flavoured) leptogenesis (see for instance~\cite{Antusch:2006gy},
where it has been shown that although all the three complex angles
enter the flavoured CP-asymmetry, the r\^ole of $\theta_1$ is
indirect, manifest via increasing (decreasing) the
$\theta_2$$-$$\theta_3$ parameter space associated with a BAU
compatible with current observation).

The interplay of LFV  (and EDMs)  and  leptogenesis in constraining
a type-I SUSY seesaw has been addressed, for 
instance, in~\cite{Ellis:2002xg, Petcov:2006pc, Joaquim:2007sm,
  Davidson:2008pf, Blanchet:2010td}.

Although in the numerical analysis of Section~\ref{sec:results} we 
will conduct general surveys of the seesaw parameter space, 
we will also consider the following leptogenesis inspired ranges for
the $R$ matrix complex angles:  
$\operatorname{Re}(\theta_2) , \operatorname{Re}(\theta_3)\in
[-\pi/4,0[ \cup ]0, +\pi/4]$ (for example). Complying with the
(severe) reheating temperature constraint suggests that the arguments
of the latter complex angles should have modulus in the range 
$[\pi/16, \pi/4]$. This corresponds to a conservative choice of 
volume in the $\theta_2$$-$$\theta_3$ parameter space.

\section{LFV at the LHC: slepton mass 
splittings and flavour violating $\chi_2^0$ decays}\label{lfv:lhc}

As mentioned in Section~\ref{sec:LHC}, the different experiments at
the LHC have the potential to measure with high precision the
kinematical edges of the di-lepton invariant mass spectrum, so that
one can potentially study the slepton mass differences.
In what follows we discuss the different sources of slepton mass
splittings, and also derive, for some simple limiting cases,
approximate relations for $\frac{\Delta m_{\tilde \ell}}{m_{\tilde
\ell}} (\tilde \ell_i, \tilde \ell_j)$.

\subsection{Charged slepton mass differences in the type-I SUSY
  seesaw} 
Within the cMSSM, and in the absence of flavour mixing angles, there
are only two sources of non-universality for the masses of left- and
right-handed sleptons: 
(i) RGE effects proportional to $({Y^l})^2_{ij}$  (see
Eqs.~(\ref{eq:slepton:RGE:LLog})); (ii) $LR$ mixing effects, also 
proportional to the lepton masses ($ m^l_{i}\ \tan \beta $).  The
mass difference of the first generations of sleptons is thus extremely
small: neglecting RGE corrections, and considering only $LR$ mixing
for the smuons, the mass splitting between the left-handed selectron and
the heaviest smuon is approximately given by
\begin{equation}\label{eq:MS:emu:cMSSM}
\frac{\Delta m_{\tilde \ell}}{m_{\tilde \ell}} (\tilde e_L, \tilde\mu_L)
\, \approx \, 
\frac{m^2_\mu}{2 m_{\tilde \ell}^2 }\,\,\left|
\frac{(A_0-\mu \tan \beta)^2}{\,0.35\  M_{1/2}^2 +M_Z^2 \cos 2 \beta
  (-1/2+2 \sin^2 \theta_W)}\right|\,, 
\end{equation}
where $m_{\tilde \ell}$ denotes an averaged slepton mass, in this case
$m_{\tilde \ell} \approx 1/2 (|m_{\tilde L}^2)_{_{11}}|^{1/2}+
|(m_{\tilde L}^2)_{_{22}}|^{1/2})$.
The cMSSM mass differences between the first two families are thus 
extremely small implying that, to a large extent, the left- and
right-handed selectrons and smuons are nearly degenerate, the mass
splitting typically lying at the per mille level.

For the stau sector, $LR$ mixing effects and loop contributions are
more important and to a very good approximation, the mass difference
of the heaviest (mostly left-handed) stau and left-handed smuon is
related to that of $\tilde e_L- \tilde \mu_L$ as
 \begin{equation}\label{eq:MS:emu:mutau:cMSSM}
\frac{\Delta m_{\tilde \ell}}{m_{\tilde \ell}} (\tilde e_L, \tilde\mu_L)
\, \approx \,
\frac{m_\mu^2}{m_\tau^2}\, 
\frac{\Delta m_{\tilde \ell}}{m_{\tilde \ell}} (\tilde \mu_L, \tilde \tau_2)\,.
\end{equation}

\bigskip
When mixings are present in the lepton sector, flavour violation also
occurs in the slepton sector. The radiative corrections introduced by
the neutrino Yukawa couplings
induce both flavour conserving and flavour violating  contributions to
the slepton soft masses: in addition to generating LFV effects, the
new terms proportional to $Y^\nu$ will also break the approximate
universality of the first two generations. An augmented mixing between
$\tilde e $, $\tilde \mu$ and $\tilde \tau$ translates into larger 
mass splittings for the mass eigenstates. In particular, as 
noticed in~\cite{Buras:2009sg}, large mixings involving the third generation
can lead to sizable values of the mass splitting between slepton mass
eigenstates, while avoiding the stringent BR($\mu \to e \gamma$) constraint.

In the presence of seesaw-induced contributions to the distinct
$(\Delta m_{\tilde L}^2)_{ij}$ and $(\Delta A_{l})_{ij}$, see
Eqs.~(\ref{eq:LFV:LLog}), an analytical approach to the problem
becomes extremely complicated.  Even neglecting $LR$ mixings for the
two first generations, a numerical diagonalization is required to
obtain the different mass eigenstates, as given in
Eqs.~(\ref{eq:slepton:Rmatrix}). However, one can consider
interesting limiting cases that provide useful information (and also
help in understanding the numerical analysis of Section~\ref{sec:results}). 
Disentangling $LR$- from RGE-induced mixings, one
then has for the mass difference $\tilde \ell_i - \tilde \ell_j$
\begin{equation}\label{eq:MS:ij:seesaw}
\frac{\Delta m_{\tilde \ell}}{m_{\tilde \ell}} (\tilde \ell_i, \tilde \ell_j)
\approx 
\frac{1}{2 m_{\tilde \ell}^2 }\,\left|
\frac{m^2_i\,(A_0-\mu \tan \beta)^2}{\,0.35 M_{1/2}^2 +M_Z^2 \cos 2 \beta
  (-1/2+2 \sin^2 \theta_W) \, + (\Delta m_{\tilde L}^2)_{ii}}
\pm 2\,|(\Delta m_{\tilde L}^2)_{ij}|
\right|,
\end{equation}
where $m_i$ denotes the mass of the heaviest lepton and where we have
again neglected the RGE contributions proportional to the charged
lepton Yukawa coupling.

If the seesaw scale is sufficiently high, 
large values of the neutrino Yukawa couplings are possible, and 
hence large off-diagonal entries can be generated. Assuming that a
particular $(\Delta m_{\tilde L}^2)_{ij}$ constitutes the dominant
source of LFV, one can approximate Eq.~(\ref{eq:MS:ij:seesaw}) as
\begin{equation}\label{eq:MS:ij:seesaw:approx}
\frac{\Delta m_{\tilde \ell}}{m_{\tilde \ell}} (\tilde \ell_i, \tilde \ell_j)
\,\approx \left|
\frac{(\Delta m_{\tilde L}^2)_{ij}}{(m_{\tilde L}^2)}
\right|\, .
\end{equation}
In particular, large flavour violating entries 
involving the second and third generation can be easily induced. 
In this case, and further assuming that the stau mass eigenstates are
strongly dominated by either the left- or the right-handed state, the
diagonalization of the $\tilde \mu_L-\tilde \tau_2$ mixing matrix
(for non-vanishing $(\Delta m_{\tilde L}^2)_{23}$) leads to the
following approximate relation 
\begin{equation}\label{eq:seesaw:DM23}
\frac{\Delta m_{\tilde \ell}}{m_{\tilde \ell}} (\tilde \mu_L, \tilde
\tau_2)
\,\approx \left|
\frac{(\Delta m_{\tilde L}^2)_{23}}{(m_{\tilde L}^2)_{33}}
\right|\, ,
\end{equation}
where the quantities on the right-hand side can be found in 
Eqs.~(\ref{eq:slepton:RGE:LLog}- \ref{eq:LFV:LLog}), and where
we have also neglected cMSSM-like mass differences 
$\sim \mathcal{O} ((m_{\tilde  L}^2)_{22} - (m_{\tilde L}^2)_{33})$.
Rewriting the left-handed smuon mass in terms of the above mass
splitting further allows to relate the $\tilde e_L-\tilde \mu_L$
and the $\tilde \mu_L-\tilde \tau_2$ mass differences
in the $R=1$ seesaw limit
\begin{equation}\label{eq:MS:mutau:emu:R1seesaw}
\frac{\Delta m_{\tilde \ell}}{m_{\tilde \ell}} (\tilde e_L, \tilde
\mu_L) \, \approx \, \frac{1}{2}\,
\frac{\Delta m_{\tilde \ell}}{m_{\tilde \ell}} (\tilde \mu_L, \tilde
\tau_2)\,. 
\end{equation}

Although one can derive approximate relations that translate the
dependence of the mass splittings on the mSUGRA parameters, it is
important to stress that the conditions
to ensure that the slepton masses can indeed be reconstructed 
(see Section~\ref{sec:LHC}) imply that  mSUGRA parameters cannot
be independently varied. Under the approximations above referred,
one can nevertheless obtain a simple illustrative expression for the 
mass splittings, which we write below for the case of $\tilde
\mu_L-\tilde \tau_2$ 
\begin{equation}\label{eq:seesaw:DM23:open}
\frac{\Delta m_{\tilde \ell}}{m_{\tilde \ell}} (\tilde \mu_L, \tilde
\tau_2)
\,\approx \,
\frac{1}{8 \pi^2}\, \frac{L_{33}\,M_{N_3}}{v^2 \sin^2 \beta}\,
\frac{3 m_0^2+A_0^2}{m_0^2+0.5 M_{1/2}^2} \,
 \left|\sum_{ij} U_{2i}^\text{MNS} {U_{3j}^\text{MNS}}^* R^*_{3i}
   R_{3j} \sqrt{m_{\nu_i} m_{\nu_j}}\right|\, . 
\end{equation}
In the above equation, we have considered a strongly hierarchical right-handed
neutrino spectrum, only keeping the contribution associated with the
heaviest state $N_3$.

It is also interesting to investigate the relation between two flavour
violating observables strongly affected by the same LFV entry. For
instance, let us again consider $\tilde \mu_L-\tilde \tau_2$ mass
splittings and the BR($\tau \to \mu \gamma$). Comparing the previous
expression with Eq.~(\ref{eq:BR:MIA:LL}), one has, in the limit 
where $\theta_{13} \approx 0$ and $R=1$,
\begin{equation}\label{eq:seesaw:BR32:DM23}
\frac{
\text{BR}(\tau\to\mu \gamma) 
}{
\text{BR}(\tau \to \mu \,\nu_\tau\, \bar \nu_\mu)
}
\,\approx\,
\frac{\alpha^3}{16\,\pi^2\,G_F^2}\,
\frac{m_0^2+ 0.5 M_{1/2}^2}{v^2\,\cos^2 \beta\,
  m_\text{SUSY}^8}
\left(3\, m_0^2+ A_0^2\right)\, L_{33}\,M_{N_3}\,m_{\nu_3}
\,\sin 2\theta_{23}\,\times \,
\frac{\Delta m_{\tilde \ell}}{m_{\tilde \ell}} (\tilde \mu_L, \tilde
\tau_2)
\,.
\end{equation}

\bigskip
Finally, it is important to stress that depending on the amount of
flavour violation, a type-I SUSY seesaw can lead to scenarios where two 
non-degenerate mass eigenstates have almost identical flavour content
(maximal flavour mixing). As an example, one can have mass eigenstates
whose composition is approximately given by
\begin{equation}
\tilde \ell_{i,j}\,  \sim (\sqrt{2}/2 +\varepsilon )\,  \tilde \mu_L \, 
\pm\,  (\sqrt{2}/2 - \varepsilon)\, \tilde \tau_L \, + \, 
\varepsilon \tilde \tau_R\, , \nonumber
\end{equation}
where $\varepsilon$ ($\varepsilon \ll 1$) accounts  for the $LR$
mixing. To correctly
interpret a mass splitting between sleptons with quasi-degenerate
flavour content (QDFC), one has to introduce an ``effective'' mass
\begin{equation}\label{eq:effective:mass}
m^{\text{(eff)}}_i \equiv \sum_{X = \tilde{\tau}_2 \, , \tilde{\mu}_L \, ,
\tilde{e}_L} m_{\tilde{l}_X} \left( | R^{\tilde{l}}_{X i_L} |^2 + |
R^{\tilde{l}}_{X i_R} |^2 \right)\ , 
\end{equation}
which in turn provides the notion of ``effective'' mass splitting, 
\begin{equation}\label{eq:effective:MS}
\left( \frac{\Delta m}{m} \right)^{\text{(eff)}}(\tilde{l}_i,
\tilde{l}_j)\, \equiv \,\frac{\,2 \,| m^{\text{(eff)}}_i - m^{\text{(eff)}}_j
  |}{m^{\text{(eff)}}_i + m^{\text{(eff)}}_j} \,. 
\end{equation}
For mass splittings involving QDFC and non-QDFC sleptons, one should
then use the ``effective'' mass splittings,
cf. Eq.~(\ref{eq:effective:MS}); in the case where mass
splittings involving two QDFC sleptons or two non-QDFC sleptons, the
real mass splitting (see e.g. Eq.~(\ref{eq:MS:ij:seesaw})) can be employed.

\subsection{Di-lepton invariant masses from flavour violating
  $\chi_2^0$ decays}\label{subsec:di-lepton-cmssm}

In the cMSSM, the decays of the $\chi_2^0$ into a di-lepton final
state $\chi_2^0 \to \ell_i^\pm \,\ell_i^\mp\, \chi_1^0$ are flavour
conserving, implying that if measurable, the kinematical edges of a
di-lepton mass distribution, $m_{\ell_i \ell_i}$ necessarily lead to
the reconstruction of intermediate sleptons of the same flavour,
$\tilde \ell^i_{L,R}$.

SUSY models violating strict lepton flavour symmetry may leave
distinct imprints on the di-lepton mass distribution, depending on
whether the soft-breaking slepton terms are non-universal (but flavour
conserving) or truly flavour-violating.  In the first case, the most
significant effect will be a visible displacement of the kinematical
edges in each 
of the di-lepton distributions: for instance, the edge corresponding
to $\tilde e_L$ in $m_{ee}$ will not appear at the same values as that
of $\tilde \mu_L$ in $m_{\mu \mu}$, thus implying that $m_{\tilde e_L}
\neq m_{\tilde \mu_L}$.
 
The second case will lead to far richer imprints: as discussed in the
previous subsection, flavour violation has the potential to induce
significant mass differences for the sleptons, so that one
should again  observe a relative displacement of the $\tilde \ell_X$ in the
corresponding $m_{\ell_i \ell_i}$ distributions. Nevertheless, the
most striking effect is the appearance of new edges in a given
di-lepton mass distribution: provided there is a large flavour mixing
in the mass eigenstates (and that all the decays are kinematically
viable), one can have
\begin{equation}
\chi_2^0 \to \left\{
\begin{array}{l}
\tilde \ell^i_L\, \ell_i \vspace*{1mm}\\
\tilde \ell^i_R\,\ell_i \vspace*{1mm}\\
\tilde \ell^j_X \,\ell_i
\end{array}
\right\}
\to \chi_1^0 \,\ell_i\, \ell_i 
\end{equation}
so that in addition to the two $\tilde \ell_{L,R}^i$ edges, 
a new one would appear due to the exchange of $\tilde
\ell_{X}^j$.

\section{Numerical results and discussion}\label{sec:results}
We start our analysis by first considering the cMSSM parameter space,
looking for regions where one can fulfil the necessary conditions to
have reasonably large BRs for the decay 
$ \chi_2^0\to \chi_1^0 \ell \ell$, with
sufficiently hard outgoing leptons. After identifying some
representative (benchmark) points, we analyse the prospects for the
LHC (production cross sections and decay rates).  The second part of
our analysis will be devoted to slepton mass splittings and flavour
violation in the  type-I  SUSY seesaw: we briefly discuss the cMSSM case
and then study the different high- and low-energy observables in
the seesaw case. This will also allow to draw some conclusions on the
viability of a type-I SUSY seesaw as the underlying mechanism of LFV.

\bigskip
For the numerical computation, we have used the public code
{\sc SPheno} (v3.beta.47)~\cite{Porod:2003um} to carry out the
numerical integration of the RGEs of the cMSSM (extended by three
right-handed neutrino superfields).  With the exception of light
neutrino data (masses and mixing angles) which is set as a low-energy
input, all the parameters of the model are defined at the GUT
scale. The low-energy parameters are then computed by running first
the full set of RGEs to the seesaw scale, where the boundary
conditions of Eq.~(\ref{eq:seesaw:casas}) are imposed, and at which
the heavy RH neutrinos decouple at their corresponding thresholds. We
notice that we do not take into account separate thresholds for
right-handed neutrinos and sneutrinos, and that we also neglect
$B_\nu$ - see Eq.~(\ref{eq:Lsoftslepton:def}) - which is valid,
provided that one considers $B_\nu \ll M_N$.  Below $M_{N_1}$, the
cMSSM RGEs are run to the EW scale, at which the low-energy Lagrangian
(masses~\footnote{We notice that {\sc SPheno} uses the
$\overline{\text{DR}}$ scheme.  Also, 2-loop RGEs are used for the
running of the slepton masses, while the actual pole masses are
calculated at the one-loop level, with all running parameters set at
the SUSY scale. We have also verified that self-energies (and
associated uncertainties) provide a negligible source of slepton mass
splittings.} and couplings) is
determined and the different observables (such as the LFV 
BRs and CR~\cite{Arganda:2005ji}, as well as lepton EDMs) are computed.  
The dark matter relic density is evaluated using a link to
{\sc{micrOMEGAs}} v2.2~\cite{Belanger:2008sj}.

The production cross sections at LHC operating at c.o.m. energy 
of $7 ~\text{TeV}$ and $14 ~\text{TeV}$ have been computed 
using {\sc{Prospino}}2.1~\cite{Prospino2.1}.
To obtain the di-lepton invariant mass distributions $\frac{d
\Gamma}{d m_{i{j}}}({\chi}^0_2 \to {\chi}^0_1 \, \ell_i \,
{\ell}_j)$, we have used {\sc{Cuba}}'s {\sc{Divonne}}
algorithm~\cite{Hahn:2004fe} to integrate numerically over the
$\ell\!-\!{\chi}^0$ angle in the c.o.m. frame of the two leptons.

In what concerns the experimental constraints applied to the Higgs boson and
sparticle spectrum, we have imposed that all SUSY particles comply
with LEP and Tevatron bounds~\cite{PDG}.  Throughout the analysis, and
except if otherwise stated, we will always be imposing the bound for a
SM-like Higgs boson to the lightest scalar: $m_h \gtrsim 114$
GeV~\cite{Barate:2003sz}.  Finally, the LSP relic density is required to
lie within a $3 \sigma$ interval (extrapolated from WMAP 7-year data
taking~\cite{WMAP}, and  assuming a gaussian distribution):
\begin{equation}\label{exp:dm:wmap}
0.0941\, \lesssim \,\Omega h^2 \,\lesssim \, 0.1277\,.
\end{equation}

\subsection{Di-lepton final states: neutralino production and cascade
  decays in the cMSSM}\label{results:di-lepton:cMSSM}

We begin by studying the cMSSM (without implementing a type-I 
seesaw), looking for regions in the mSUGRA parameter space where 
the  requirements of a ``standard window'' can be met:
\begin{itemize}
\item[(i)] the spectrum is such that the decay chain 
$ \chi_2^0\to \tilde  \ell  \ell \to \chi_1^0 \ell \ell$, with intermediate
  real sleptons, is allowed;
\item[(ii)] it is possible to have sufficiently hard outgoing 
leptons: $m_{\chi_2^0}-m_{\tilde  \ell_L, \tilde \tau_2} > 10$ GeV. 
\end{itemize}
Notice that the above requirements automatically ensure that the
sparticle spectrum complies with current  experimental bounds. 
Once these regions are identified, we then impose the requirements of
a correct relic density, cf. Eq.~(\ref{exp:dm:wmap}). Naturally, 
in order to maximise the prospects for observing  the above
processes at the LHC, the SUSY spectrum should not be
excessively heavy (as to have a sufficiently large production cross
section) and the BRs of the $\chi^0_2$ decay into slepton-lepton
pairs (neutral or charged) also have be large (as to render these decays
observable).  Here we will systematically consider two centre of mass
energies for the LHC, $\sqrt s=7$ TeV and 14 TeV (correspondingly,  
we consider either $\mathcal{L}= 1\ \text{fb}^{-1}$ or
$100\ \text{fb}^{-1}$ for the integrated luminosity~\cite{LHC:Luminosity}).

Before starting the discussion, we remark that throughout the analysis, and
except if otherwise stated, we will always denote the flavour corresponding
to an electron or a muon by $\ell$  
to distinguish it from the $\tau$ flavour (except in inset figure labels). 

\medskip
In Fig.~\ref{fig:msugra.space1} we present the $m_0 - M_{1/2}$
parameter space, for $\mu >0$~\footnote{Throughout the numerical 
analysis we will always be considering positive values
of $\mu$.} and two combinations of $A_0$ and $\tan \beta$.  On the
left we take $A_0=-1$ TeV and $\tan \beta= 10$, while on the right
$A_0=0$ and $\tan \beta = 40$.  All the points presented are in
agreement with current LEP and Tevatron constraints~\cite{PDG} on the
sparticle and Higgs boson spectra, and the region where the LSP relic
density is in agreement with WMAP observations (within a 3$\sigma$
interval, Eq.~(\ref{exp:dm:wmap})) is denoted by a black band across
the parameter space. The excluded (shaded) areas correspond to a
charged LSP and to kinematically disfavoured regimes, while the white
region in the centre corresponds to the requirements of a ``standard
window''. We superimpose the contour lines for BR($ \chi_2^0\to
\chi_1^0 \ell \ell$) and BR($ \chi_2^0\to \chi_1^0 \tau
\tau$). Approximately horizontal lines denote different values of the
production cross section of (at least) one $\chi_2^0$.

\begin{figure}[h!]
\begin{center}
\hspace*{-10mm}
\begin{tabular}{cc}
\epsfig{file=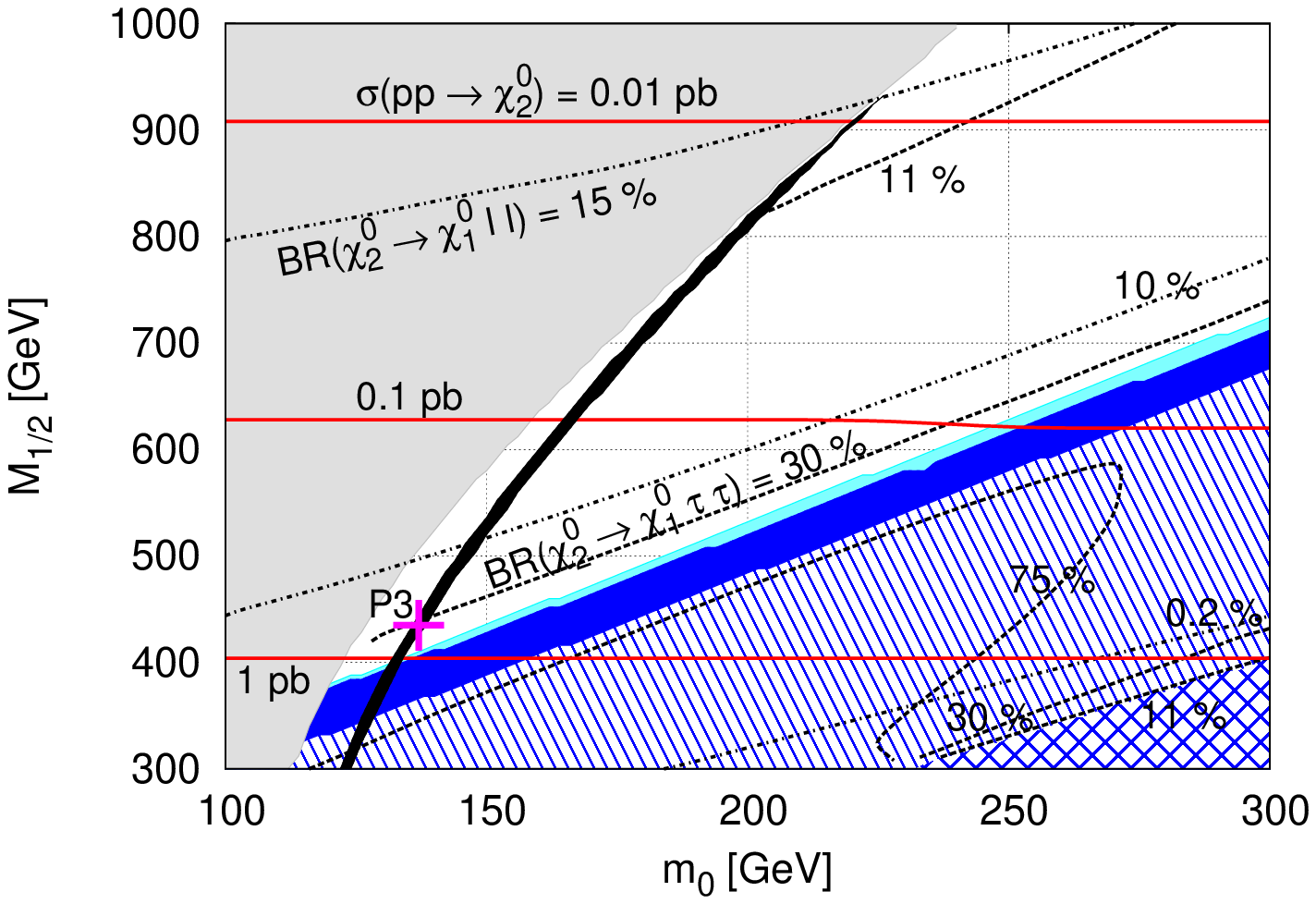, clip=, angle=0, width=85mm}&
\epsfig{file=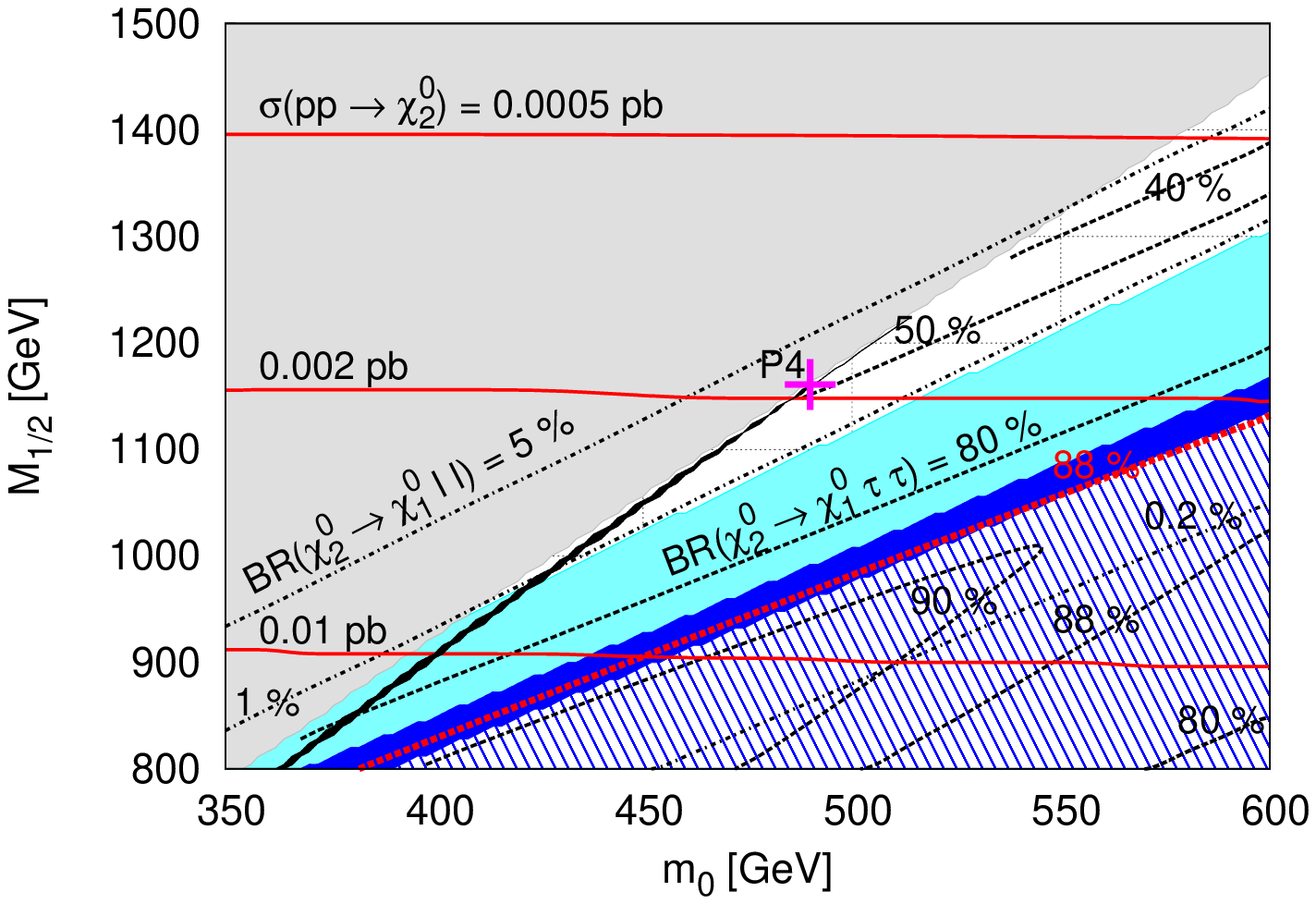, clip=, angle=0, width=85mm}
\end{tabular}
\caption{$m_0-M_{1/2}$ plane (in GeV), for $A_0=-1$ TeV and $\tan
  \beta=10$ (left); the same but with $A_0=0$ and $\tan
  \beta=40$ (right).
  In both figures, the shaded region on the left is excluded due to the
  presence of a charged LSP. The full black region corresponds to a
  WMAP compatible $\chi_1^0$ relic density. 
  Likewise, on the dashed region on the
  bottom, the spectrum does not fulfil the kinematical requirements
  described in the text: the solid regions correspond to having
  $m_{\chi_2^0} < m_{\tilde \ell_L} +10$ GeV (cyan), 
  $m_{\chi_2^0} < m_{\tilde \tau_2} +10$ GeV (blue), 
  $m_{\chi_2^0} < m_{\tilde \ell_L, \tau_2}$ (dashed blue), and 
  $m_{\chi_2^0} < m_{\tilde \tau_1} + m_\tau$ (blue crosses).
  The centre (white) region denotes the parameter space
  obeying the ``standard window'' constraints. 
  The dotted and dashed lines respectively denote isosurfaces for 
  BR($ \chi_2^0\to \chi_1^0 \ell \ell$) and BR($ \chi_2^0\to \chi_1^0
  \tau \tau$). Full red lines denote the contours of $\chi_2^0$
  production cross sections. Superimposed crosses (pink) correspond
  to benchmark  points P3 and P4 (see Table~\ref{table:points1}).  }
\label{fig:msugra.space1}
\end{center}
\end{figure}

The left panel of Fig.~\ref{fig:msugra.space1} corresponds to a
scenario of a relatively light SUSY spectrum (with slepton masses
between 110 GeV and 730 GeV, and 
$230 \text{ GeV} \lesssim m_{\chi_2^0} \lesssim $
805 GeV).  The region compatible with the ``standard window''
constraints is quite large, and the correct LSP relic density can be
easily obtained (the dominant channel being $\chi_1^0 - \tilde \tau_1$
co-annihilation). Having a light neutralino spectrum further implies
that the production cross section of at least one $\chi_2^0$ at the
LHC (with $\sqrt s =14$ TeV, via direct and indirect processes - see
Section~\ref{sec:LHC}) is expected to be $0.01 \text { pb}\lesssim
\sigma (pp \to \chi_2^0) \lesssim 1$ pb.  In the ``standard window'',
the probability of having opposite-sign di-leptons in the final state
ranges between 11\% and 30\% for $\tau \tau$, and between 10\% and
15\% for $\ell \ell$ (i.e.~$ee, \,\mu\mu$) final states. 
It is worth noticing that larger values of BR($ \chi_2^0\to \chi_1^0
\tau \tau$) could be found for smaller $M_{1/2}$, since $\chi_2^0\to
\chi_1^0 \tilde \tau_1 \to \chi_1^0 \tau \tau$ becomes one of the few
kinematically opened decay channels due to heavier LH
sleptons. Nevertheless, no edges would be observable in this regime.
Although the
processes $\chi_2^0\to \chi_1^0 \ell \ell$ and $\chi_2^0\to \chi_1^0
\tau \tau$ are mostly dominated by the exchange of intermediate left-
and right-handed real sleptons, there are other channels leading to
the same final states, e.g. via the direct decay of the $\chi_2^0$
into an LSP and the lightest Higgs boson or the $Z$.
Throughout the experimentally viable $m_0-M_{1/2}$ parameter space,
the BR($\chi_2^0\to \chi_1^0 Z \to \chi_1^0 \tau \tau (\ell \ell)$)
never exceeds the level of 0.03\%, while the BR($\chi_2^0\to \chi_1^0
h$) is at most $\mathcal{O}(12\%)$ inside the ``standard window'',
growing to 25\% when softer outgoing leptons are allowed (solid blue
bands). In turn, this induces a contribution to BR($\chi_2^0\to
\chi_1^0 \tau \tau$) ranging from 1.3\% to 3\%.

On the right panel of Fig.~\ref{fig:msugra.space1}, we illustrate the
parameter space for larger values of $\tan \beta$ (and a heavier
spectrum). 
Having a substantially heavier gaugino and squark spectra when
compared to that of the sleptons implies that the available phase
space for $\chi_2^0$ decays is much enlarged so that one 
can have sizable BR($\chi_2^0\to \chi_1^0 \tau \tau$).
However, the region strictly corresponding to the requirements of a
``standard window'' is somehow smaller, despite having the increased
$LR$ mixing compensated by heavier gauginos. As in the case of lower
$\tan \beta$, intermediate $h$ and $Z$ states only marginally
contribute to the final BRs.  Finally, as expected from the
significantly heavier SUSY spectrum, the production of at least one
$\chi_2^0$ at the LHC has a cross section that now varies between $5
\times 10^{-4}$ pb and 0.01 pb.

From Fig.~\ref{fig:msugra.space1}, we extract two points in mSUGRA
parameter space that we will use in the analysis of the slepton mass
splittings (especially when studying the SUSY seesaw).  
Thus, points P3 and P4 (superimposed on the left and right panels, 
respectively) are points which in addition to complying 
with observational and experimental
constraints, and being inside the corresponding ``standard window'',
also have sizable BR($\chi_2^0\to \chi_1^0 \ell \ell$) and
BR($\chi_2^0\to \chi_1^0 \tau \tau$).

Other analyses of different regimes in mSUGRA parameter space have led
us to identify two additional points P1 and P2 (with $A_0=0$ and $1$ TeV, 
respectively), whose features complement points P3 and P4. To these
points we further add two LHC benchmark points: P5-HM1 (from
CMS~\cite{Ball:2007zza}) and P6-SU1~\footnote{Although the 
P6-SU1 benchmark point does not fully fulfil  the ``standard window''
requirements, we nevertheless consider it in our analysis, to study 
the flavour prospects of one of the ATLAS benchmark 
points.} (from ATLAS~\cite{ATLAS}). This allows
to establish a connection with the already conducted simulations and
to study the flavour prospects at high energy.  The most important
features of these six points (mSUGRA parameters, spectra, production
cross sections and BRs) are summarised in Tables
\ref{table:points1}-\ref{table:points:BRX2decay}.\\

\begin{table}[h!]
\begin{center}
\begin{tabular}{|c|c|c|c|c|}
\hline
Point & $m_0$ (GeV)& $M_{1/2}$ (GeV)& $A_0$ (TeV)& $\tan \beta$ \\
\hline
P1& 110 & 528& 0& 10\\
\hline 
P2& 110& 471& 1& 10\\
\hline 
P3& 137& 435& -1& 10\\
\hline 
P4& 490& 1161& 0& 40\\
\hline 
P5-HM1~\cite{Ball:2007zza}& 180&850 &0 &10 \\
\hline 
P6-SU1~\cite{ATLAS}& 70& 350& 0& 10\\
\hline 
\end{tabular}
\end{center}
\caption{mSUGRA benchmark points selected for the  LFV
    analysis: $m_0$, $M_{1/2}$ (in GeV) and $A_0$ (in TeV), as well as $\tan
    \beta$. For all points we take $\mu >0$. Points P5-HM1 and P6-SU1 are
    LHC CMS- and ATLAS-proposed benchmark points.
}\label{table:points1}
\end{table}

After summarising the mSUGRA coordinates of each point in
Table~\ref{table:points1}, we present part of the corresponding SUSY 
spectrum on Table~\ref{table:points2}.  Among the six points we find
distinct hierarchies for the slepton sector, which will have an impact
regarding the di-lepton mass distributions:
(a) $m_{\tilde \tau_2} \gtrsim m_{\tilde \ell_L}$;
(b) $m_{\tilde \ell_L}\gtrsim m_{\tilde \tau_2} $. 
For instance, points P1 and P6-SU1 are examples of (a) 
while all the others fall in (b). 
The hierarchy in the right-handed sleptons is always $m_{\tilde
\ell_R}\gtrsim m_{\tilde \tau_1}$,  the stau being the NLSP. A common feature
to all these proposed points (and an indirect consequence of the
``standard window'') is that the correct relic density of the LSP is
always obtained from $\tilde \tau_1 - \chi_1^0$ co-annihilation,
as already noticed in {\cite{Carquin:2008gv, Carvalho:2002jg,
Buras:2009sg}. We also notice that P2 and P6-SU1 lead to a value of $m_h
\sim 111$ GeV using the {\sc SPheno} code (which is still in agreement
with data if one allows for  a theoretical error of $\pm 3$
GeV~\cite{Allanach:2004rh}).\\

\begin{table}[h!]
\begin{center}
\begin{tabular}{|c|c|c|c|c|c|c|c|c|}
\hline
Point \phantom{\large$y_{\chi_p}$}& $m_{\chi_2^0}$ & $m_{\chi_1^0}$ &
$m_{\tilde \ell_L}$  & $m_{\tilde \ell_R}$  &  
$m_{\tilde \tau_2}$ & $m_{\tilde \tau_1}$  & $<m_{\tilde q}>$ & $m_h$  \\
\hline
P1 & 410 & 217 & 374 & 231 & 375& 224& 1064 & 115.1\\
\hline
P2 & 356 & 191 & 338 & 212 & 335 & 198 & 963 & 111.4\\
\hline
P3 & 342 & 179 & 327 & 218 & 325 & 186 & 877 & 117.6\\
\hline
P4 & 938 & 499 & 911 & 653 & 877 & 499 & 2189 & 121.6\\
\hline
P5-HM1 & 676 & 358 & 595 & 368 & 594 & 360 & 1641 & 118.6\\
\hline
P6-SU1 & 262 & 140 & 251 & 156 & 254 & 147 & 733 & 111.8\\
\hline 
\end{tabular}
\end{center}
\caption{Part of the neutralino and slepton spectra for the 
 benchmark points, as well as the average squark mass (in
GeV). For completeness we include $m_h$ as obtained from {\sc SPheno}.}
\label{table:points2}
\end{table}

Regarding the prospects for production at the LHC, we present in
Table~\ref{table:points3} the NLO production cross sections in fb
(obtained using {\sc Prospino}2.1~\cite{Prospino2.1})
for c.o.m. energies 
of 7 TeV and 14 TeV. We separately display the production of
at least one and exactly two $\chi_2^0$
states. For illustrative purposes, we also
detail  in Table~\ref{table:points4} the production cross section for
at least one $\chi_2^0$, identifying the dominant production modes:
direct $\chi_2^0$ production, from squark decay, or from $\tilde g
\tilde g$~(see Section \ref{sec:LHC}).
\begin{table}[ht]
\begin{center}
\begin{tabular}{|c|r|r|r|r|}
\hline
\multirow{2}{*}{Point}  & \multicolumn{2}{c|}{$\sigma(pp \rightarrow
  \tilde{\chi}^0_2)$ (fb)}
& \multicolumn{2}{c|}{$\sigma(pp \rightarrow \tilde{\chi}^0_2
  \,\tilde{\chi}^0_2)$ (fb)} \\
\cline{2-5}
& \multicolumn{1}{c|}{7 TeV}    & \multicolumn{1}{c|}{14 TeV}   &
\multicolumn{1}{c|}{7 TeV}      & \multicolumn{1}{c|}{14 TeV} \\
\hline
P1 & $17.5$     & $278.7$       & $1.0$ & $19.1$        \\
\hline
P2 & $38.8$     & $513.9$       & $2.2$         & $32.6$        \\
\hline
P3 & $60.6$     & $806.9$       & $3.8$ & $52.1$        \\
\hline
P4 & $0.04$     & $1.87$ & $\sim0.00$    & $0.13$ \\
\hline
P5-HM1 & $0.57$  & $16.50$        & $0.02$        & $1.24$ \\
\hline
P6-SU1 & $239.0$        & $2485.8$      & $15.1$        & $158.0$ \\
\hline
\end{tabular}
\end{center}
\caption{
Production cross sections for at least one $\chi_2^0$, $\sigma(pp \rightarrow
  \tilde{\chi}^0_2)$ (in fb), and exactly two $\chi_2^0$,
$\sigma(pp \rightarrow \tilde{\chi}^0_2 \,\tilde{\chi}^0_2)$ (in fb), 
for the  benchmark points, with $\sqrt s =7$ TeV and 14 TeV.}
\label{table:points3}	
\end{table}

\begin{table}[h!]
\begin{center}
\begin{tabular}{|c|c|r|r|r|r|r|r|}
\hline
\multirow{2}{*}{Primary prod. mode}
& \multirow{2}{*}{$\sqrt{s}$ (TeV)}     &\multicolumn{6}{c|}{$\sigma$
  (fb) for the production of at least one
  $\tilde{\chi}^0_2$} \\
\cline{3-8}&
& \multicolumn{1}{c|}{P1} & \multicolumn{1}{c|}{P2}
& \multicolumn{1}{c|}{P3} & \multicolumn{1}{c|}{P4}
& \multicolumn{1}{c|}{P5-HM1} & \multicolumn{1}{c|}{P6-SU1}     \\
\hline
\multirow{2}{*}{``Direct'' -- $\sum_X \tilde{\chi}^0_2 \, X$}
& 7     & $11.1$        & $23.1$        & $28.8$        & $0.04$
& $0.53$ & $101.8$
\\ \cline{2-8}
& 14    & $69.0$        & $124.4$       & $154.5$       & $1.11$
& $6.50$        & $447.8$
\\ \hline
\multirow{2}{*}{``Squarks'' -- $\sum_{Y} \tilde{q}_L \, Y$}
& 7     & $6.3$ & $15.3$        & $31.0$        & $\sim0.00$
& $0.04$        & $129.6$
\\ \cline{2-8}
& 14    & $194.2$       & $356.4$       & $602.5$       & $0.75$
& $9.70$        & $1758.3$
\\ \hline
\multirow{2}{*}{$\tilde{g} \, \tilde{g}$}
& 7     & $0.1$ & $0.4$ & $0.8$ & $\sim0.00$
& $\sim0.00$        & $7.6$                                     \\
 \cline{2-8}
& 14    & $15.5$        & $33.1$        & $49.9$        & $0.01$
& $0.30$        &  $279.7$
\\ \hline
\end{tabular}
\end{center}
\caption{Primary production modes and corresponding
cross sections for at least one $\chi_2^0$ (in fb)
for the  benchmark points, for $\sqrt s =7$ TeV and 14 TeV.}
\label{table:points4}	
\end{table}

Finally, in Table~\ref{table:points:BRX2decay} we summarise the
information regarding $\chi_2^0$ decays into a di-lepton final state. 
In each case we present the
specific BR($\chi_2^0 \to \tilde l^i_X \ell_i \to \ell_i \ell_i$),
corresponding to the contribution of a given intermediate $\tilde
l^i_X$ ($X$ denoting $L,R$) and the total sum over $\tilde l^i_X$
states.\\

\begin{table}[ht]
\begin{center}
\begin{tabular}{|c|c|r|r|r|r|r|r|}
\hline
\multirow{2}{*}{$\ell_i \,\ell_i$} & \multirow{2}{*}{$\tilde{l}^i_X$} &
\multicolumn{6}{c|}{BR($\chi_2^0\to \tilde l_X^i l_i \to l_i l_i
  \chi_1^0$) (\%)}  	\\ \cline{3-8}
& &  \multicolumn{1}{c|}{P1} 	& \multicolumn{1}{c|}{P2} &
\multicolumn{1}{c|}{P3} & \multicolumn{1}{c|}{P4} 
& \multicolumn{1}{c|}{P5-HM1}	& \multicolumn{1}{c|}{P6-SU1}	\\ \hline
\multirow{3}{*}{$\tau \tau$} & $\sum_{\tilde{l}}$    & 15.2  & 19.2  &
30.2  & 1.7   & 9.4   & 25.6  \\ \cline{2-8} 
        & $\tilde{\tau}_2$      & 7.9   & 7.6   & 4.0   & 1.7   & 9.4
        & 2.4   \\ \cline{2-8} 
        & $\tilde{\tau}_1$      & 7.3   & 11.6  & 26.2  & ---   & ---
        & 23.2  \\ \hline 
\multirow{3}{*}{$\mu \mu$}      & $\sum_{\tilde{l}}$    & 12.6  & 8.7
& 6.1   & 3.1   & 15.2  & 6.5   \\ \cline{2-8} 
        & $\tilde{\mu}_L$       & 12.2  & 7.3   & 5.8   & 3.0   & 15.1
        & 4.6   \\ \cline{2-8} 
        & $\tilde{\mu}_R$       & 0.4   & 1.4   & 0.3   & 0.1   & $6.5
        \times 10^{-2}$  & 1.9   \\ \hline 
\multirow{3}{*}{$e e$}  & $\sum_{\tilde{l}}$    & 12.5  & 8.7   & 6.0
& 3.0   & 15.3  & 6.5   \\ \cline{2-8} 
        & $\tilde{e}_L$ & 12.2  & 7.3   & 5.8   & 3.0   & 15.2  & 4.6
        \\ \cline{2-8} 
        & $\tilde{e}_R$ & 0.3   & 1.4   & 0.2   & $3.2 \times 10^{-2}$
& $5.7 \times 10^{-2}$  & 1.9   \\ \hline    
\end{tabular}
\end{center}
\caption{Branching ratios BR($\chi_2^0\to \tilde l_X^i l_i \to l_i l_i
  \chi_1^0$) (in \%) for a given di-lepton final state, 
  isolating specific intermediate sleptons and summing over all
  exchanged (slepton) states.} \label{table:points:BRX2decay}	
\end{table}

The decay chains considered in this study, with charged leptons in the
final state and missing energy from the escaping $ \chi^0_1$, ensure
that a large signal to background ratio is likely to be
obtained. Notice that we will not address background estimation in
the present analysis. For the points P5-HM1 and P6-SU1, estimations of
the corresponding backgrounds can be found in
Refs.~\cite{Ball:2007zza,ATLAS}, respectively.
Since at least one of the sleptons will always be lighter
than the  $\chi^0_2$, the distribution of the di-lepton invariant mass
will be (double-) triangular with an endpoint given by Eq.~(\ref{eq:mll}) of
Section~\ref{sec:LHC}.   

In Figs.~\ref{fig:msugra.di-leptonx}, we illustrate the di-muon
invariant mass ($m_{\mu \mu}$) versus the BR($\chi_2^0 \to \mu \mu
\chi^0_1$) for the mSUGRA points proposed in
Table~\ref{table:points1}. We also display
the expected number of events for $\sqrt s = 7$~TeV and 14~TeV and
corresponding expected integrated luminosities of $\mathcal{L}=1\
\text{fb}^{-1}$ and $\mathcal{L}=100\ \text{fb}^{-1}$. In general, in
our analysis, we will only study di-muon (di-electron) mass
distributions. It is expected that the edges of di-muon mass
distributions will be successfully reconstructed to an edge splitting
resolution of around 1 GeV~\cite{Hinchliffe:1996iu}.  Although
di-tau mass distributions are equally rich in the information they
might convey on the edges, the experimental reconstruction of the
decay chains can be more complicated: if decaying hadronically, the
taus can still be identified, but the associated signal is plagued by
an important SM background so that the reconstruction of its momentum
can be comparatively more difficult.

\begin{figure}[ht!]
\begin{center}
\begin{tabular}{c}
\epsfig{file=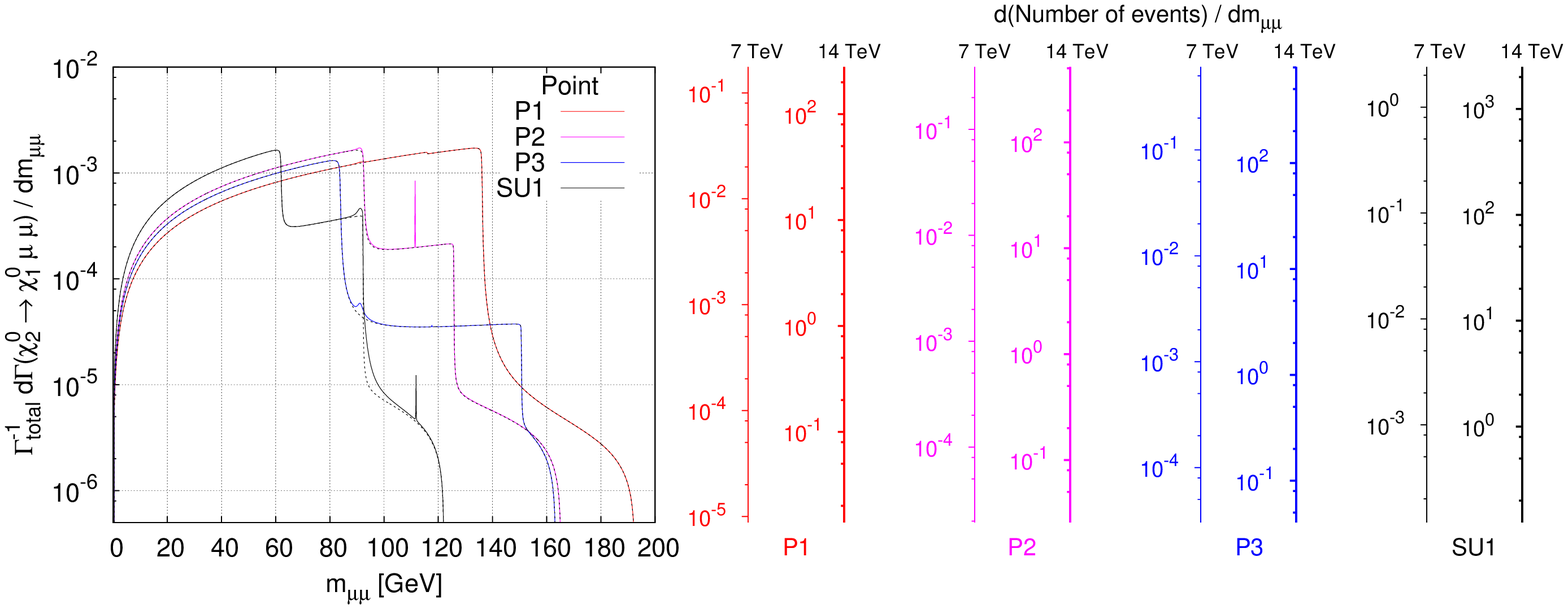, clip=, angle=0, width=165mm}\\
\epsfig{file=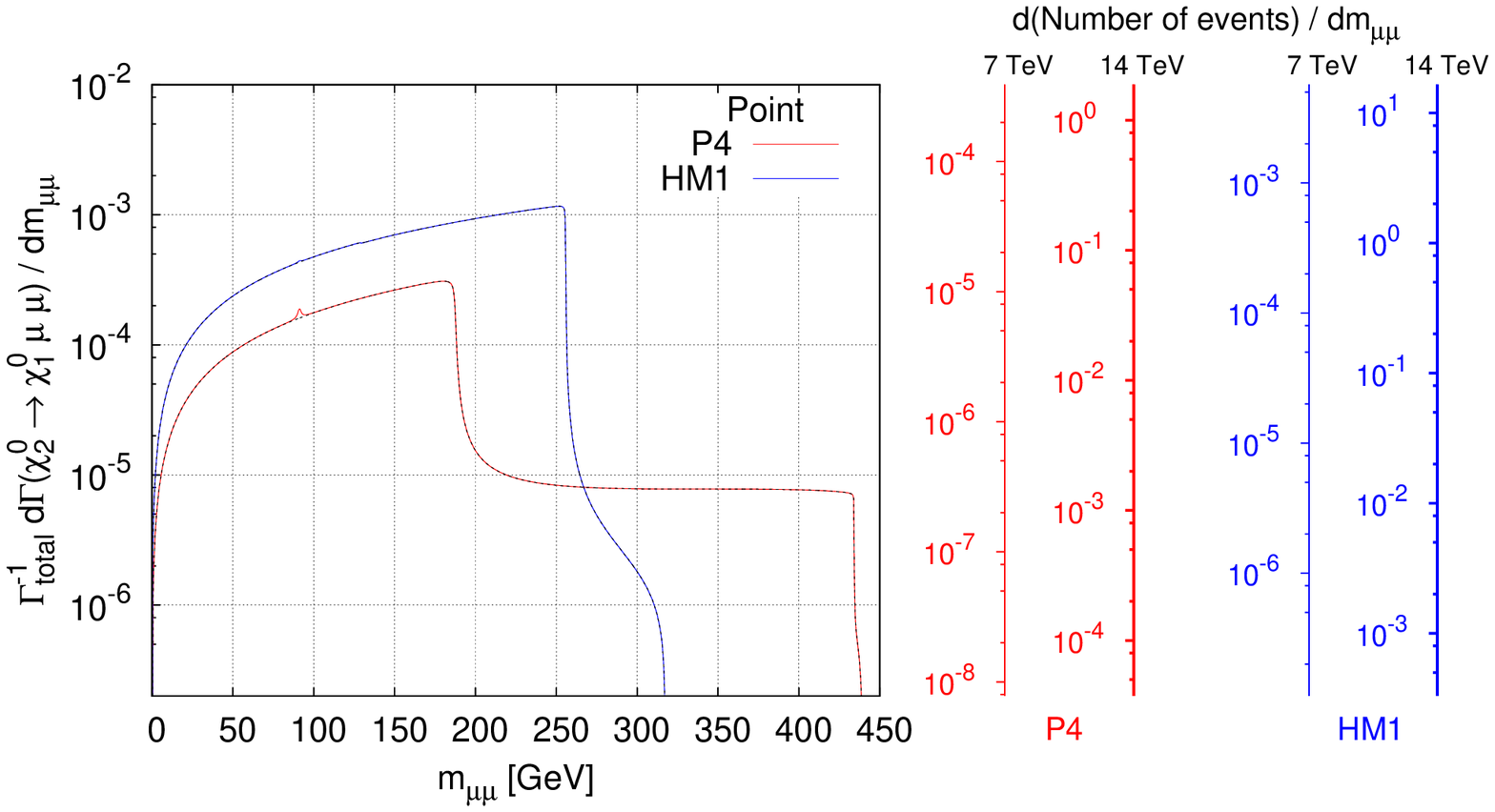, clip=, angle=0, width=115mm}\\
\end{tabular}
\caption{Di-lepton invariant mass ($m_{\mu \mu}$) 
versus BR($\chi_2^0 \to \mu \mu \chi^0_1$)  for the  benchmark 
points (Table~\ref{table:points1}). Upper panel: 
P1 (red), P2 (magenta), P3 (blue) and P6-SU1 (black); lower panel:
P4 (red) and P5-HM1 (blue). 
Secondary-right y-axes denote the corresponding expected number of events for 
$\sqrt s = 7$  TeV and 14 TeV, with 
$\mathcal{L}=1\ \text{fb}^{-1}$ and $\mathcal{L}=100\ 
\text{fb}^{-1}$, respectively.}
\label{fig:msugra.di-leptonx}
\end{center}
\end{figure}

As expected from the spectrum of the benchmark points
(in particular  from the slepton hierarchy), points 
P2, P3 and P6-SU1 have
a  double triangular distribution  for the invariant di-muon
mass. This is confirmed by the upper panel of
Fig.~\ref{fig:msugra.di-leptonx}, where two edges are visible in the
different distributions, each corresponding to the intermediate left-
and right-handed smuons in the chain. We summarise the numerical
values of the kinematical edges in
Table~\ref{table:cmssm:di-lepton:values}.  The lowest edge of P1
(corresponding to $\tilde \mu_R$) is hardly visible, while that of P2
appears superimposed on the $Z$ peak. The same distribution shape is
present for points P4 and P5-HM1. However,  in the latter the lowest
edge (around 130 GeV) is almost invisible to the naked eye and the 
values of the edges $m_{\mu\mu}(\tilde \mu_{L,R})$  are in agreement with those
obtained using Eq.~(\ref{eq:mll}).

\begin{table}[ht]
\begin{center}
\begin{tabular}{|c|r|r|r|r|r|r|}
\hline
\multirow{2}{*}{$\tilde{l}_X$}  &
\multicolumn{6}{c|}{$m_{ll}(\tilde{l}_X)$ (GeV)}
\\ \cline{2-7}
& \multicolumn{1}{c|}{P1}       &  \multicolumn{1}{c|}{P2}
& \multicolumn{1}{c|}{P3}       &
\multicolumn{1}{c|}{P4}         & \multicolumn{1}{c|}{P5-HM1}    &
\multicolumn{1}{c|}{P6-SU1}
\\ \hline
$\tilde{e}_R$   & 116.1 & 125.9
& 150.8 & 434.3 & 129.2 & 92.3
\\ \hline
$\tilde{e}_L$   & 136.2 &  92.5
& 83.8 & 187.2 & 255.7 & 62.0
\\ \hline
$\tilde{\mu}_R$ & 116.0 & 125.7
& 150.7 & 434.2 & 129.0 & 92.2
\\ \hline
$\tilde{\mu}_L$ & 136.2 & 92.5
& 83.8 & 187.5 & 255.7 & 62.0
\\ \hline
$\tilde{\tau}_1$ & 82.6 & 77.5
& 78.4 & 16.2 & 56.0 & 67.7
\\ \hline
$\tilde{\tau}_2$ & 134.4 & 98.0
& 87.9 & 274.4 & 256.4 & 54.2
\\ \hline
\end{tabular}
\end{center}
\caption{$m_{ll}(\tilde{l}_X)$ (GeV) where $l$ is any of the 
charged leptons and $X$ stands for left- and right-handed sleptons 
(all families).}\label{table:cmssm:di-lepton:values} 
\end{table}

In all points (upper and lower panels) the $Z$ peak is visible,
although in some cases, such as P1 and P5-HM1, the relative height of
the peak (as given by the corresponding BR) is very small compared to
its width. The peak of the lightest Higgs boson is only visible for some of
the points - P2, and P6-SU1 - since for the others the width is tiny
when compared to the corresponding height.  In general, the expected
number of events renders these processes visible only for a high
centre of mass energy (i.e. $\sqrt s \approx 14$ TeV), as can be seen
from the secondary y-axes on the right. Notice, however, that a proper
study of the background has to be taken into account.

Although we will not display it here, a  comparison of di-electron and
di-muon distributions for different benchmark points would 
confirm the superposition of the kinematical edges of both distributions - see
exact values in Table~\ref{table:cmssm:di-lepton:values} -, 
the only significant difference 
between them being the disappearance of the Higgs boson peaks.

\subsection{Slepton mass splittings and BR($\chi^0_2 \to
  \chi^0_1 l_i l_i $) in the cMSSM}

As mentioned in Section~\ref{sec:LHC}, one expects that the LHC will
measure the kinematical edges of the di-lepton distributions with a
precision of $\mathcal{O}(0.1\%)$. 
Although it has been claimed~\cite{Allanach:2000kt} that a
$\tilde e - \tilde \mu$ relative mass difference as small as $10^{-4}$
could be measurable, in the discussion of our numerical results we
will always adopt a  conservative view, assuming maximal
sensitivities of $\mathcal{O}(0.1\%)$ for $\Delta m_{\tilde
\ell}/m_{\tilde \ell} (\tilde e ,\tilde \mu)$ and $\mathcal{O}(1\%)$
for $\Delta m_{\tilde \ell}/m_{\tilde \ell} (\tilde \mu , \tilde
\tau)$.

\bigskip
We begin the numerical analysis of slepton mass splittings by  
a brief overview of the cMSSM case (no flavour mixing in the lepton
and slepton sectors). 

In Fig.~\ref{fig:cmssm1:BR:MS} we display the correlation between 
the BR of the neutralino cascade decay, 
BR$(\chi_2^0 \to \tilde \ell_{L,R} \, \ell \to \chi_1^0
\,\ell \,\ell)$ and the different slepton mass differences. In
particular, we present the numerical results for the 
mass splittings $\tilde e_{L,R} - \tilde \mu_{L,R}$ and 
$\tilde \mu_{L,R} - \tilde \tau_{2,1}$, where the heaviest/lightest
staus are dominated by the left-/right-handed component. 
Here, as throughout the remaining analysis, we 
normalise the slepton mass splittings 
to the corresponding average slepton masses (cf. Eq.~(\ref{eq:MS:def})).
Fixing $\tan\beta = 10$ and taking $\mu > 0$, we have scanned the
remaining mSUGRA parameters as follows: 
$300 \text{ GeV} \leq M_{1/2} \leq 1.2\text{ TeV}$, 
$-1 \text{ TeV} \leq A_0  \leq  1\text{ TeV} $, 
$m_0$ being determined in each point by the requirements of a
``standard window'' 
(leading to $50 \text{ GeV} \lesssim m_0 \lesssim 550 \text{ GeV}$).
In this case, and for simplicity, we have relaxed the 
requirement of compatibility with the WMAP
bound of Eq.~(\ref{exp:dm:wmap}). To illustrate 
the mass splittings associated with the proposed benchmark 
points (see Table~\ref{table:points1}), we
superimpose the corresponding predictions on the different panels. 

\begin{figure}[h!]
\begin{center}
\begin{tabular}{cc}
\epsfig{file=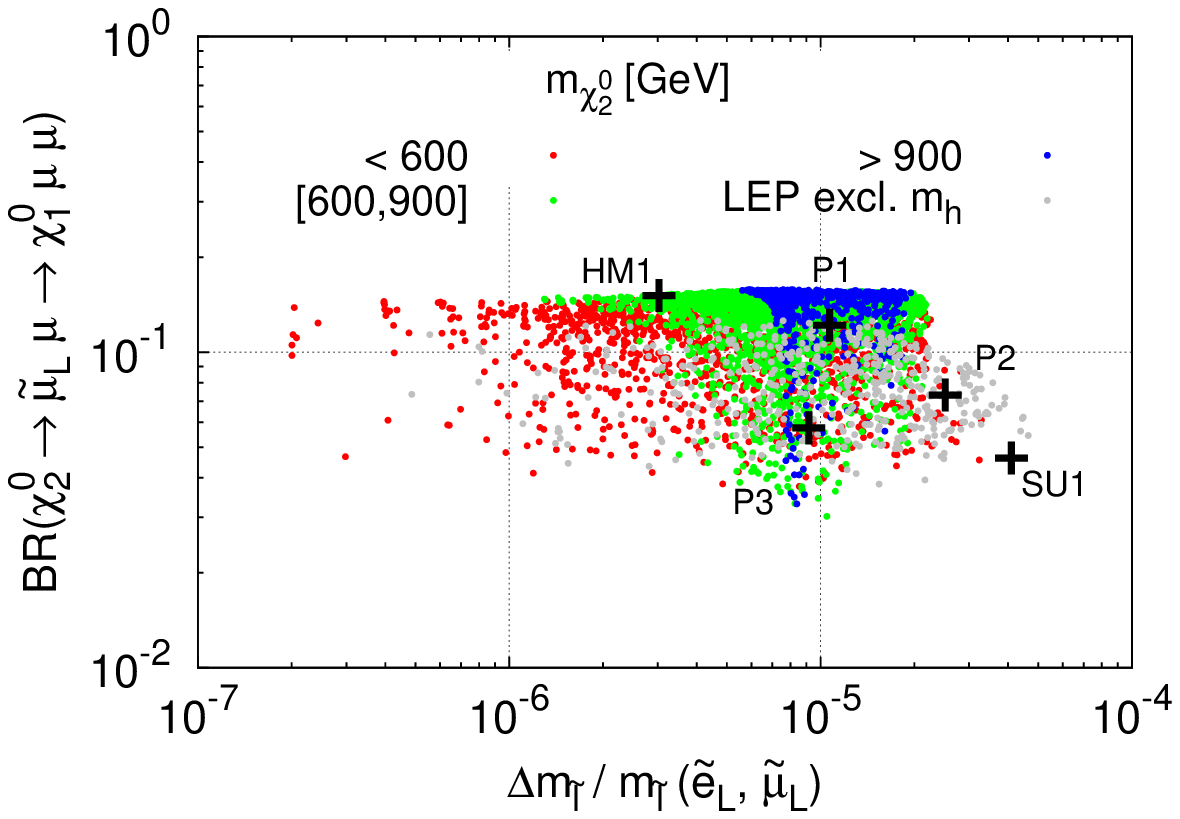, clip=, angle=0, width=80mm}&
\epsfig{file=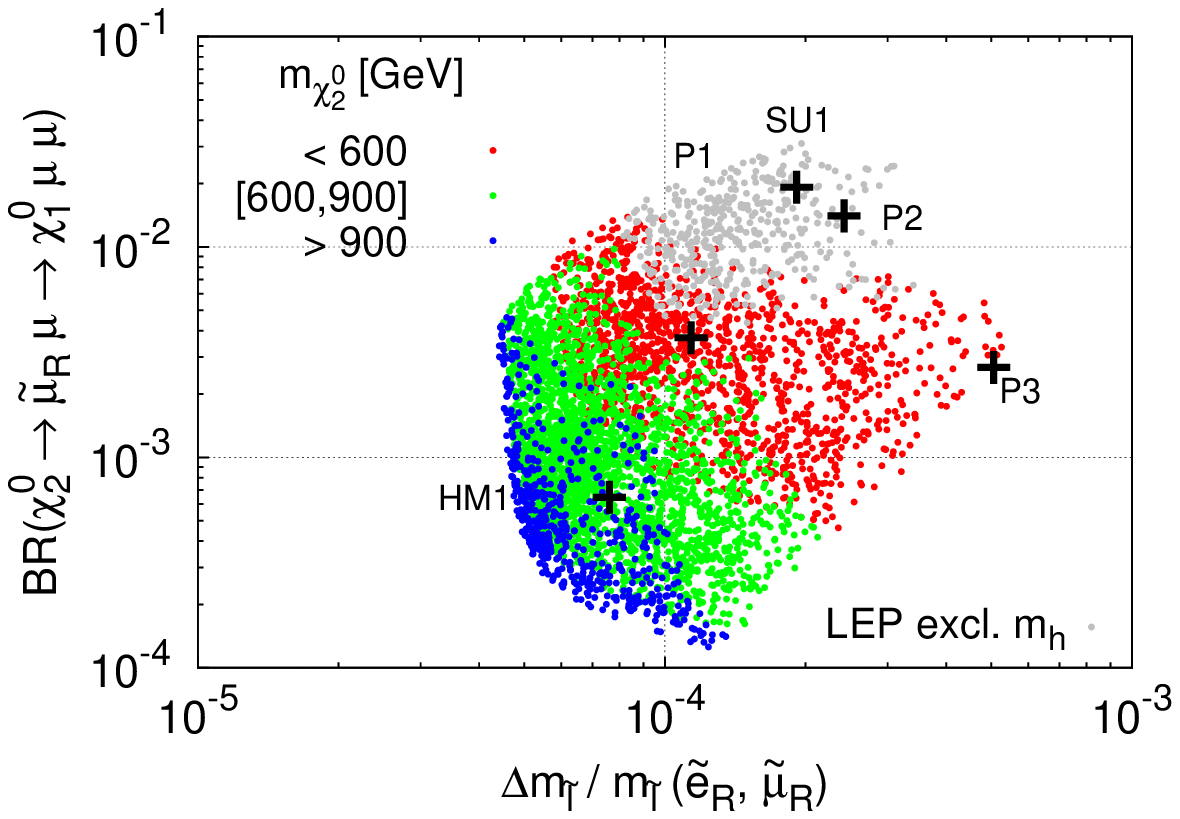, clip=, angle=0, width=80mm}\\
\epsfig{file=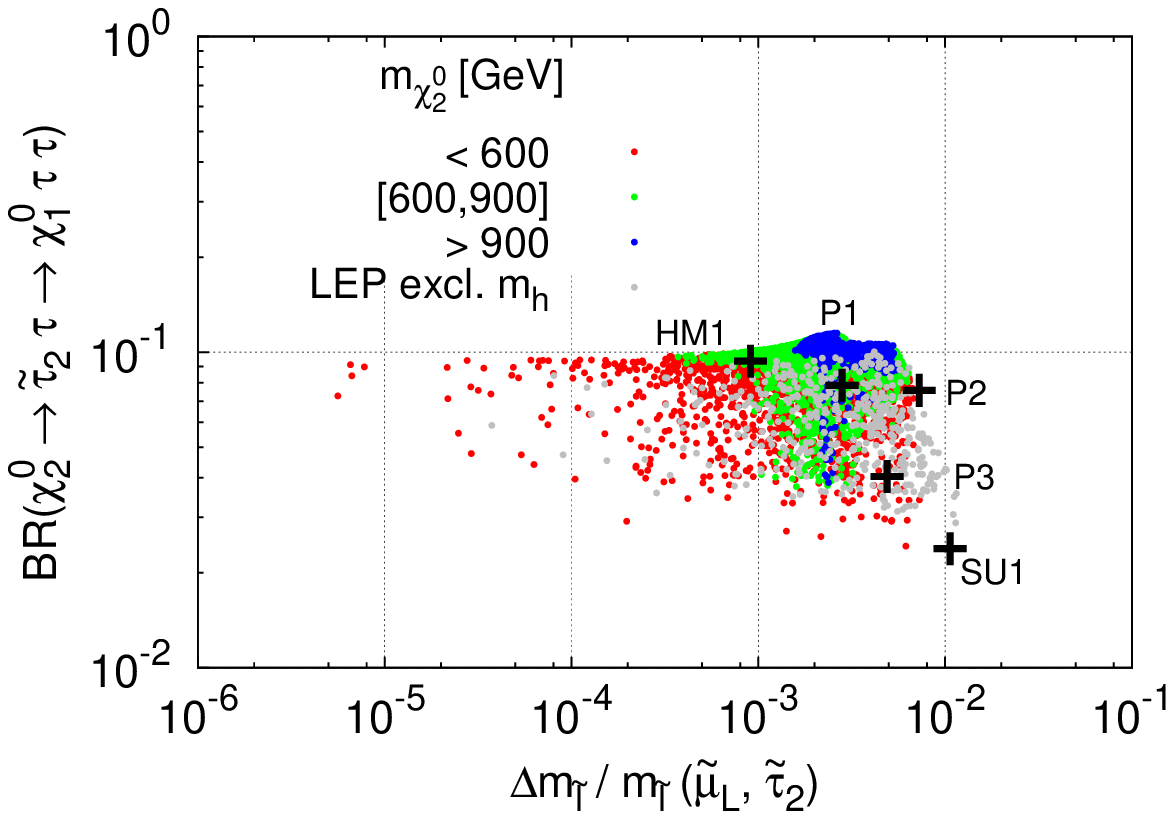, clip=, angle=0, width=80mm}&
\epsfig{file=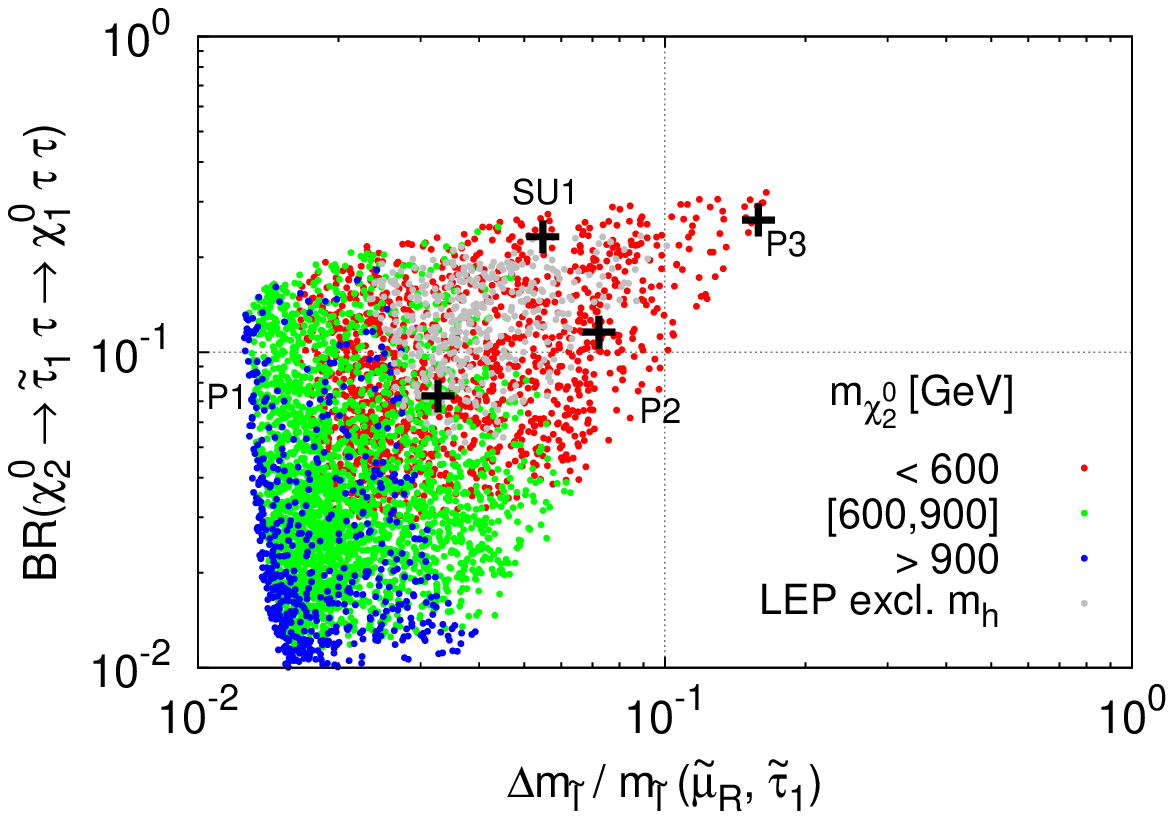, clip=, angle=0, width=80mm} 
\end{tabular}
\caption{
BR$(\chi_2^0 \to \tilde l_{L,R} \, l_i \to \chi_1^0
\,l_i \,l_i)$ as a function of $\Delta m_{\tilde l}/m_{\tilde l}$ for
the cMSSM. Upper panels: $\Delta m_{\tilde l}/m_{\tilde l}({\tilde
  e_{L,R}}, {\tilde \mu_{L,R}})$; 
lower panels: 
$\Delta m_{\tilde l}/m_{\tilde l}({\tilde \mu_{L,R}}, {\tilde \tau_{2,1}})$.
We take $\tan\beta = 10$, $\mu > 0$, and scan over  $-1 \text{ TeV}
\leq A_0  \leq  1\text{ TeV} $, 
$300 \text{ GeV} \leq M_{1/2}
\leq 1.2\text{ TeV}$, $m_0$ determined as to account for the
``standard window'' ($50 \text{ GeV} \lesssim m_0 \lesssim 550 \text{ GeV}$). 
The different coloured regions 
illustrate regimes for the decaying neutralino mass.
Gray points correspond to cases
in which $m_h\lesssim 114$ GeV. Crosses denote some of the
benchmark points defined in Table~\ref{table:points1}.}
\label{fig:cmssm1:BR:MS}
\end{center}
\end{figure}

As stated in Section~\ref{lfv:lhc}, in the absence of flavour
violation, the mass degeneracy between the first two slepton families
is only lifted by tiny RGE-running and $LR$ mixing effects. Since both
are proportional to the corresponding Yukawa couplings,  
the ${\tilde e_{L,R}} - {\tilde \mu_{L,R}}$
mass differences are expected to be very small 
(see Eq.~(\ref{eq:MS:emu:cMSSM})). 
This can be observed in Fig.~\ref{fig:cmssm1:BR:MS}, where 
one confirms that  
both $\Delta  m_{\tilde{\ell}}/m_{\tilde{\ell}} 
({\tilde e_L},{\tilde \mu_L})$ and 
$\Delta  m_{\tilde{\ell}}/m_{\tilde{\ell}} ({\tilde e_R},{\tilde
  \mu_R})$ lie in the range $10^{-7} - 10^{-3}$. 
Both $LR$ mixing and RGE-induced effects are more important for the
stau sector, so that the splittings 
$\Delta  m_{\tilde{\ell}}/m_{\tilde{\ell}} 
({\tilde \ell_L}, {\tilde \tau_2})$ and 
$\Delta  m_{\tilde{\ell}}/m_{\tilde{\ell}} 
({\tilde \ell_R},{\tilde \tau_1})$ are somewhat larger, 
typically above $10^{-3}$. 
Mass splittings involving third generation sleptons 
strongly depend on $\tan \beta$: as an example, for $\tan\beta = 40$,
with $A_0$ being varied as in Figs.~\ref{fig:cmssm1:BR:MS} and
$m_0$, $M_{1/2}$ randomly varied as to fulfil the standard window
requirement -- which for this strong $\tan\beta$ regime corresponds to
$900 \text{ GeV} \leq M_{1/2} \leq 2\text{ TeV}$, and
$380 \text{ GeV} \lesssim m_0 \lesssim 1\text{ TeV}$ --, 
we find $3 \% \leq \frac{\Delta
m_{\tilde{\ell}}}{m_{\tilde{\ell}}} ( \tilde{\mu}_L, \tilde{\tau}_2)
\leq 6.5 \%$, as can be read from the right panel of
Fig.~\ref{fig:cmssm:deltaMSratio}.
Nevertheless, it should be stressed that increasing $\tan \beta$ (both
in the cMSSM and in its right-handed neutrino extensions) lowers the
lightest stau mass,  so that in the large $\tan \beta$ regime
$\chi^0_2$ predominantly decays via
an intermediated $\tilde \tau_1$ ($\sim\tilde \tau_R$), with
BR$(\chi_2^0 \to \tilde \tau_{R} \,\tau) \sim 1$.
Fig.~\ref{fig:cmssm1:BR:MS} also summarises the prospects of the
different benchmark points regarding production at the LHC
(notice that since the spectrum of P5-HM1 
kinematically forbids $\tilde \tau_1 \to \chi_1^0 \tau$ decays, 
this point is absent from the lower right panel).

From the comparison of each of the upper panels of
Figs.~\ref{fig:cmssm1:BR:MS} to the corresponding lower one, it can
also be observed  that in the cMSSM the
ratio of the $\tilde \mu - \tilde \tau$ and $\tilde e - \tilde \mu$
mass splittings indeed goes as   
$\Delta m({\tilde \mu_{L,R}}, {\tilde \tau_{2,1}})/
\Delta m({\tilde e_{L,R}} , {\tilde\mu_{L,R}}) \sim (m_\tau^2 /
m_\mu^2)$ (see Eq.~(\ref{eq:MS:emu:mutau:cMSSM})). This can be further
confirmed in  
Figs.~\ref{fig:cmssm:deltaMSratio}, where we display 
$\tilde  \mu_L - \tilde \tau_2$ versus 
$\tilde e_L - \tilde \mu_L$  mass differences for two values of 
$\tan \beta$. The full line denotes the $m^2_{ \tilde \tau}
/m^2_{\tilde\mu} $ slope. 
For larger $\tan \beta$ (as displayed on the right panel 
of Fig.~\ref{fig:cmssm:deltaMSratio}) the increased $LR$ mixing
effects for the staus induce a deviation to the simple approximation
of Eq.~(\ref{eq:MS:emu:mutau:cMSSM}). 

\begin{figure}[h!]
\begin{center}
\begin{tabular}{cc}
\epsfig{file=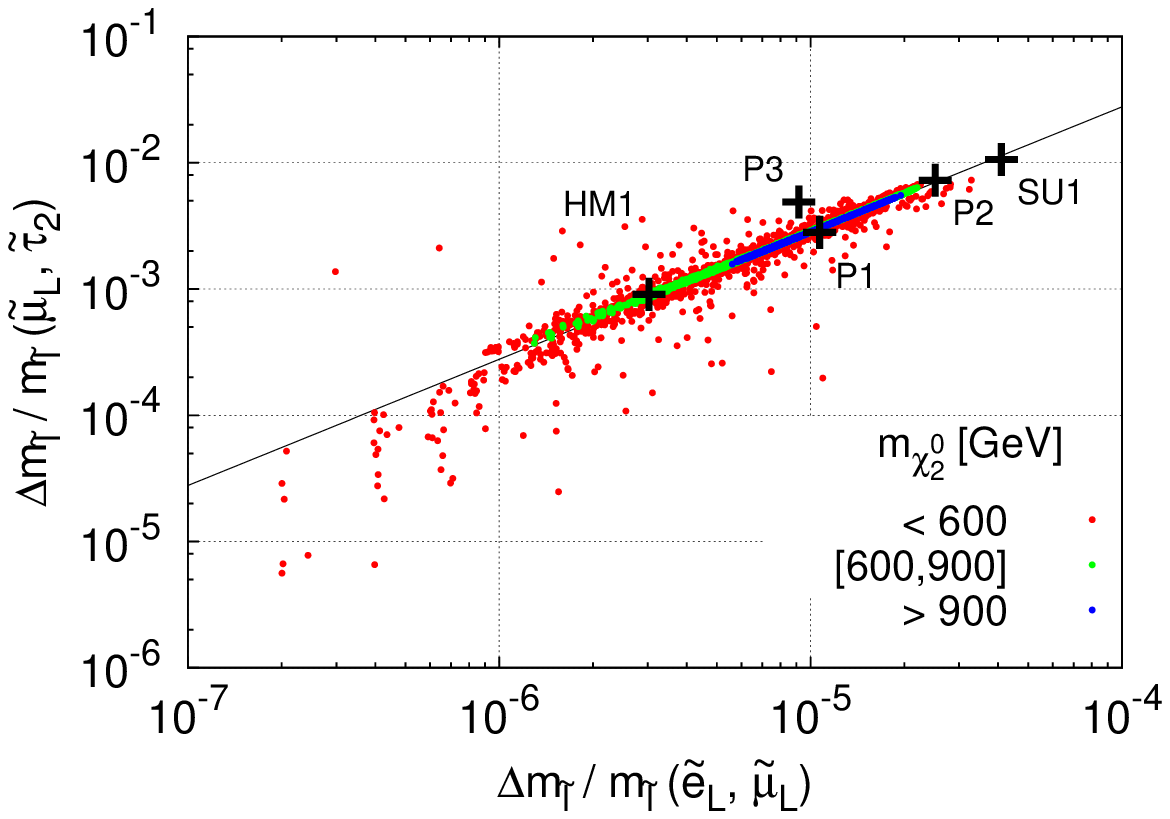, clip=, angle=0, width=80mm}
\epsfig{file=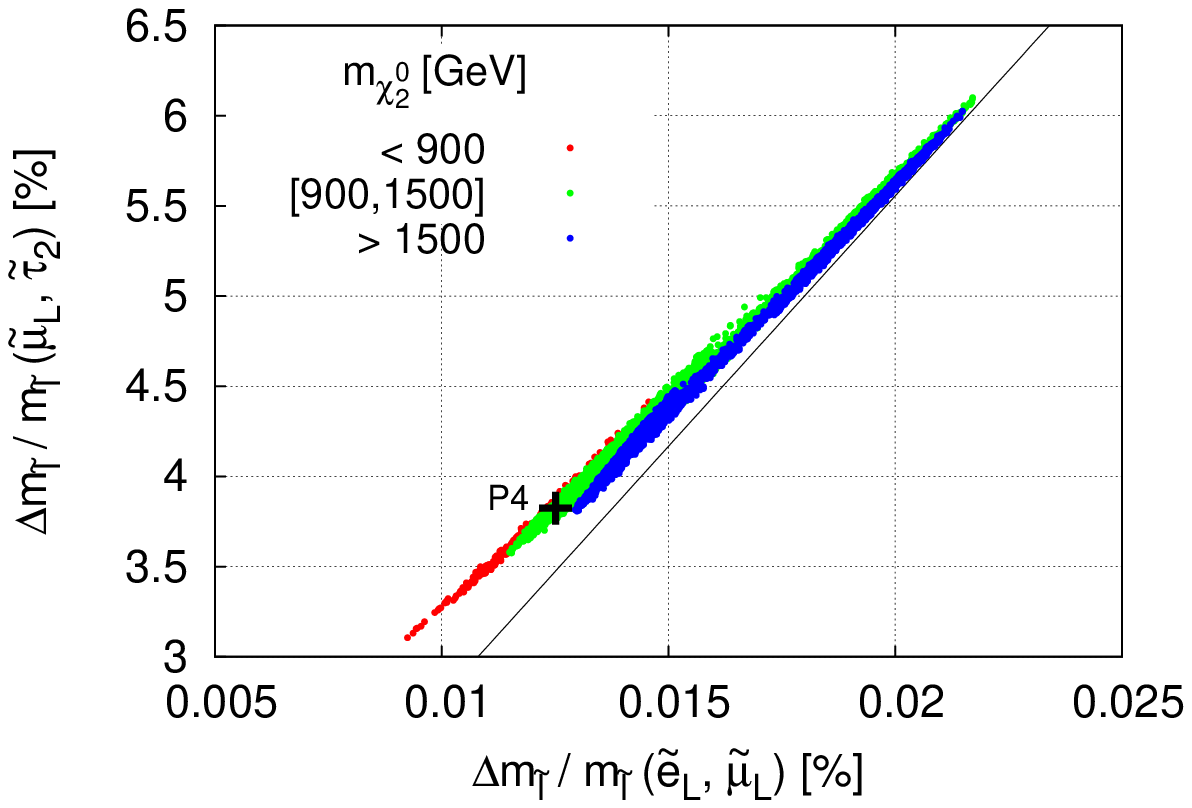, clip=, angle=0, width=80mm}
\end{tabular}
\caption{Mass differences  $\tilde  \mu_L - \tilde \tau_2$ versus 
$\tilde e_L - \tilde \mu_L$ (both normalised to an average slepton
mass) for the cMSSM. On the left $\tan \beta=10$, while on the right 
$\tan \beta=40$ (notice that in this case the mass differences are
given in \%). In the left panel, scan and colour code as in
Fig.~\ref{fig:cmssm1:BR:MS}, while in the right panel ($\tan\beta=40$)
we scan over $900 \text{ GeV} \leq M_{1/2} \leq 2 \text{ TeV}$ with
$m_0$ determined as to account for the ``standard window'' ($380
\text{ GeV} \lesssim m_0 \lesssim 1 \text{ TeV}$), and $A_0$ varied as
in Fig.~\ref{fig:cmssm1:BR:MS}.
Crosses denote some of the benchmark points defined in 
Table~\ref{table:points1}.}
\label{fig:cmssm:deltaMSratio}
\end{center}
\end{figure}

In Figs.~\ref{fig:deltams:scan}, we display a comprehensive scan of
the $\tilde  \mu_L - \tilde \tau_2$ mass difference in the cMSSM (the
corresponding predictions for $\tilde e_L - \tilde \mu_L$ can be
inferred from the previous discussion of
Fig.~\ref{fig:cmssm:deltaMSratio}). 
For three different values of the trilinear soft term ($A_0=-1, 0, 1$
TeV), we scan the mSUGRA parameter space to ensure an
optimal survey of the volumes complying with the ``standard
window'' requirement. More precisely, we 
have taken a range $0 < ( M_{1/2} - M^\text{(min)}_{1/2}(\tan\beta) ) \lesssim 1.4$ TeV, with 
$M^\text{(min)}_{1/2}(\tan\beta)$ being the 
minimal $M_{1/2}$ for a given $\tan\beta$ that provides 
$m_{\chi_2^0} > m_{\tilde \ell, \tilde \tau_2}$ and a $\chi_1^0$ LSP.
We present the resulting mass splitting (in percentage) as a function
of $\tan \beta$, identifying also distinct regimes for the $\chi_2^0$
mass (and hence $M_{1/2}$, implicitly understood from the GUT relation 
$m_{\chi_2^0} \approx 0.8 M_{1/2}$). For completeness, we also display regions
corresponding to a relaxation of the energy of the outgoing leptons 
($0 < m_{\chi_2^0}-m_{\tilde  \ell_L, \tilde \tau_2} < 10$ GeV).
Finally, we provide complementary information about the
corresponding ranges for the lightest Higgs boson mass, which can
severely constrain the explored parameter space, especially in the low
$\tan \beta$ and $A_0 \gtrsim 0$ regimes.

\begin{figure}[ht!]
\vspace*{-7mm}
\begin{center}
\begin{tabular}{c}
\epsfig{file=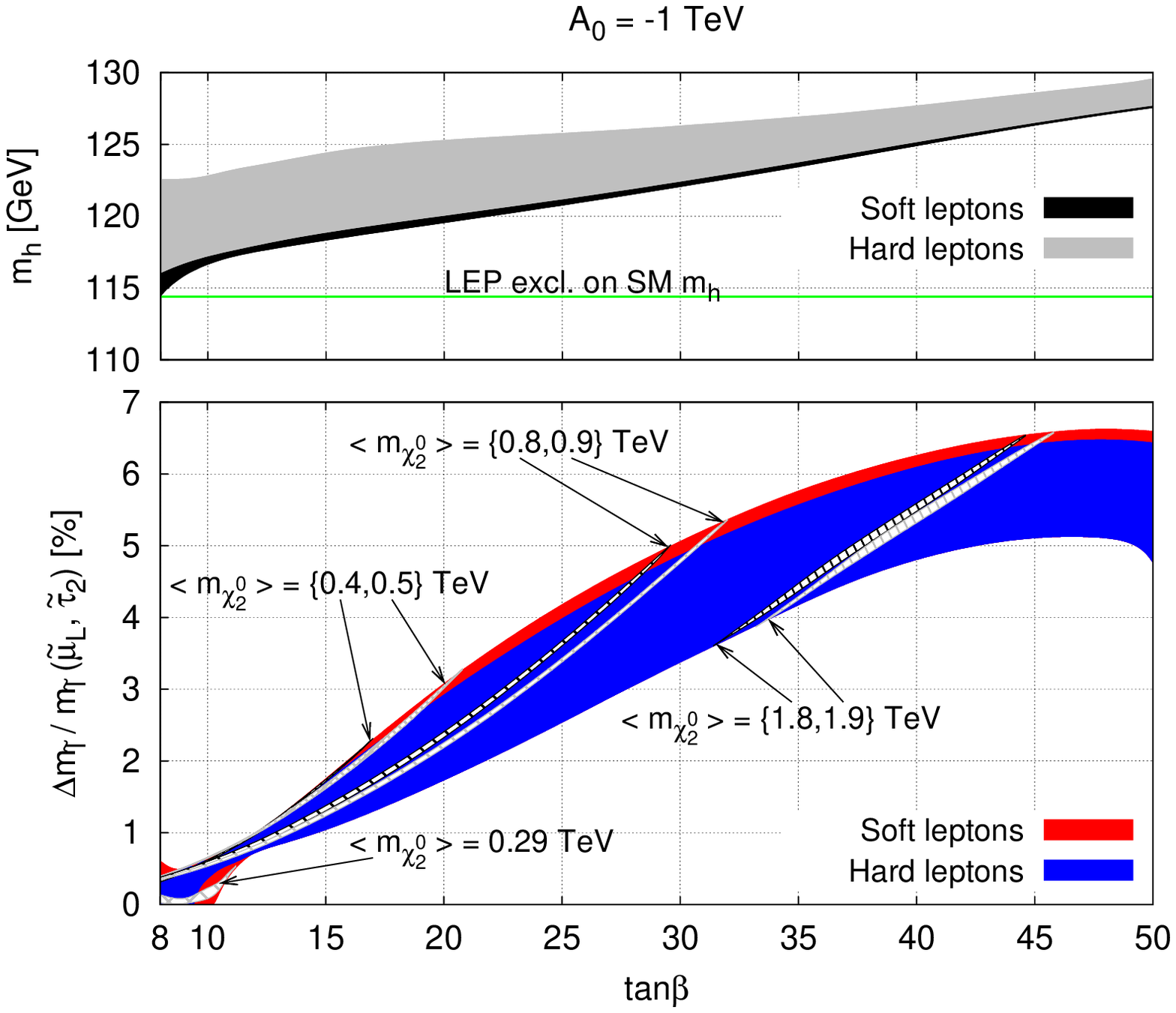, clip=, angle=0, width=80mm, height=60mm}
\vspace*{3mm}\\
\epsfig{file=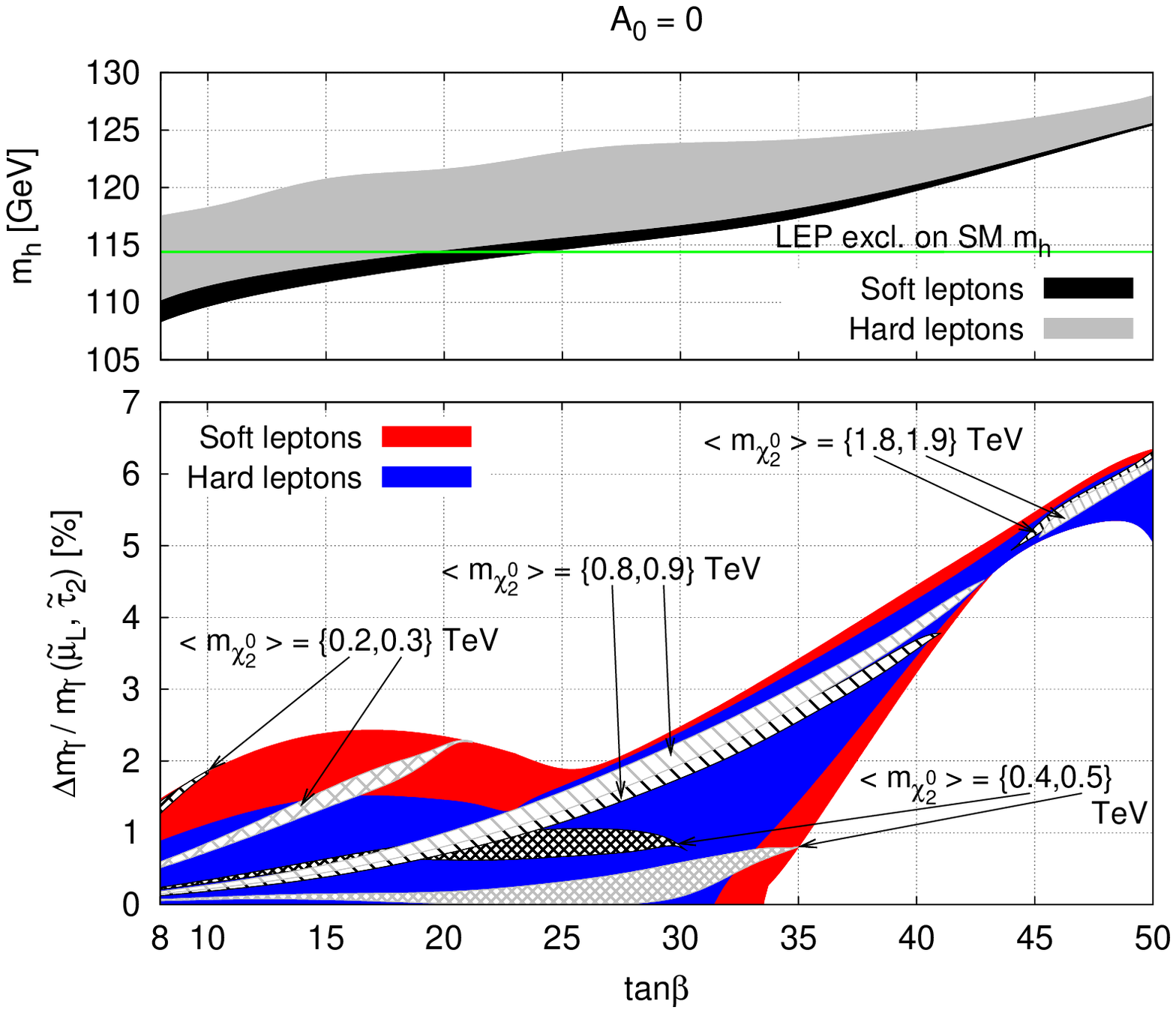, clip=, angle=0, width=80mm, height=60mm}
\vspace*{3mm} \\
\epsfig{file=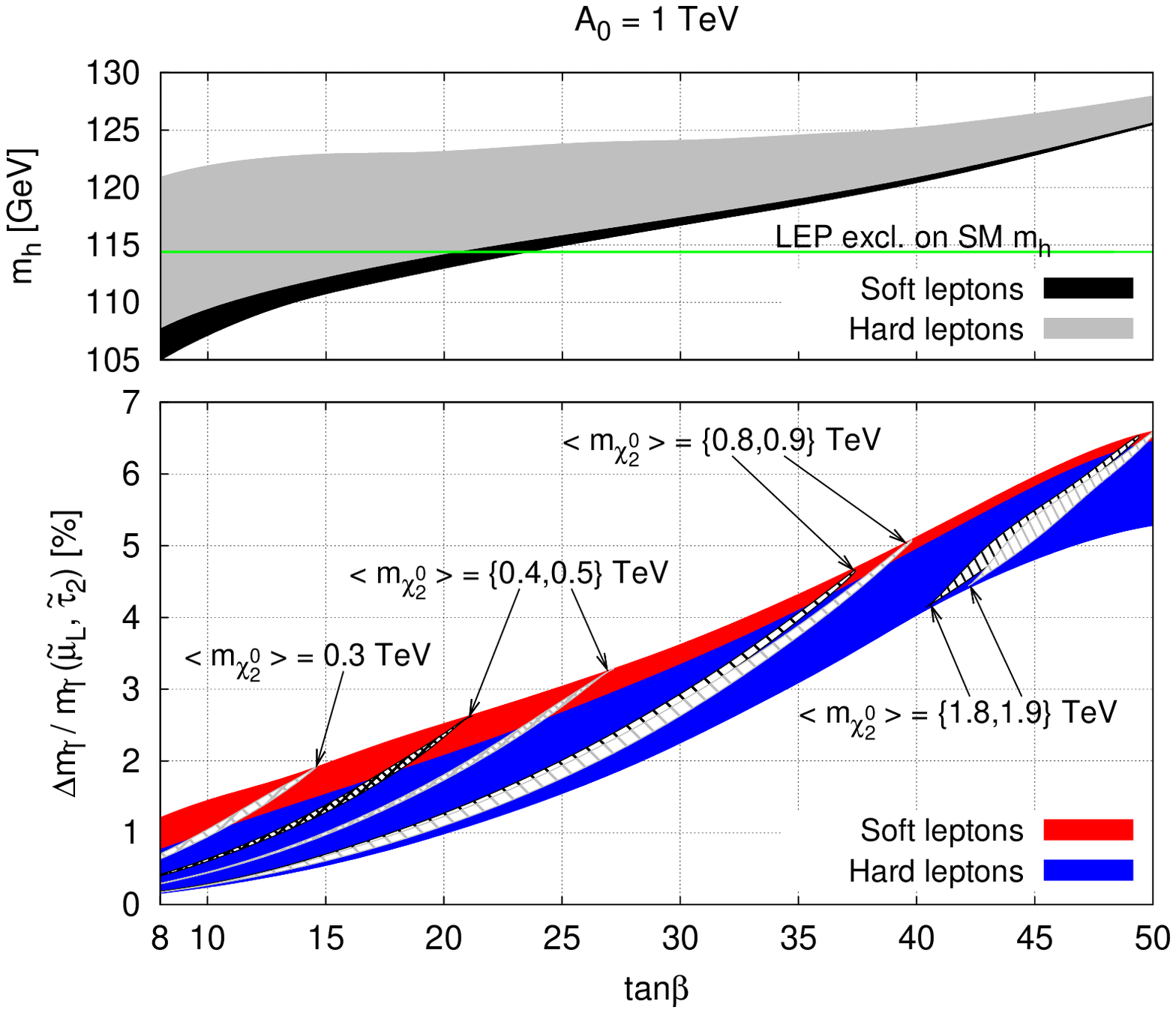, clip=, angle=0, width=80mm, height=60mm}
\end{tabular}
\caption{
  Mass difference  
  $\tilde  \mu_L - \tilde \tau_2$ (normalised to the average 
  $\tilde  \mu_L, \tilde \tau_2$ masses) in the cMSSM as a
  function of $\tan \beta$, for different values of $A_0$ (from top to
  bottom, $A_0=-1, 0, 1$ TeV).
  The subplots above each panel denote the corresponding variation of 
  $m_h$. The different solid regions correspond to hard (blue, gray)
  or soft (red, black) leptons in the final state. Inset are 
  bands corresponding to different regimes for $m_{\chi_2^0}$ (in TeV).}
\label{fig:deltams:scan}
\end{center}
\end{figure}

The most important conclusion to be drawn from
Fig.~\ref{fig:deltams:scan} is that in the cMSSM $\tilde \mu_L -
\tilde \tau_2$ mass splittings are at most $\mathcal{O}(7\%)$ (if
$|A_0| \lesssim 1$ TeV), and this occurs for regimes of very large
$\tan \beta$.  With increasing $\tan \beta$, the lowest vertex of the
region complying with the ``standard window'' constraints is pushed
towards larger values of both $m_0$ and $M_{1/2}$ (as can be seen from
the displacement of the triangular-shape central regions in
Figs.~\ref{fig:msugra.space1}).  This in turn implies that regions in
mSUGRA parameter space associated with the largest values of the
$\tilde \mu_L - \tilde \tau_2$ mass splittings will have poor
prospects for production at the LHC (smaller cross sections),
rendering them likely unobservable. 

For intermediate regimes of $\tan \beta$, one expects $\Delta
m_{\tilde \ell}/ m_{\tilde \ell} (\tilde \mu_L, \tilde \tau_2)$ to lie
in the range 2\% - 5\%, the latter corresponding to large (and
negative) $A_0$. This $A_0$ regime increases $LR$ mixing in the stau
sector, thus augmenting the cMSSM mass difference between the
left-handed smuons and the heaviest (dominantly left-handed) stau.

Following the discussion of Section~\ref{lfv:lhc} concerning the
correct definition of mass splittings for QDFC sleptons, we present
here the ``effective'' (according to Eqs.~(\ref{eq:effective:mass},
\ref{eq:effective:MS})) and ``real'' $\tilde e_L - \tilde \mu_L$ mass
differences. Since in the cMSSM the sleptons have a
well-defined flavour content, ``real'' and ``effective'' approaches coincide
to a very good extent as can be seen from
Fig.~\ref{fig:cmssm:deltaMS:effective}. Hereafter, and when addressing
seesaw-induced slepton flavour mixings, we will always use the
``effective'' mass splitting for the first two slepton generations.

\begin{figure}[h!]
\begin{center}
\epsfig{file=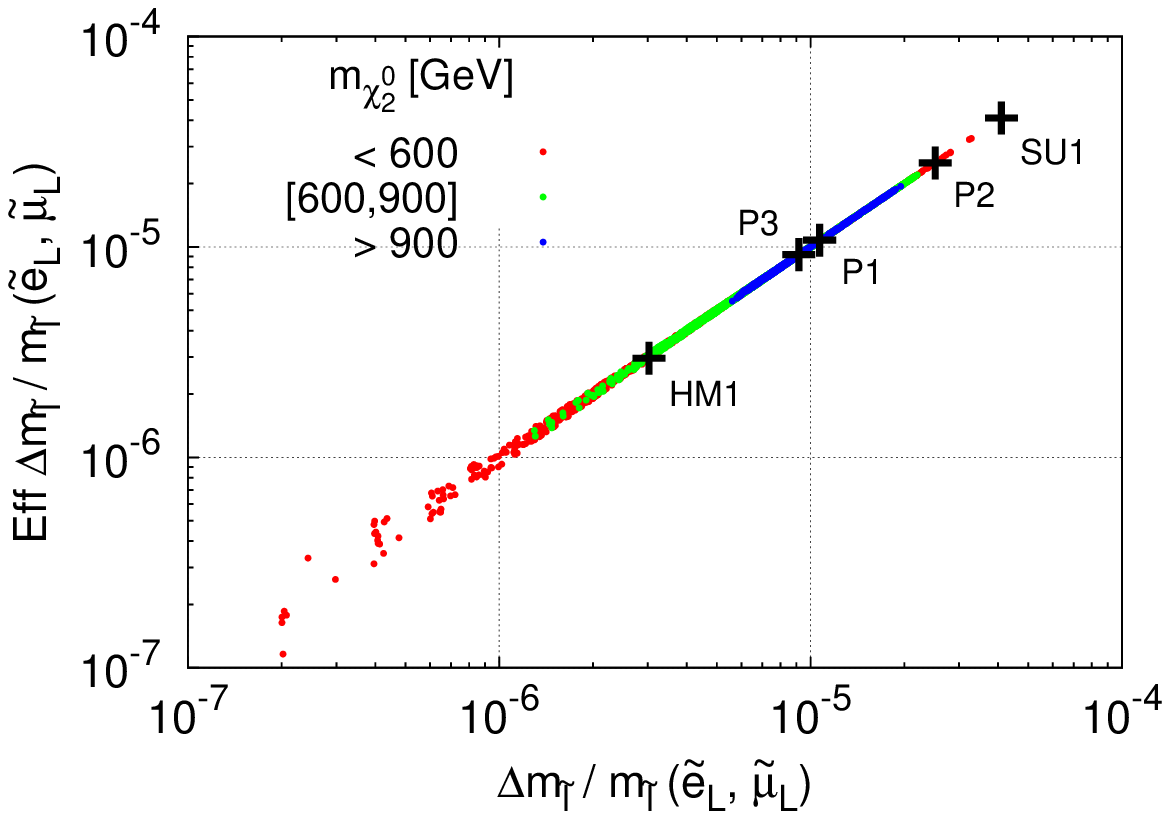, clip=, angle=0, width=100mm}
\caption{
``Effective'' parametrization of 
$\tilde e_L - \tilde \mu_L$ mass difference
 versus the ``real'' $\tilde e_L - \tilde \mu_L$ mass
 difference. Scan and colour code as in Fig.~\ref{fig:cmssm1:BR:MS}
 (taking $\tan \beta=10$). 
Crosses denote some of the benchmark points defined in 
Table~\ref{table:points1}.}
\label{fig:cmssm:deltaMS:effective}
\end{center}
\end{figure}

\subsection{Slepton mass splittings in the type-I SUSY seesaw}

As seen in the previous subsection, in the absence of flavour violation in the
lepton sector, the mass splittings between the sleptons of the first
two families are extremely small. The situation changes
if interactions that violate lepton flavour are switched
on:  either schemes where the  SUSY-breaking  parameters for the sleptons  are
flavour violating (or at least non-universal) or mechanisms that
account for both neutrino masses and lepton mixings, could induce
significantly larger slepton mass splittings,  
large enough to be observed at the LHC. If  flavour violating 
interactions in the slepton sector are indeed present, 
then other LFV processes are likely to occur: in addition to 
radiative and three body decays, lepton flavour can also be 
directly violated in sparticle decays, e.g. 
$\chi_2^0 \to \tilde \ell_i \ell_j$. 

Recent studies~\cite{Buras:2009sg} have addressed the complementarity of
high- and low-energy LFV adopting an effective approach.
Here we will consider the framework of the type-I SUSY seesaw,
studying the implications of having a unique source of flavour
violation: the neutrino Yukawa couplings.
Parametrizing $Y^\nu$ according to Eq.~(\ref{eq:seesaw:casas}),
flavour violation can arise both from the observed low-energy neutrino
mixing pattern, or from mixings involving the (heavy) right-handed
neutrino sector. 
Even though very little data is available to efficiently constrain each
$Y^\nu_{ij}$, there are several experimental bounds and theoretical 
arguments that should be  taken into consideration:
\begin{itemize}
\item data on  light neutrino mass-squared differences and leptonic
  mixing angles 
(cf. Eqs.~(\ref{eq:mixingangles:data}, \ref{eq:lightmasses:data}));
\item bounds on LFV  BRs and CRs (see Table~\ref{table:LFV:bounds}), as well as
  lepton EDMs;
\item perturbativity of the Yukawa couplings, $|Y^\nu_{ij}|^2 < 4 \pi$;
\item under the hypothesis that the BAU is explained via a mechanism
  of thermal leptogenesis, the requirement of a sufficiently large CP
  asymmetry (while avoiding the gravitino problem) leads to 
  bounds on $M_{N_1}$ and to constraints on
  combinations of the complex $R$ matrix angles $\theta_i$.
\end{itemize}
Aside from the perturbativity bounds, the most important
constraints on the seesaw parameters will arise from the
non-observation of LFV processes: 
since both flavour violating BRs and slepton mass splittings 
originate from the same unique source ($Y^\nu$), compatibility with
current bounds, in particular on BR($\mu \to e \gamma $) and BR($\tau
\to \mu \gamma$), may preclude sizable values for the slepton mass
splittings. This is in contrast with other scenarios of (effective) flavour
violation in the slepton sector where the
different off-diagonal elements of the slepton mass matrix can be
independently varied~\cite{Buras:2009sg}.

\bigskip
We begin by considering a minimal implementation of the SUSY seesaw, 
where flavour violation arises solely from the $U^\text{MNS}$
mixing angles. This corresponds to taking $R=1$ (i.e.~$\theta_i=0$) 
in the Casas-Ibarra parametrization of Eq.~(\ref{eq:seesaw:casas}), and 
translates into a ``conservative'' limit for flavour violation:
apart from possible cancellations, and for a fixed
seesaw scale (i.e. $M_N$), this limit provides in
general a lower bound for the amount of LFV. 
(Notice however that leptogenesis is not viable in this case.) 
In the subsequent numerical analysis we will consider first 
strict normal hierarchies for both heavy and
light neutrinos, commenting  at a later stage on the effect 
of different mass schemes.

As can be inferred from the analytical discussion in
Section~\ref{lfv:lhc} (based on the LLog approximation),  
$\Delta m_{\tilde \ell}/m_{\tilde \ell}$ are strongly dependent on
the RH neutrino mass scale, $M_N$. 
In the limit $R=1$, the overall magnitude of the flavour violating
entries is dominantly driven by $M_{N_3}$ (see Eq.~(\ref{eq:seesaw:DM23:open})).

We begin by revisiting the correlation between the
$\tilde  \mu_L - \tilde \tau_2$ and $\tilde e_L - \tilde \mu_L$
slepton mass splittings. We conduct a similar  scan over  the mSUGRA parameters
as in the previous subsections (see discussion leading to
Fig.~\ref{fig:cmssm1:BR:MS}), considering a regime of low $\tan
\beta=10$, but now requiring compatibility with the WMAP bound on 
$\Omega h^2$ (cf. Eq~({\ref{exp:dm:wmap})).
Here we take very small values of the reactor angle $\theta_{13}$,
also setting the CPV Dirac phase $\delta=0$.  
The impact of $\theta_{13}$ will be addressed at a later stage.  

\begin{figure}[h!]
\begin{center}
\epsfig{file=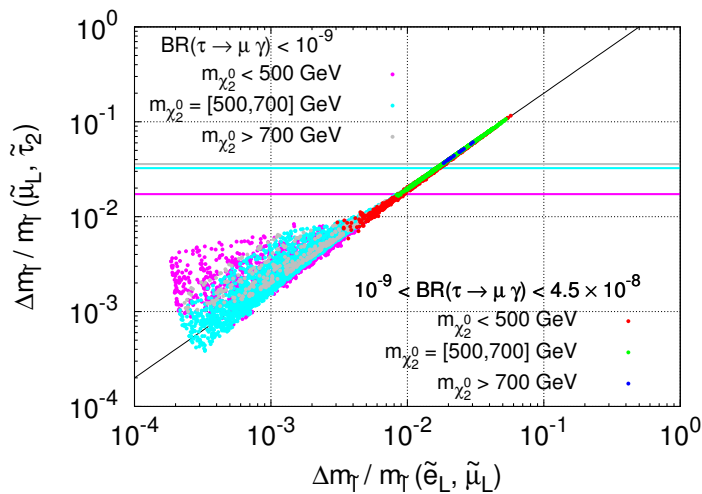, clip=, angle=0, width=100mm}
\caption{Mass differences $\tilde  \mu_L - \tilde \tau_2$ versus 
$\tilde e_L - \tilde \mu_L$ (both normalised to an average slepton
mass) in the type-I SUSY seesaw. Leading to the scan, we set $\tan \beta=10$, 
and randomly vary the remaining mSUGRA parameters (with $|A_0| \lesssim 1$ TeV, 
satisfying the ``standard window'' constraint and requiring consistency with 
the dark matter and Higgs boson mass bounds). For the seesaw parameters we have 
taken $R=1$, $\theta_{13}=0.1^\circ$ (with $\delta=\varphi_{1,2}=0$), and 
$M_{N_1}=10^{10}$ GeV, $M_{N_2}=10^{11}$ GeV, varying 
$10^{13} \lesssim M_{N_3} \lesssim 10^{15}$ GeV. All points shown comply
with present bounds on LFV observables.
We highlight in a different colour scheme points whose associated
prediction for BR($\tau \to \mu \gamma$) lies in the interval
delimited by current experimental bounds and future sensitivities
(red, green, blue), corresponding to $m_{\chi_2^0}$ regimes. The
magenta / cyan / gray lines denote the maximal value of $\Delta
m_{\tilde \ell}/ m_{\tilde \ell} (\tilde \mu_L,\tilde \tau_2)$
attainable for the magenta / cyan / gray points.
} 
\label{fig:seesawR1:deltaMS:emu:mutau:effective}
\end{center}
\end{figure}

The effect of implementing a type-I seesaw for the slepton mass
splittings is clearly visible in 
Fig.~\ref{fig:seesawR1:deltaMS:emu:mutau:effective}. It becomes even
more striking noticing  that this is the $R=1$ seesaw version of the
cMSSM case shown in the left panel of Fig.~\ref{fig:cmssm:deltaMSratio}.
Firstly, one observes that both  
$\tilde  \mu_L - \tilde \tau_2$ and 
$\tilde e_L - \tilde \mu_L$ mass splittings become much larger, with
values respectively up to 10\% and 6\%, well within the sensitivity
range of the LHC. Recall that in the cMSSM case
one typically had values $\mathcal{O}(10^{-3}, 10^{-5})$.
Furthermore, notice that the points whose BR($\tau \to \mu \gamma$) is
in the sensitivity range of future experiments are in general
associated to observable $\Delta m_{\tilde \ell}/m_{\tilde \ell}$,
especially for $m_{\chi_2^0} > 500$ GeV. 
(The maximum value of the $\tilde  \mu_L - \tilde \tau_2$ splitting 
for points whose BR($\tau \to \mu \gamma$) lies
beyond experimental capabilities is marked by horizontal lines, with a
colour code matching the corresponding spectrum colour code.)
One can also observe an 
important deviation from the correlated behaviour of both mass
splittings (see Eq.~(\ref{eq:MS:mutau:emu:R1seesaw})), 
symbolically depicted by the 
full dark line, with a slope given by
$| m_{\tilde\mu_L}-m_{\tilde \tau_2} | / | m_{\tilde e_L}-m_{\tilde
  \mu_L} | \approx 2$.
This deviation towards the pure mSUGRA limit of $m_\tau^2 / m_\mu^2$
(see Eq.~(\ref{eq:MS:emu:mutau:cMSSM})) occurs especially for points
associated to both smaller mass splittings and smaller BR($\tau \to \mu
\gamma$), starting at an intermediate seesaw scale of about $M_{N_3}
\lesssim 2 \times 10^{13}$ GeV for small $| A_0 |$ and approaching the
mSUGRA limit for $M_{N_3} \lesssim 10^{10}$ GeV.
For these regions in parameter space, even for comparatively smaller
flavour violating entries, the seesaw induces corrections to flavour
conserving $LR$ terms, which in turn imply larger 
$\tilde  \mu_L - \tilde \tau_2$ splitting when compared to   
$\tilde e_L - \tilde \mu_L$.

In Figs.~\ref{fig:seesawR1:deltaMS.A0.M3}
we display the variation of $\tilde e_L - \tilde \mu_L$ 
and $\tilde \mu_L - \tilde \tau_2$ mass differences as a function 
of $A_0$, showing also the comparison with the cMSSM case. 
First of all, notice that both mass splittings are substantially
larger, and for most of the $A_0$ interval considered, well within the
expected sensitivity of the LHC. Recall however that the overall
enhancement in $\Delta m_{\tilde \ell}/m_{\tilde \ell}$ is a
consequence of having taken very large values of $M_{N_3}$, 
close to the perturbativity limit of the neutrino Yukawa couplings
(especially $Y^\nu_{32}$ and $Y^\nu_{33}$).
Nevertheless, the r\^ole
of $A_0$ in the SUSY seesaw is clearly manifest in
Figs.~\ref{fig:seesawR1:deltaMS.A0.M3}, and especially in the
comparison of the former with the $\tan \beta =10$ band of
Figs.~\ref{fig:deltams:scan} (where discrete values of $A_0$ were
taken). While in the cMSSM the effect of $A_0$ was manifest 
through $LR$ mixing (and
via $m_h$ constraints on the parameter space), in the seesaw case the
dominant impact of $A_0$ on the mass splittings occurs via the
RGE-induced contributions to the $LL$ block (and $LR$, to a smaller
extent) of the slepton mass matrix, as given in
Eqs.~(\ref{eq:LFV:LLog}). In other words, the dominant contribution to 
$\Delta m_{\tilde \ell}/m_{\tilde \ell} $ now clearly arises from the second term
on the right-hand side of Eq.~(\ref{eq:MS:ij:seesaw}). 
This is substantiated by the approximate
symmetric dependence of $\Delta m_{\tilde \ell}/m_{\tilde \ell}$ on $A_0$. 
As expected, the regions of large positive $A_0$ (where small $LR$
mixing effects in the squark sector reduce the supersymmetric radiative
contributions
to the Higgs boson mass) are disfavoured due to conflict with the LEP bounds
on $m_h$. For very large negative values of $A_0$, the RGE-induced
amount of flavour violation is such that points associated with the
largest mass splittings have corresponding predictions to 
BR($\tau \to \mu \gamma$) already excluded by experiment.

\begin{figure}[ht!]
\begin{center}
\begin{tabular}{cc}
\epsfig{file=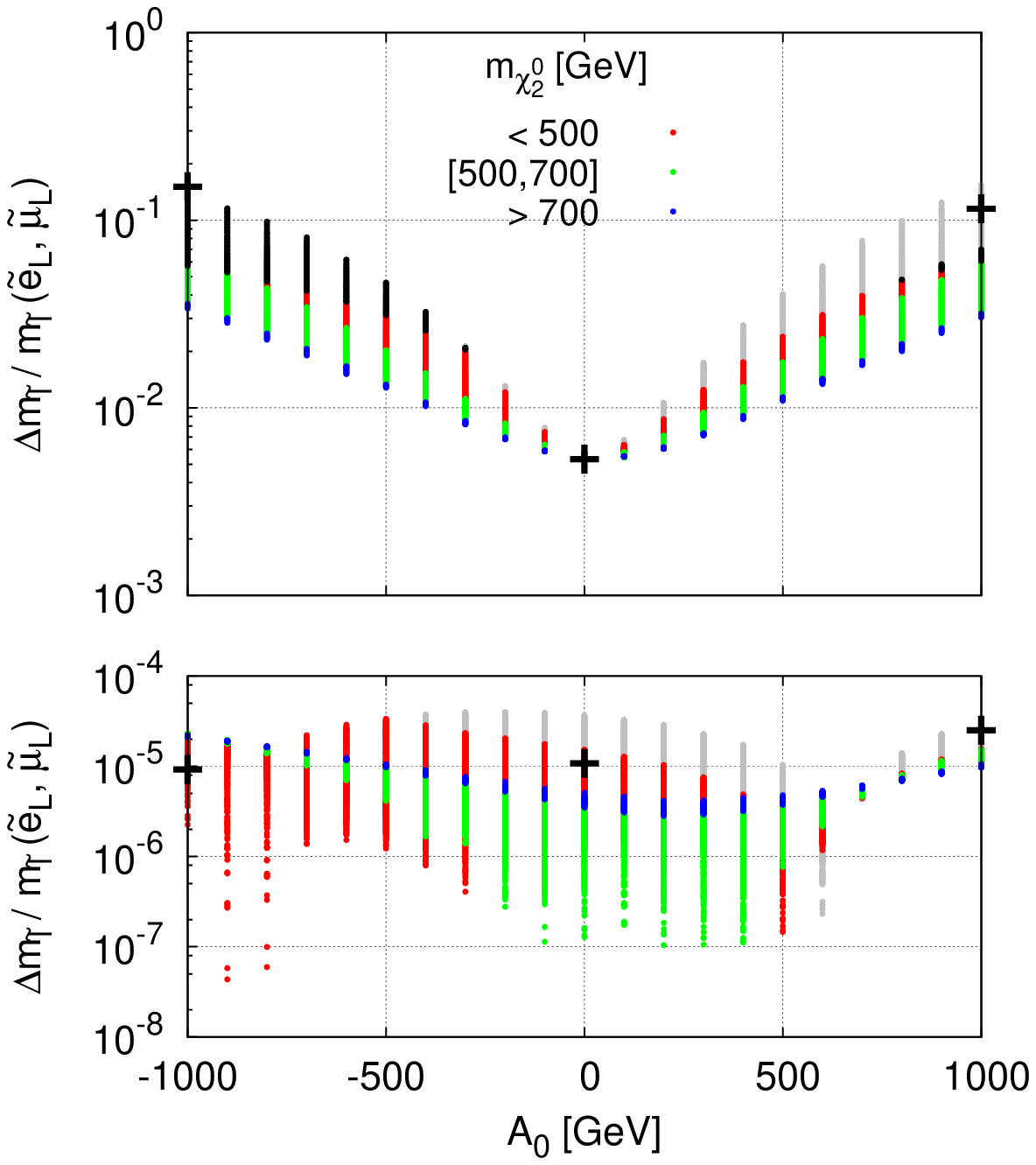, clip=, angle=0, width=80mm}
 &
\epsfig{file=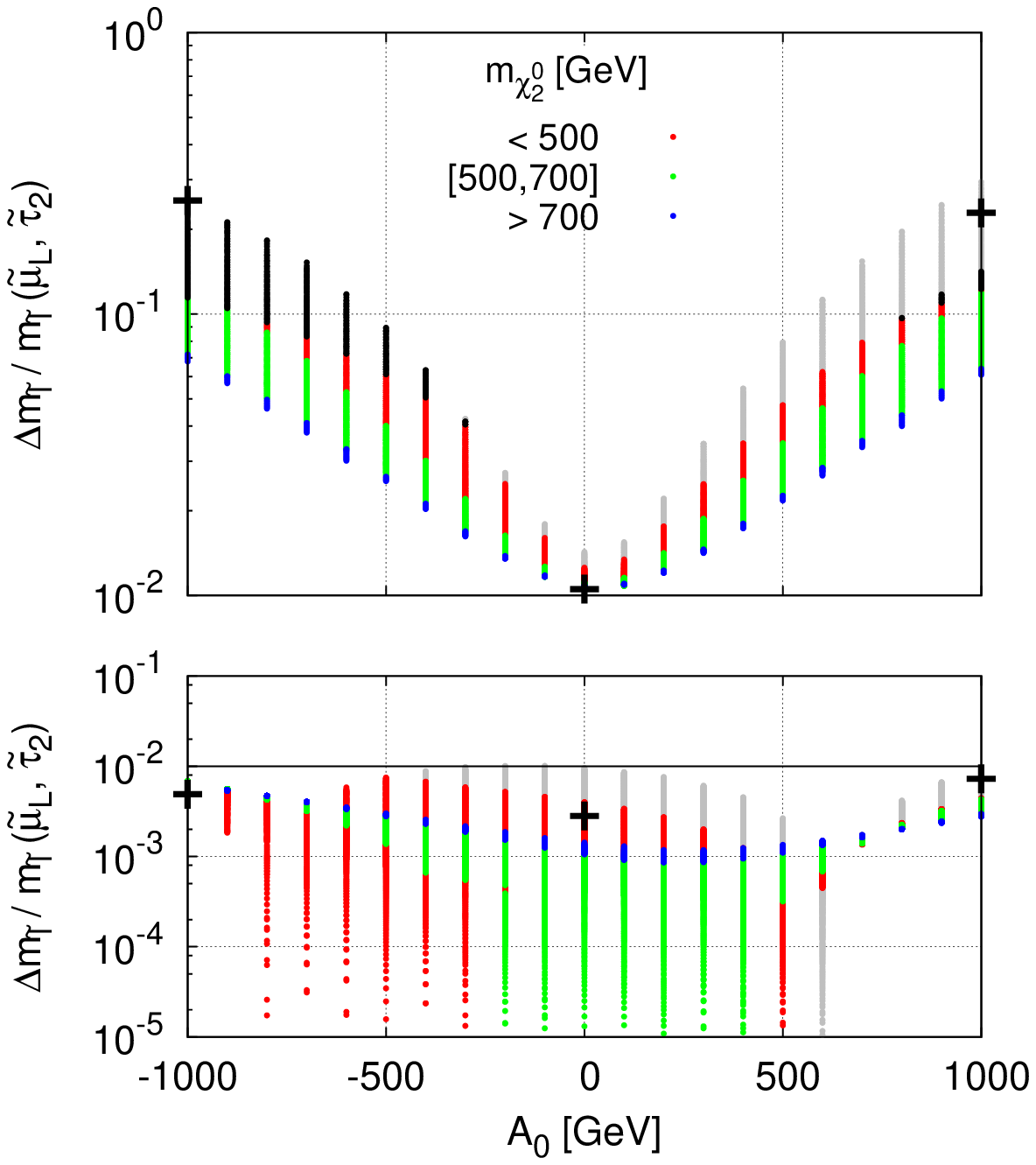, clip=, angle=0, width=80mm}
\end{tabular}
\caption{Mass differences $\tilde e_L - \tilde \mu_L$ (on the left) 
and $\tilde \mu_L - \tilde \tau_2$ (on the right) as 
a function of $A_0$ (in GeV). 
We have taken $\tan \beta=10$, and scanned over $m_0$
and $M_{1/2}$ as to satisfy the ``standard window'' and  the dark
matter constraints. The seesaw parameters have been set as $R=1$, 
$\theta_{13}=0.1^\circ$ (with $\delta=\varphi_{1,2}=0$), and 
$M_{N_i}=\{10^{10},10^{11},10^{15} \}$ GeV. The colour code denotes
different ranges for $m_{\chi_2^0}$ (black points denote violation of
at least one experimental bound - in these cases BR($\tau \to \mu 
\gamma$) -, while gray correspond to $m_h < 114$ GeV).
Crosses denote the benchmark points P1, P2 and P3 as defined 
in Table~\ref{table:points1}.
The lower panels illustrate the corresponding cMSSM study (same 
mSUGRA parameters, with $Y^\nu_{ij}=0$).}
\label{fig:seesawR1:deltaMS.A0.M3}
\end{center}
\end{figure}

In Fig.~\ref{fig:seesawR1:deltaMS.A0.M3} (as in all seesaw cases), we
have displayed the ``effective'' $\tilde e_L - \tilde \mu_L$ mass
difference, as justified by the discussion in
Section~\ref{lfv:lhc}. We recall that for the cMSSM, and as
emphasised by Fig.~\ref{fig:cmssm:deltaMS:effective}, both approaches
coincided. However when FV interactions are switched on, 
one should use the  ``effective'' mass
splitting. This can be confirmed in 
Fig.~\ref{fig:seesawR1:Mslepton.M3:effective.MS}, where we compare 
``real'' and ``effective'' $\Delta m_{\tilde \ell}/m_{\tilde \ell}$ as
a function of $A_0$. Leading to this figure, we have chosen the mSUGRA 
point P1, and allowed for variations of the trilinear coupling, $|A_0|
\leq 1.2$ TeV (recall that for P1, $A_0=0$). Regarding the seesaw
parameters, we have taken $R=1$, $\theta_{13}=0.1^\circ$, and
considered three distinct right-handed neutrino spectra for
illustrative purposes.

\begin{figure}[ht!]
\begin{center}
\epsfig{file=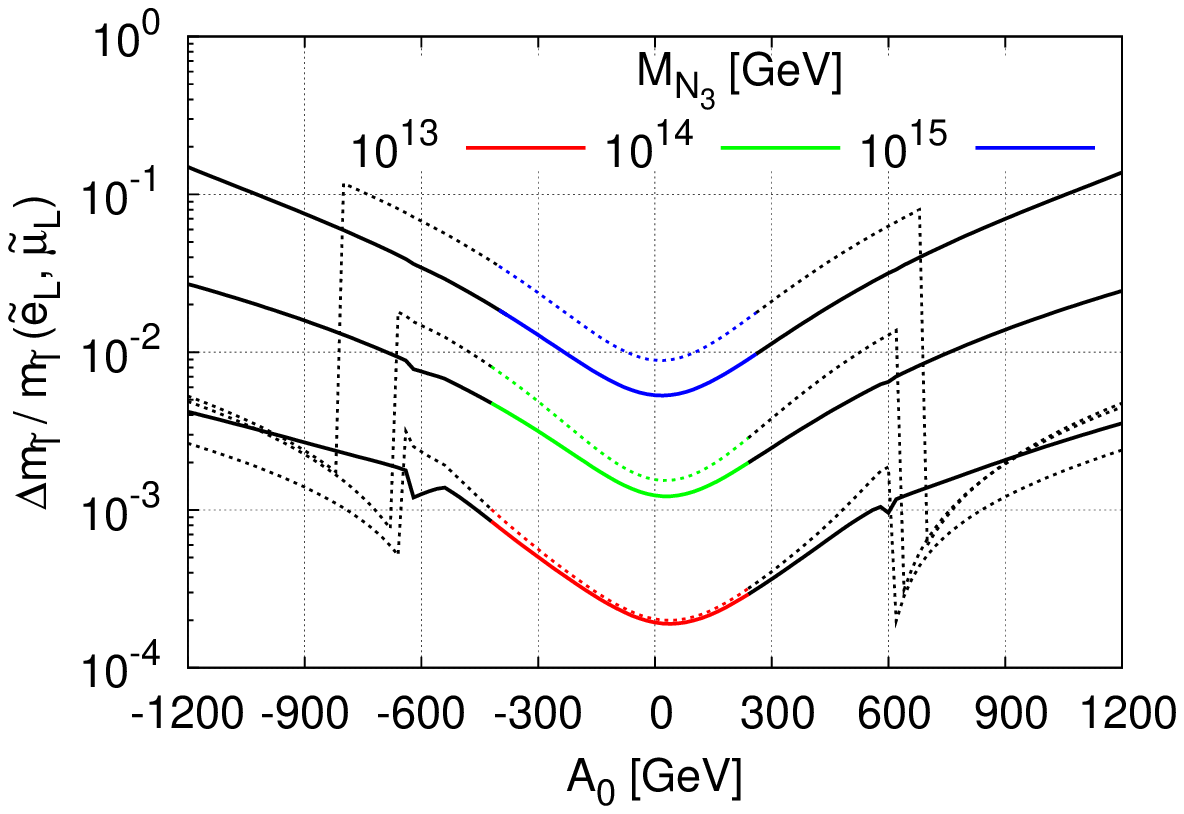, clip=, angle=0, width=80mm}
\caption{Comparison between ``real'' (dashed lines) and 
``effective'' (full) 
slepton mass differences ($\tilde e_L - \tilde \mu_L$), normalised 
to the average $\tilde e_L,\tilde \mu_L $ mass, as a 
function of $A_0$ (in GeV). We have considered 
$R=1$, $\theta_{13}=0.1^\circ$, $M_{N_1}= 10^{10}$ GeV, $M_{N_2}=
10^{11}$ GeV, taking three distinct values for 
$M_{N_3}= 10^{13}$ GeV (red), $M_{N_3}= 10^{14}$ GeV (green) and 
$M_{N_3}= 10^{15}$ GeV (blue).
The mSUGRA parameters have been set as for point P1 (except for 
$|A_0|\leq 1.2$ TeV). Black lines denote points excluded due to the
violation of at least one experimental and/or observational constraint.
}\label{fig:seesawR1:Mslepton.M3:effective.MS}
\end{center}
\end{figure}

For a comparatively light seesaw scale (i.e. $M_{N_3} \sim
\mathcal{O}(10^{13} \text{ GeV})$) minimising the amount of flavour
violation, and taking small $|A_0|$, which minimises (diagonal)
non-universality effects for the first two generations (see
Eqs.~(\ref{eq:slepton:RGE:LLog}, \ref{eq:slepton:RGE:y})), one
verifies that both approaches nearly coincide.  As the seesaw effects
become more important, and flavour mixing increases, one clearly
verifies that ``effective'' mass difference provides the
phenomenologically reliable $\tilde e_L - \tilde \mu_L$ splittings.

Notice that the ratio ``effective''/``real'' mass splitting is always
$\gtrsim 1/2$. For small values of $| A_0 |$ and/or low seesaw scales
we have a ratio of $\sim 1$. For increasing $| A_0 |$, approaching the
turning point $m_{\tilde{\tau}_2} > m_{\tilde{\mu}_L} \rightarrow
m_{\tilde{\tau}_2} < m_{\tilde{\mu}_L}$ the ``effective''/``real''
mass splitting ratio approaches the $\sim 1/2$ limit (this also
implies that, in this region in parameter space,
 taking the ``real'' splitting could lead to a considerable
overestimation of $\Delta m_{\tilde \ell}/m_{\tilde \ell}$).  For even
higher values of $| A_0 |$ and high seesaw scales the
``effective''/``real'' mass splitting ratio can be greater than 1
order of magnitude, however this typically corresponds to scenarios
excluded by current bounds on LFV observables.

\begin{figure}[ht!]
\begin{center}
\begin{tabular}{cc}
\epsfig{file=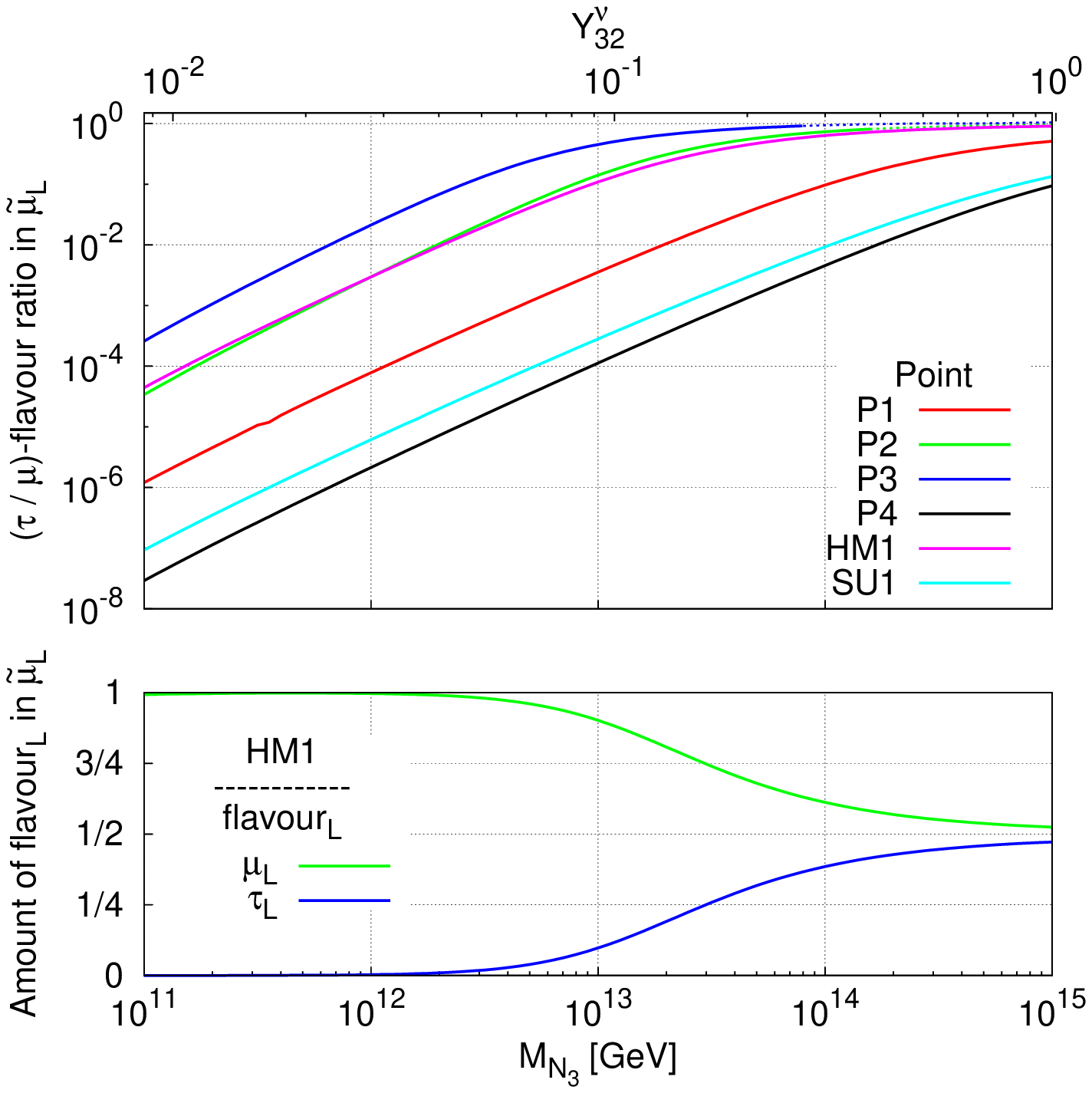, clip=, angle=0, width=80mm}
&
\epsfig{file=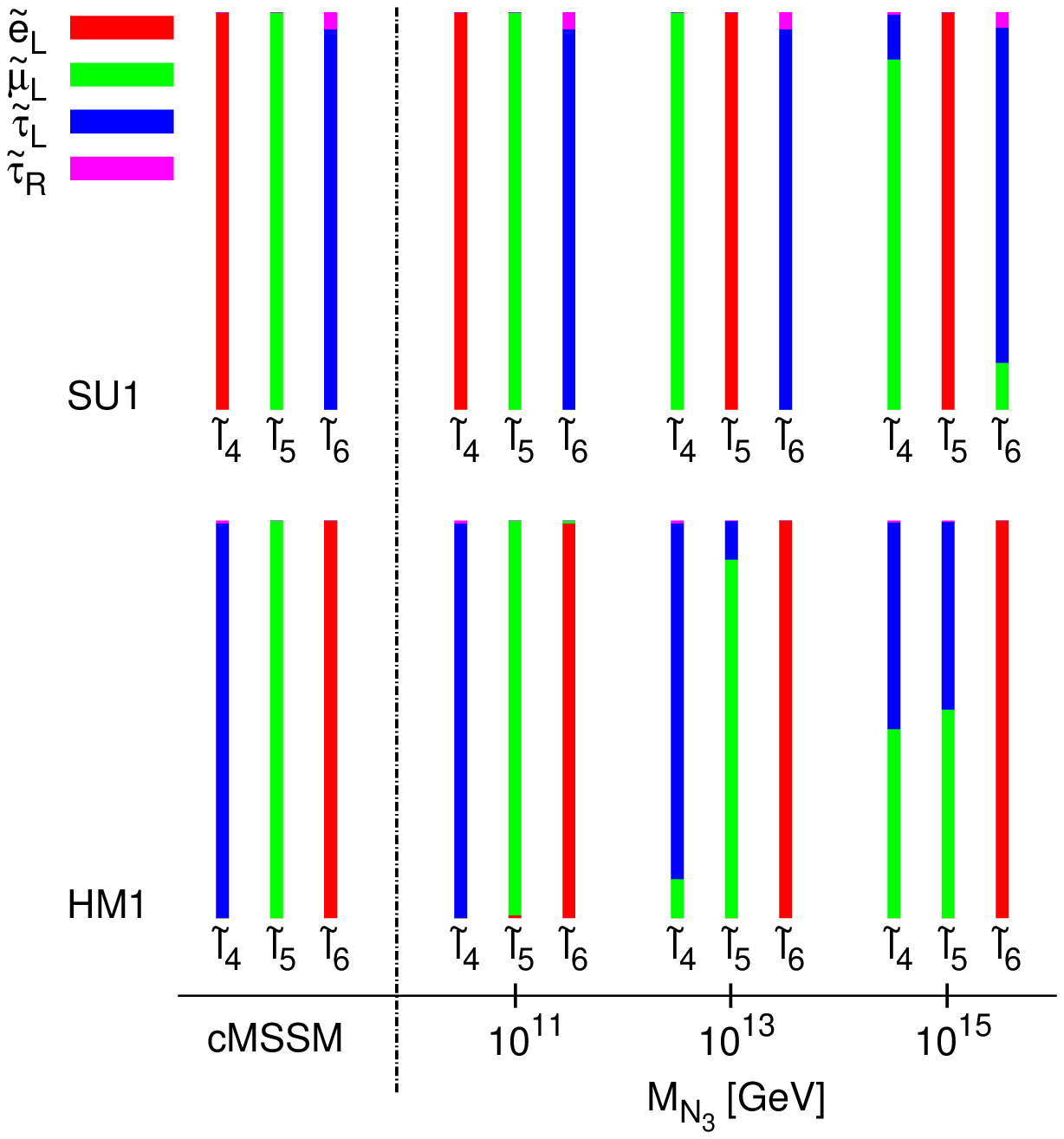, clip=, angle=0, width=70mm}
\end{tabular}
\caption{On the left, $\tau/\mu$ flavour ratio in $\tilde \mu_L$ mass eigenstate
 as a function of $M_{N_3}$
(in GeV). For the mSUGRA benchmark points of
Table~\ref{table:points1}, we set $R=1$, $\theta_{13}=0.1^\circ$ (with
$\delta=\varphi_{1,2}=0$), and take $M_{N_1}=10^{10}$ GeV, $M_{N_2}=10^{11}$
GeV. On the upper axis we display the values of $Y^\nu_{32}$.
The secondary panel illustrates $|R^{\tilde l}_{5 \mu_L}|^2$ and 
$|R^{\tilde l}_{5 \tau_L}|^2$ for the same $M_{N_3}$ interval.
On the right we depict the flavour content of the 3 heavier 
mass eigenstates: red - $\tilde e_L$, green - $\tilde \mu_L$, 
blue (magenta) - $\tilde \tau_{L (R)}$, for P5-HM1 and P6-SU1, 
illustrating both the cMSSM case (on the far left) and the type-I
seesaw, for three values of $M_{N_3}$ (with the remaining seesaw
parameters as before). 
}\label{fig:seesawR1:flavour:content}
\end{center}
\end{figure}

To better illustrate the evolution of the flavour
content of a given slepton eigenstate in the presence of the 
seesaw (even for the conservative $R=1$ case), we display on the
left panel of 
Fig.~\ref{fig:seesawR1:flavour:content} a simultaneous analysis of the 
variation of the flavour content of a 
slepton mass eigenstate, in particular of the  $\tau/\mu$
flavour ratio of the $\tilde \mu_L$ mass eigenstate as a function of
$M_{N_3}$. We present this for the different mSUGRA points,  
also showing the evolution of $Y^\nu_{32}$ on the upper axis. 

For very low seesaw scales (i.e. $M_{N_3} \sim 10^{11}$ GeV), 
flavour and mass eigenstates coincide to a very good approximation. 
As $M_{N_3}$ increases, and especially for points like P2 and P3 with
large $| A_0 |$ (enhancing the seesaw effects, see
Eq.~(\ref{eq:seesaw:DM23:open})) or points like P5-HM1 with $\left|
(m_{\tilde L}^2)_{23} / [(m_{\tilde L}^2)_{33} -(m_{\tilde L}^2)_{22} 
+ \delta_{LR}^{M^2} ] \right| \gg 1$,
\begin{equation}
\delta_{LR}^{M^2} = m^2_\tau \, \frac{ \left( A_0 - \mu \tan\beta \right)^2 }{ 
(m_{\tilde L}^2)_{33} - (m_{\tilde E}^2)_{33} + M_Z^2    
\cos2\beta ( -1/2 + 2\sin^2\theta_W ) } \,,
\end{equation} 
i.e., with a resonant-type enhancement, $\mu - \tau$ mixing becomes
maximal, and we are in the presence of truly QDFC sleptons. This is
confirmed by the lower left panel.

On the right panel of Fig.~\ref{fig:seesawR1:flavour:content}, we
symbolically represent (not to scale) the flavour composition of the
three heaviest mass eigenstates for the points P5-HM1 and P6-SU1, both
for the cMSSM and distinct seesaw scales. Notice that in the cMSSM
limit the slepton hierarchy is quite different in each case. For
P6-SU1, the seesaw immediately induces an overcross of the $\tilde e -
\tilde \mu$ eigenstates (no mixing involved); only for very large
$M_{N_3}$ does one observe a small mixing of the $\tilde \mu_L -
\tilde \tau_L$ components.

As expected  from the left panel of Fig.~\ref{fig:seesawR1:flavour:content},
large mixings occur for a much lower seesaw scale in the case of P5-HM1,
with a nearly maximal mixing for $M_{N_3} \sim 10^{15}$ GeV. This
further provides an excellent illustration of a configuration 
with QDFC sleptons.

\bigskip
One of the (perhaps) most illustrative ways of exploring the impact of
a type-I SUSY seesaw is to consider the correlated behaviour of
mass splittings and  flavour-violating decays. 
In Figs.~\ref{fig:seesawR1:BRCR.MSemutau}, we present the   
$\tilde e_L - \tilde \mu_L$ and  $\tilde \mu_L - \tilde \tau_2 $
mass differences versus BR($\tau \to \mu \gamma$) and
BR($\mu \to e \gamma$) (providing in this case
additional information on the CR($\mu-e$, Ti)). 
The data displayed in these figures
corresponds to $\tan \beta=10$, with the remaining 
mSUGRA parameters being randomly varied ($|A_0| \lesssim 1$ TeV), 
satisfying the ``standard window'' and 
requiring consistency with the dark matter and Higgs boson mass bounds. 
Regarding the right-handed neutrino spectrum, we have again taken 
(as for Fig.~~\ref{fig:seesawR1:deltaMS:emu:mutau:effective})
$M_{N_1}=10^{10}$ GeV, $M_{N_2}=10^{11}$ GeV, varying 
$10^{13} \lesssim M_{N_3} \lesssim 10^{15}$ GeV to ensure that mass
splittings are within the experimental sensitivity range.

One of the most interesting results of
Figs.~\ref{fig:seesawR1:BRCR.MSemutau} consists in the fact that almost 
the entire region in parameter space associated with a  
$\tilde e_L - \tilde \mu_L$ mass splitting $\sim\mathcal{O}(1\%)$ 
is also within the future sensitivity of low-energy facilities, 
especially for CR($\mu-e$, Ti) (even without the expected upgrade to 
$\mathcal{O}(10^{-18})$ for PRISM/PRIME)~\footnote{ $\Delta m_{\tilde
    \ell}/m_{\tilde \ell}(\tilde e_L,\tilde\mu_L)
  \sim\mathcal{O}(0.1\%)$ would still be associated to 
  predictions for CR ($\mu - e$, Ti) within the sensitivity of the
  future upgrade, $\mathcal{O}(10^{-18})$.}. 
Furthermore, any $\tilde e_L -
\tilde \mu_L$ mass splitting above 4\% would also be associated with a
$\mu \to e \gamma$ signal within MEG sensitivity.
A similar situation (albeit not so striking) is observed for 
$\tilde \mu_L-\tilde \tau_2 $ mass differences: as an example, mass
splittings above 3\%, 4\% and 6\% would be associated to 
low-energy signals of LFV within PRISM/PRIME, SuperB, 
and MEG reach, respectively.

As already observed  before, points with  a tiny 
$\tilde \mu_L - \tilde \tau_2$ mass splitting and small LFV BRs are
distributed in a more disperse way (fuzzy dropping region) due to
the fact that the corresponding 
mass splittings are mostly arising due to an enhanced $LR$ mixing
(large, negative values of $A_0$) and due to the diagonal Yukawa-tau
RGE contribution. The latter can even outplay $LR$ effects for large
$M_{1/2}$ and $|A_0|$, and sizable $\tan \beta$.

The most significant effect of considering larger values of 
$\theta_{13}$  would be to displace the depicted regions
towards higher values of BR($\mu \to e \gamma$) implying that points with 
smaller mass splittings could be  within MEG reach.
A regime of larger $\tan \beta$ would increase the mass differences,
as already seen in Figs.~\ref{fig:deltams:scan}, but the associated
``standard window'' would require a heavier SUSY spectrum. Although
the BRs do indeed augment with increasing $\tan \beta$
(see Eq.~(\ref{eq:BR:MIA:LL})), this would be balanced by the
suppression effects of having heavier sparticles in the loop.

\begin{figure}[ht!]
\begin{center}
\begin{tabular}{cc}
\epsfig{file=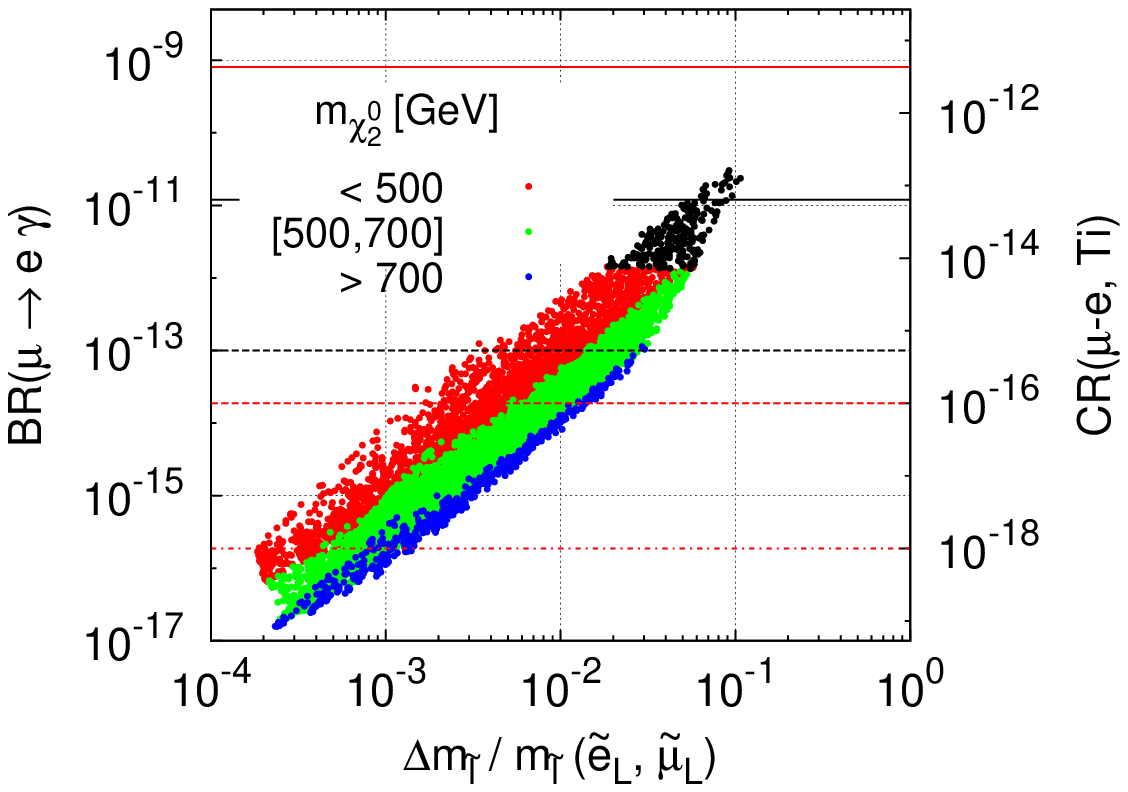, clip=, angle=0, width=80mm}
&
\epsfig{file=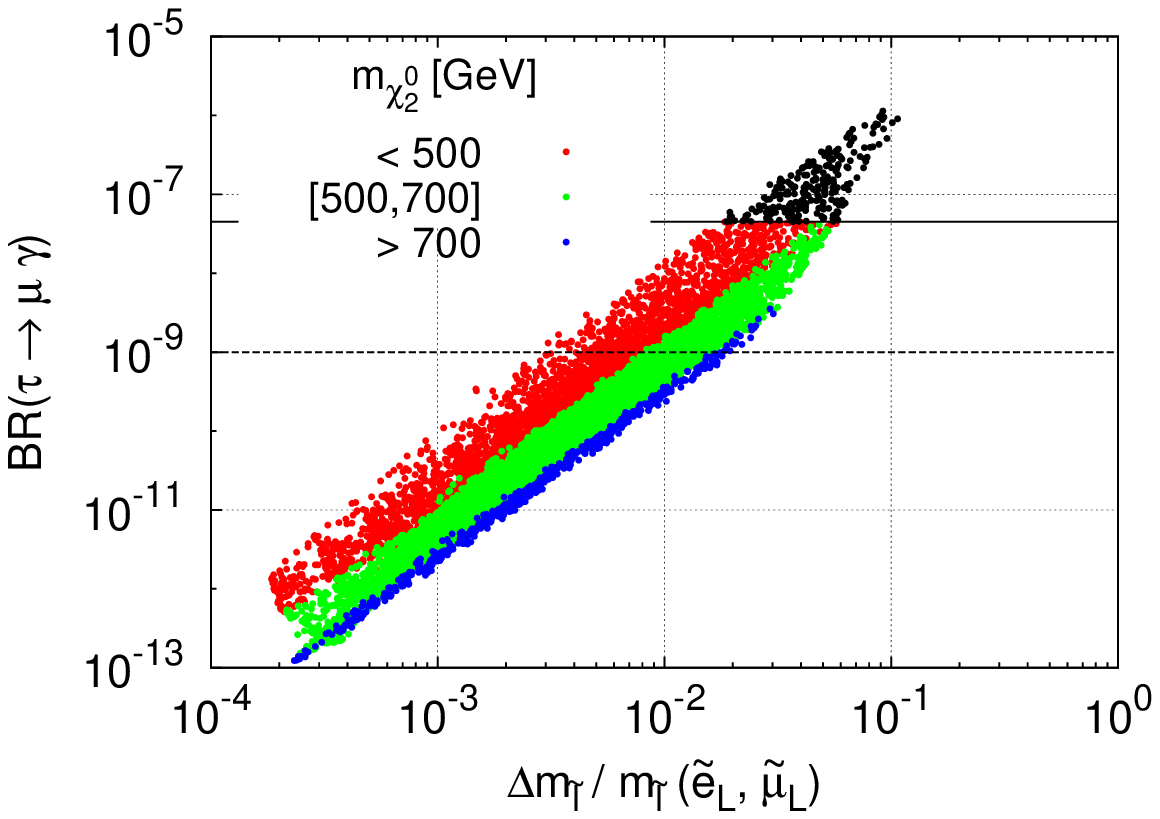, clip=, angle=0, width=80mm}\\
\epsfig{file=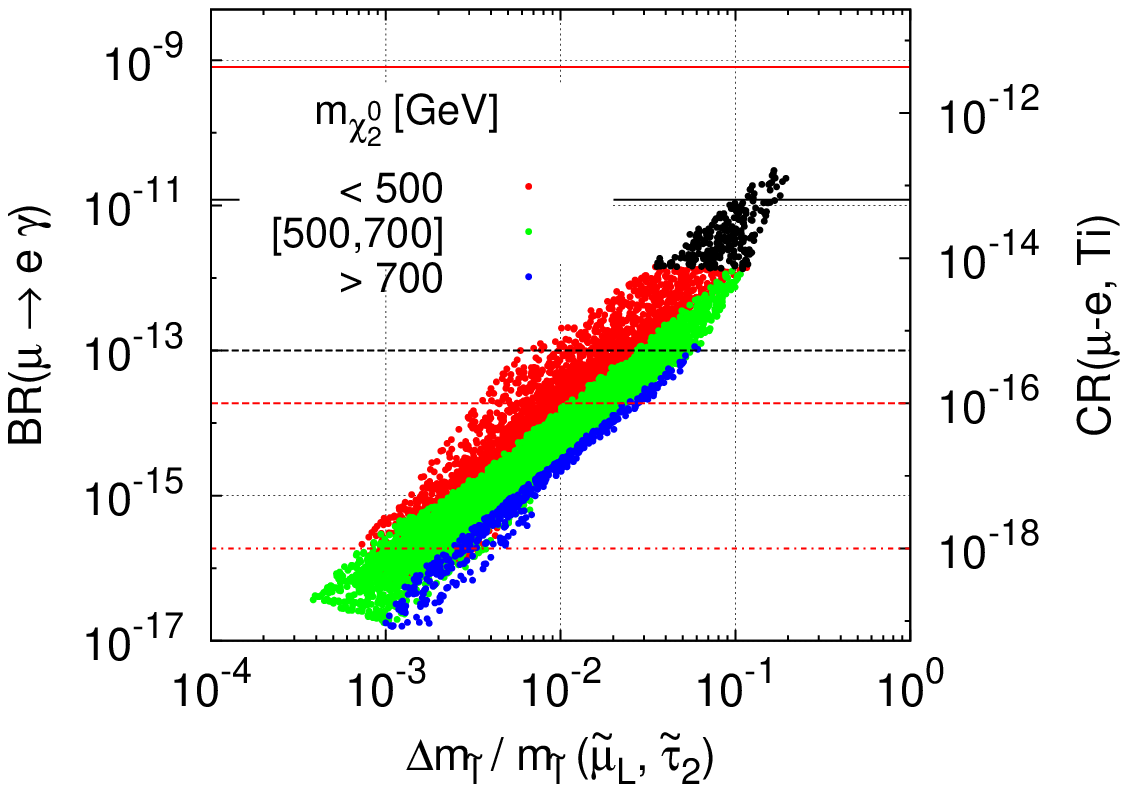, clip=, angle=0, width=80mm}
&
\epsfig{file=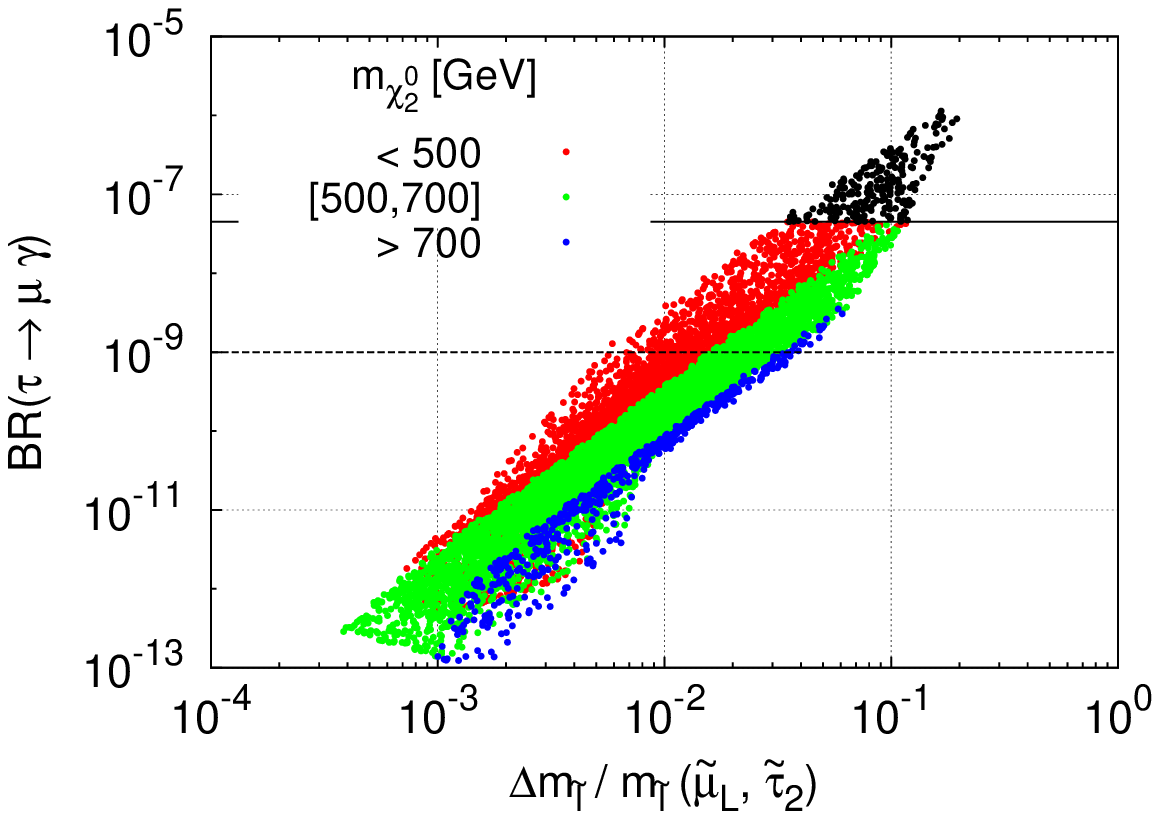, clip=, angle=0, width=80mm}
\end{tabular}
\caption{Upper left (right) panel: BR($\mu \to e \gamma$) (BR($\tau
\to \mu \gamma$)) on the left y-axis as a function of the  
mass difference $\tilde e_L - \tilde \mu_L$,
normalised to the average $\tilde e_L,\tilde \mu_L $  mass. We
display the corresponding predictions of CR($\mu-e$, Ti) on the right
y-axis. Horizontal lines denote the corresponding 
current bounds/future sensitivities.
The lower panels correspond to the mass difference $\tilde \mu_L -
\tilde \tau_2$, normalised to the average $\tilde \mu_L, \tilde
\tau_2$ mass.  
Parameters varied as in
Fig.~\ref{fig:seesawR1:deltaMS:emu:mutau:effective}. 
The colour code denotes different regimes of
$m_{\chi_2^0}$ mass, and black points are associated with the
violation of at least one experimental bound.
}\label{fig:seesawR1:BRCR.MSemutau}
\end{center}
\end{figure}

Figs.~\ref{fig:seesawR1:BRCR.MSemutau} have
been obtained in a very conservative limit for the seesaw parameters,
i.e. $\theta_i=0$~\footnote{As we will later see, non-zero values of
$\theta_i$ imply  in general larger predictions for the BRs and mass 
splittings.}, 
very small $\theta_{13}$ and hierarchical light and heavy neutrino
spectra. Nevertheless one can immediately draw some preliminary
conclusions regarding the implications of high- and low-energy LFV observables 
for the seesaw mechanism: if the LHC measures a
given mass splitting, predictions can be made regarding the associated
LFV BRs (for an already reconstructed set of mSUGRA
parameters). Comparison with current bounds (or possibly an already
existing BR measurement) may allow to derive some hints on the
underlying source of flavour violation: a measurement of a slepton mass
splitting of a few percent, together with a measurement of a
low-energy observable, for instance BR($\mu
\to e \gamma$) $\sim 10^{-12}$ at MEG
(in agreement to what
could be expected from the already reconstructed SUSY spectrum) would
constitute two signals of LFV that could be simultaneously explained through
one common origin - a type-I seesaw mechanism. 

\noindent
On the other hand, two conflicting situations can occur:
(i) a measurement of a mass splitting associated
to LFV decays experimentally excluded at present (black points in 
Figs.~\ref{fig:seesawR1:BRCR.MSemutau}) or in a region already covered by the
low-energy facilities at the time; (ii) observation of an LFV low-energy
signal, and (for an already reconstructed SUSY spectrum) approximate
slepton mass universality.
These scenarios would either suggest that
non-universal slepton masses or low-energy LFV 
would stem from a mechanism other than 
such a simple realisation of a type-I seesaw (barring accidental 
cancellations or different neutrino mass schemes). 
For instance, 
a simple explanation for the first scenario would be that
the mechanism for SUSY breaking is
slightly non-universal (albeit flavour conserving).

Although the reactor angle $\theta_{13}$ (and the Dirac phase
$\delta$) has no direct impact upon $\Delta m_{\tilde \ell}/m_{\tilde
\ell}$, its r\^ole for some LFV transitions may preclude observable
mass splittings: recall that $\Delta m_{\tilde \ell}/m_{\tilde \ell}$
is controlled by the dominant flavour violating entry of the slepton
mass matrix, which is in general $\theta_{13}$ insensitive (only the
$\tau$-$e$ and $\mu$-$e$ entries can have $\sin \theta_{13}$ as a
global factor, while for $\tau$-$\mu$, $\sin \theta_{13}$ is a second
order perturbation).  However, flavour violating transitions involving
the first generation (as is the case of $\mu (\tau) \to e \gamma$,
$\mu - e$ in nuclei, etc.) are very sensitive to
$\theta_{13}$~\cite{Antusch:2006vw}. Intermediate to large values of
the Chooz angle, $\theta_{13} \sim 5^\circ-12^\circ$, may lead to
predictions for BR($\mu \to e \gamma $) (among others) already in
conflict with current bounds.  In Fig.\ref{fig:MS:theta13}, we
consider the impact of different values of the Chooz angle
($\theta_{13}=0.1^\circ,\ 
1^\circ, \ 5^\circ$ and $12^\circ$)  for  the slepton mass splittings
and BR($\mu \to e \gamma$). From left to right,
each set of points for a given mSUGRA benchmark is associated with
increasing values of $M_{N_3}$. 

Although (and as expected) $\theta_{13}$ indeed has a very small impact for the
mass splittings, a joint measurement of flavour violation at the LHC
and at a $\mu \to e \gamma$ dedicated facility (for a given
reconstructed SUSY spectrum) strongly depends on the value of this
angle. This is readily seen from Figs.~\ref{fig:MS:theta13}, and as
an example let us notice that for P3-like spectra a $\tilde e_L -
\tilde \mu_L $ MS, in agreement with BR($\mu \to e \gamma$) bounds, is
only possible for very small $\theta_{13}$ $\lesssim
1^\circ$. Conversely, any hope of a joint signal at the LHC and at MEG
for HM1-like points requires $\theta_{13} \gtrsim 1^\circ $ (recall
that for $M_{N_3}$ much larger than $10^{15}$ GeV, the Yukawa
couplings become non-perturbative). 

\begin{figure}[ht!]
\begin{center}
\begin{tabular}{cc}
\epsfig{file=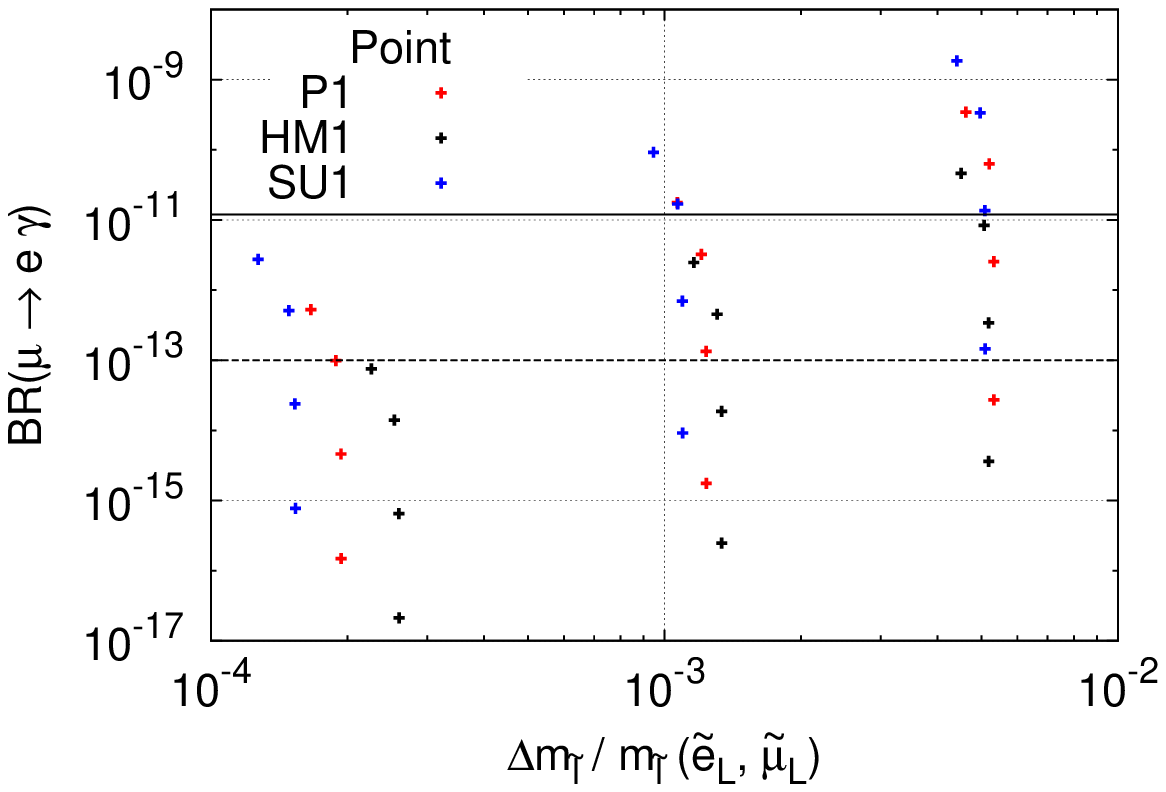, clip=, angle=0, width=80mm}
&
\epsfig{file=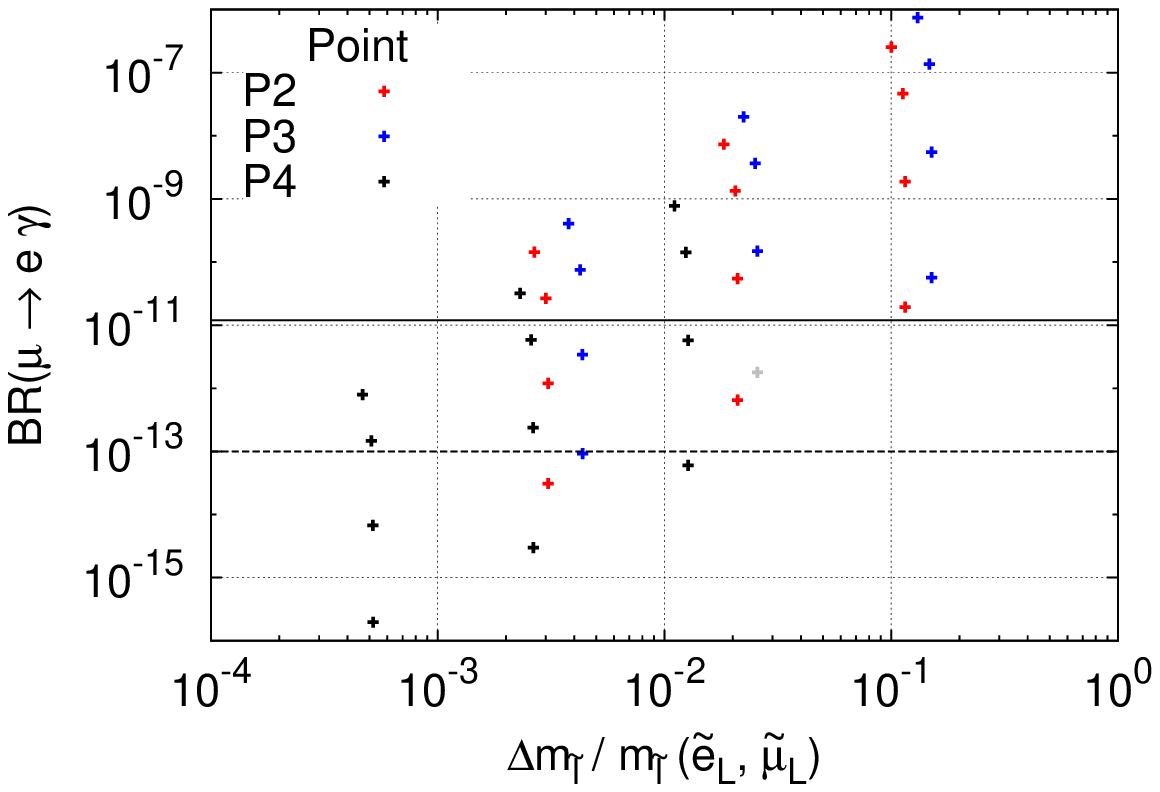, clip=, angle=0, width=80mm}
\end{tabular}
\caption{
Mass difference $\tilde e_L - \tilde \mu_L$, normalised to the average 
$\tilde e_L,\tilde \mu_L $ mass, versus BR($\mu \to e \gamma$)
for the benchmark points (Table~\ref{table:points1}) and for distinct
values of $\theta_{13}=0.1^\circ,\ 1^\circ, \ 5^\circ, \ 12^\circ$
(corresponding to increasing values of the BR). The remaining seesaw
parameters were set as $R=1$, with $M_{N_1}=10^{10}$ GeV, $M_{N_2}=10^{11}$ GeV
and $M_{N_3}=\{10^{13},10^{14},10^{15}\}$ GeV. 
Gray points are those associated with the violation of 
BR($\tau \to \mu \gamma$) and non-violation of BR($\mu \to e \gamma$).
}\label{fig:MS:theta13}
\end{center}
\end{figure}

Before addressing the impact of the additional mixing involving the
right-handed neutrinos (i.e. $\theta_i \neq 0$), let us consider how
the conclusions so far derived hold for a different hierarchy in the
heavy neutrino sector. In Fig.~\ref{fig:BRCR.MSemutau:deg}, we study
the case of degenerate right-handed neutrinos, displaying the mass
differences $\tilde e_L - \tilde \mu_L$ and $\tilde \mu_L - \tilde
\tau_2$ versus BR($\tau \to \mu \gamma$) and BR($\mu \to e \gamma$)
(also providing information on CR($\mu-e$, Ti)).

The results shown in Fig.~\ref{fig:BRCR.MSemutau:deg} should be compared to
those of Fig.~\ref{fig:seesawR1:BRCR.MSemutau} (notice that apart from
$M_{N_i}$,  all the other parameters have been identically varied with
the exception of $\theta_{13}$ which we took as $\theta_{13}= 0.1^\circ$
in the hierarchical case and $\theta_{13} = 0.1^\circ, 1^\circ,
5^\circ$ for the degenerate case. However the comparison in the lower
panel of Fig.~\ref{fig:BRCR.MSemutau:deg} is made for the same 
$\theta_{13}= 0.1^\circ$).

\begin{figure}[ht!]
\begin{center}
\begin{tabular}{cc}
\epsfig{file=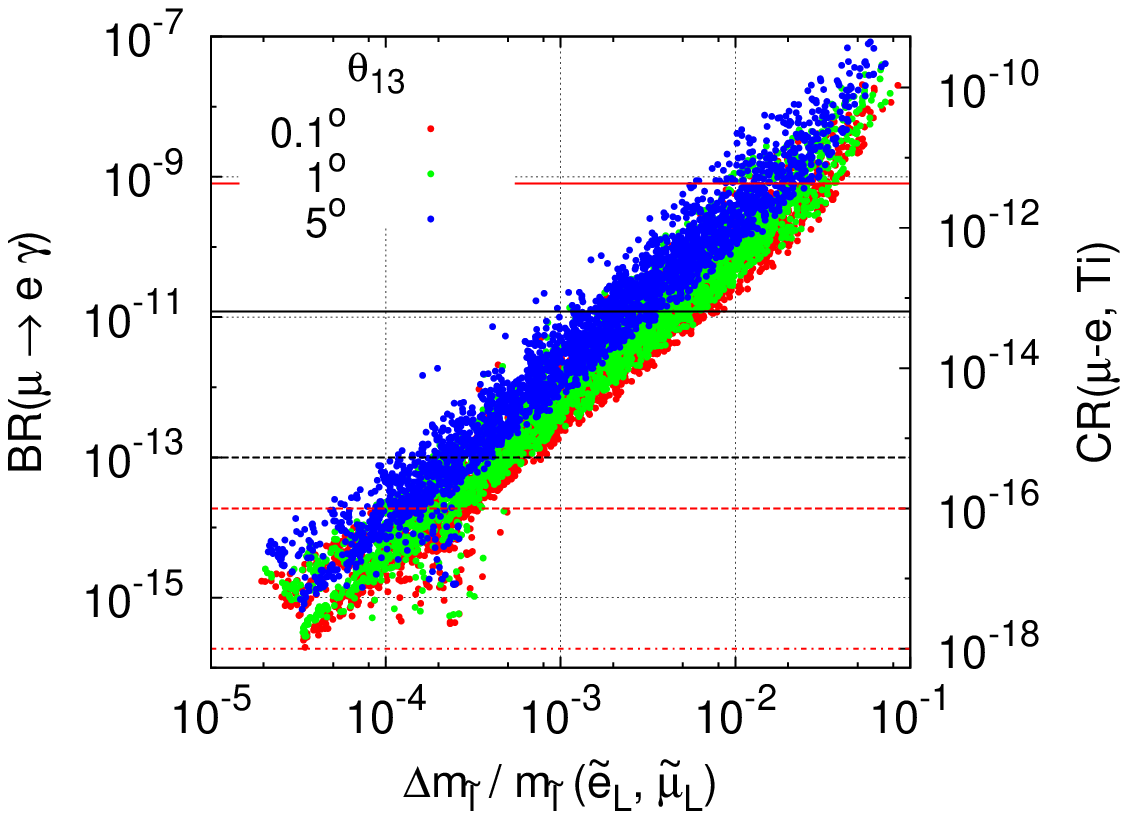, clip=, angle=0, width=80mm}
&
\epsfig{file=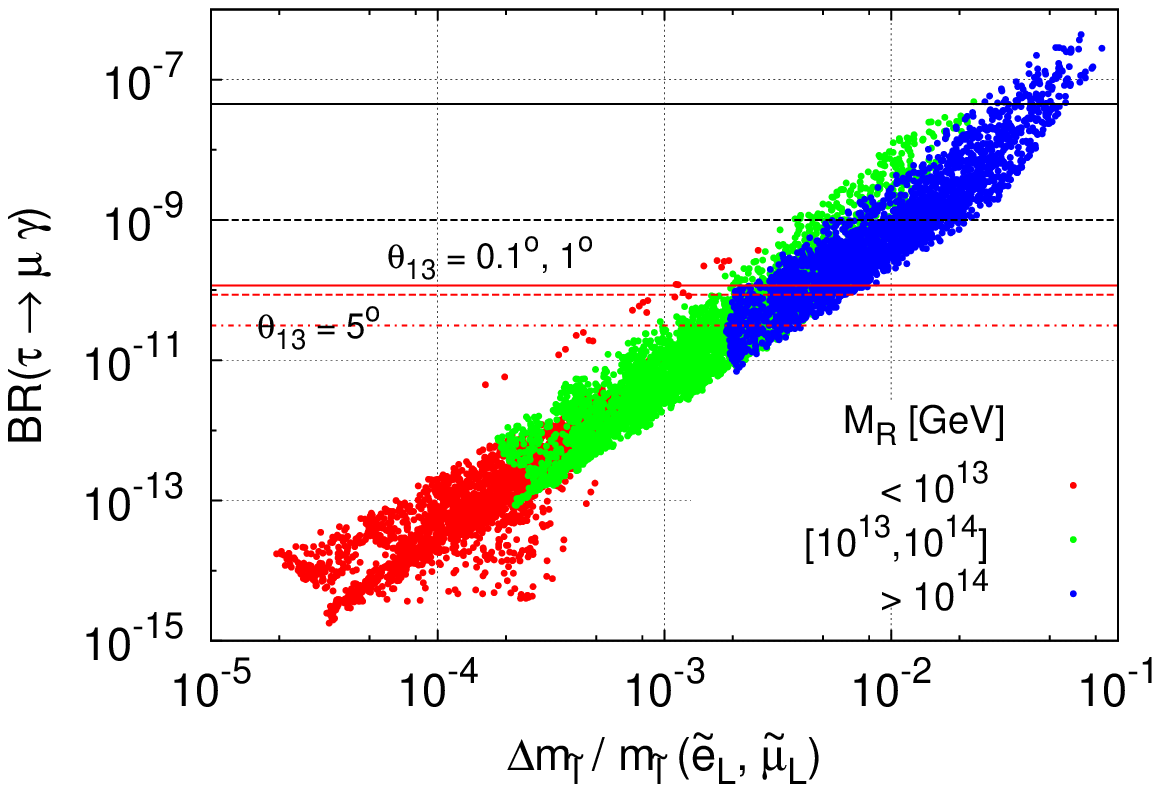, clip=, angle=0, width=80mm}\\
\end{tabular}
\epsfig{file=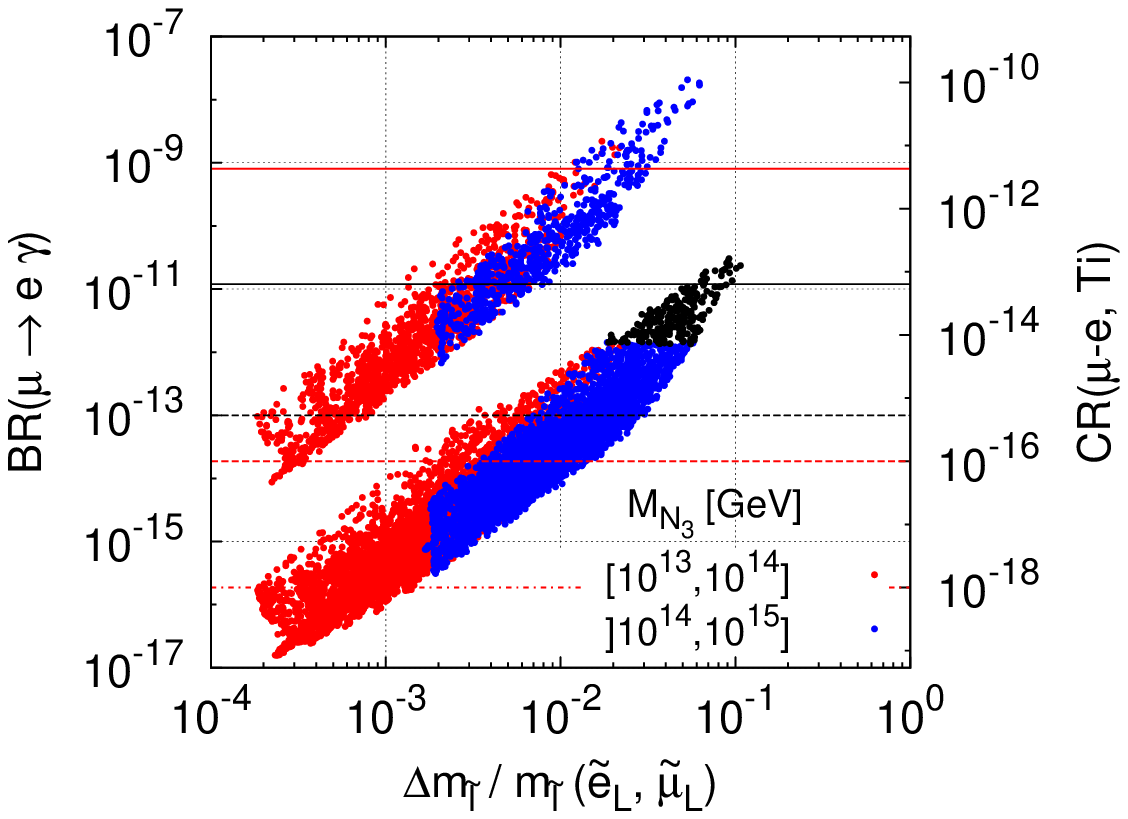, clip=, angle=0, width=80mm}
\caption{Degenerate right-handed neutrino case.
Upper left (right) panel: BR($\mu \to e \gamma$) (BR($\tau \to \mu
\gamma$)) on the left y-axis as a function of the
mass difference $\tilde e_L - \tilde \mu_L$, normalised to the average
$\tilde e_L,\tilde \mu_L $ mass. We display the corresponding
predictions of CR($\mu-e$, Ti) on the right y-axis.
Horizontal lines denote the corresponding current bounds/future
sensitivities and regimes of $\theta_{13}$ (in the upper right
panel). Leading to the scan, we
set $\tan \beta=10$, and the remaining mSUGRA parameters were randomly
varied (with $|A_0| \lesssim 1$ TeV, satisfying the ``standard
window'' constraint and requiring consistency with the dark matter and
Higgs boson mass bounds).  For the seesaw parameters we have taken
$R=1$, $\theta_{13}=0.1^\circ,\,1^\circ,\,5^\circ$ (with
$\delta=\varphi_{1,2}=0$), 
and $M_{N_1} = M_{N_2} = M_{N_3}= M_R$ being varied as $10^{12} \text{ GeV} 
\lesssim M_R \lesssim 10^{15} \text{ GeV}$.
In the upper left (right) panel colour code denotes different regimes of 
$\theta_{13}$ ($M_R$). Lower panel: comparison of degenerate (region with 
higher BR) and hierarchical (region with lower BR) spectrum. Same scan as 
before, but now taking only $\theta_{13}=0.1^\circ$ and $10^{13} \text{ GeV} 
\lesssim M_R \lesssim 10^{15} \text{ GeV}$. For the hierarchical case, same 
scan as in Fig.~\ref{fig:seesawR1:BRCR.MSemutau}. Colour code denotes different 
regimes of $M_{N_3}$ (or $M_R$ for the degenerate case) and black points are 
associated with the violation of the experimental bound on $BR(\tau
\to \mu \gamma)$.
}\label{fig:BRCR.MSemutau:deg}
\end{center}
\end{figure}

As seen from the direct comparison of the high- and low-energy flavour
violation prospects, potential measurements (and even negative
searches) can hint towards the RH neutrino hierarchy, in the 
case $R=1$. This is especially true in the limit of very small
$\theta_{13}$: if a sizable mass splitting $\tilde e_L - \tilde \mu_L$
$\sim \mathcal{O}(10^{-2})$ is measured at the LHC, then a hierarchical
spectrum appears to be the only candidate to explain such a signal.
If a $\Delta m_{\tilde \ell}/m_{\tilde\ell }(\tilde e_L, \tilde \mu_L)$ between 
$10^{-3}$ and $10^{-2}$ is reconstructed, and a $\mu \to e \gamma$
decay is observed, then both hierarchies are hard to disentangle based
on observation. For the same mass splitting range, CR($\mu-e$, Ti)
within reach of PRISM/PRIME (and a potential upgrade), would strongly
favour the hierarchical spectrum. 
Finally, should the LHC be able to measure $\Delta
m_{\tilde \ell}/m_{\tilde \ell} (\tilde e_L, \tilde \mu_L)$ $\sim
\mathcal{O}(10^{-4})$, an observation of $\mu \to e \gamma$ could be
due to either RH spectrum (although in this case larger values 
of $\theta_{13}$ would be required to accommodate the hierarchical
hypothesis). 

\bigskip
To conclude the study of the conservative limit of $R=1$ in the type-I
SUSY seesaw, we conduct a distinct analysis, explicitly focusing on
the dependence of the mass splittings on different mSUGRA parameters.
For fixed $\tan \beta=10$, a 
scan is performed over $m_0$ and $M_{1/2}$, taking several discrete
values of $A_0$ (always complying with the ``standard window''
requirements).
We fix all seesaw parameters other than $M_{N_3}$, which is varied as
to ensure that each point  has BR($\mu \to e \gamma$)
and BR($\tau \to \mu \gamma$) in agreement with current experimental bounds.
The results are shown in the left panel of
Fig.~\ref{fig:deltaMS:chi2:A0}, which clearly displays 
how a potentially measurable mass
difference (in agreement with the different low-energy LFV bounds) translates 
the interplay of $A_0$ and $M_{N_3}$. 
The two regimes (other than the nearly constant $\Delta m_{\tilde
\ell}/m_{\tilde \ell}$ for $A_0=0$) reflect the different bounds which
are effectively preventing larger values of $\Delta m_{\tilde
\ell}/m_{\tilde \ell}$: for the ascending slope, the mass splittings
are almost insensitive to the actual value of $A_0$, since in this
case the values of $Y^{\nu}_{ij}$ that saturate the current bounds on
BR($l_i \to l_j \gamma$) -- BR($\tau \to \mu \gamma$) for the
$\theta_{13} = 0.1^\circ$ regime considered -- are attainable without
violating the requirement of perturbative Yukawa couplings. On the
right-handed (descending) part of each curve, the values of $M_{N_3}$
are at the maximum value allowed by perturbative Yukawa couplings
alone while BR($l_i \to l_j \gamma$) is below current bounds. In this
latter case, $A_0$ is the discriminatory factor that enhances the
$\propto Y^{\nu \dagger} L Y^{\nu}$ radiative corrections to the soft
slepton mass matrices, in turn constraining the maximal value of
$\Delta m_{\tilde \ell}/m_{\tilde \ell}$ for a given $A_0$.
In each pair of lines, the one whose maximum mass splitting occurs for 
a lighter $\chi_2^0$ corresponds to the positive value of $A_0$.
Larger values of $|A_0|$ would lead to an increase in the 
maximal values of $\Delta m_{\tilde \ell}/m_{\tilde \ell}$, which would also
be associated with a heavier gaugino spectrum.

\begin{figure}[ht!]
\begin{center}
\hspace*{-10mm}
\begin{tabular}{cc}
\epsfig{file=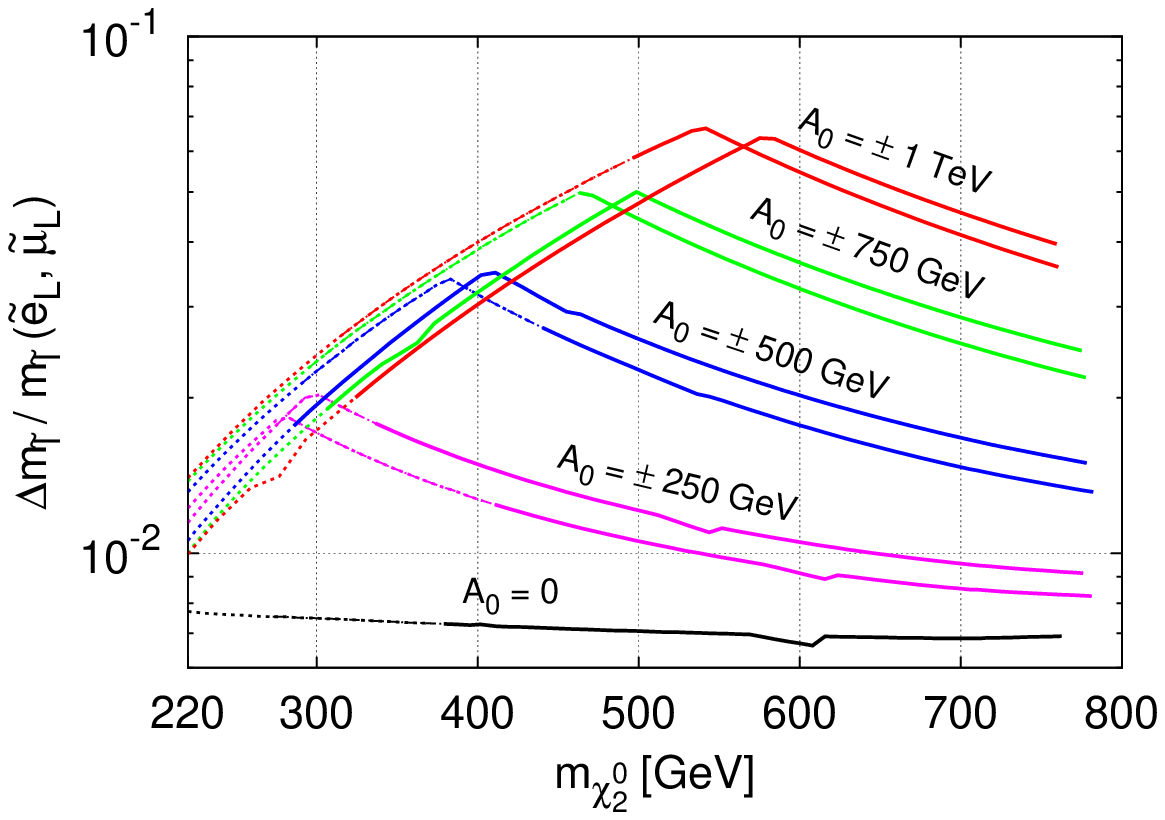, clip=, angle=0, width=85mm}
&
\epsfig{file=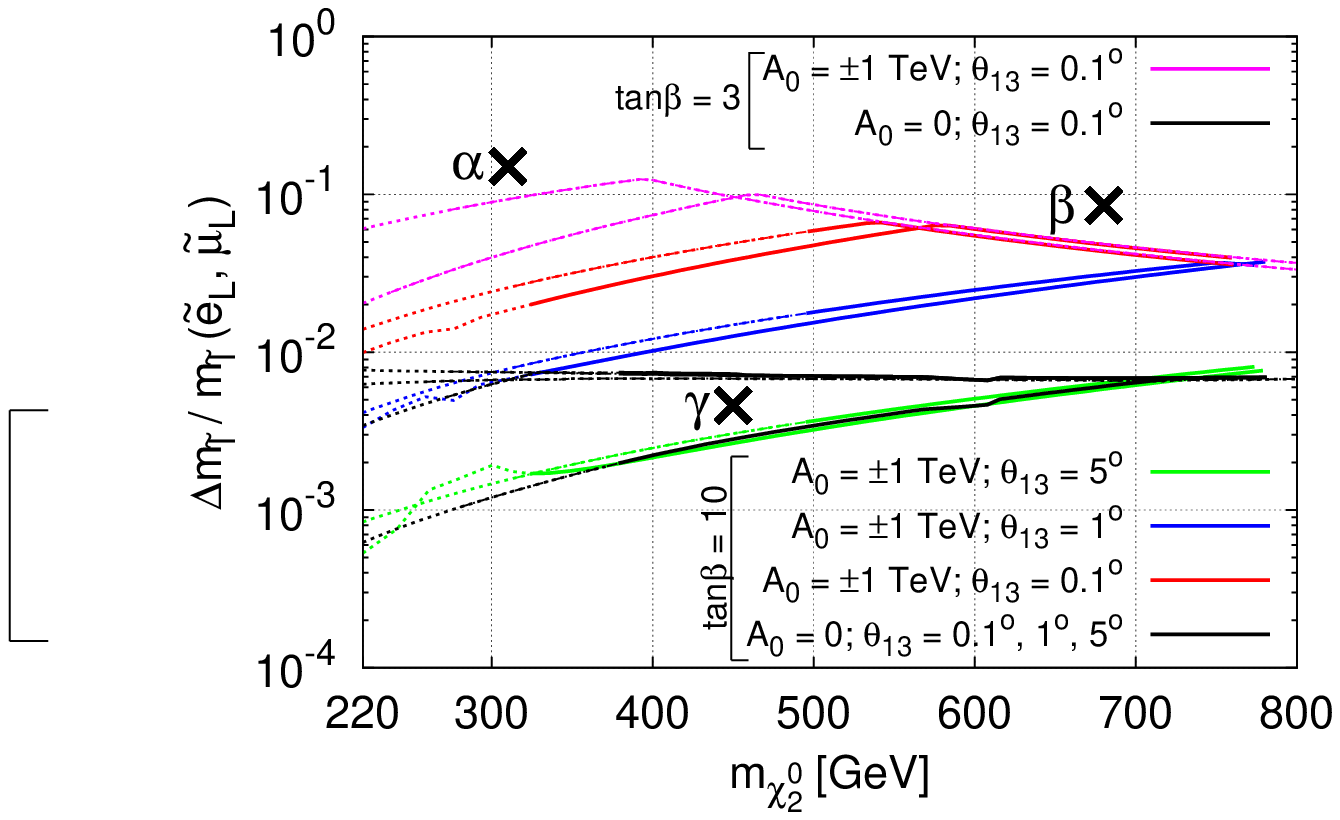, clip=, angle=0, width=85mm}
\end{tabular}
\caption{Mass difference $\tilde e_L - \tilde \mu_L$, normalised to the average 
$\tilde e_L,\tilde \mu_L$ mass, as a function of $m_{\chi_2^0}$ (in GeV).
On the left we consider different values of $|A_0|$, 
setting $\tan \beta=10$ and $\theta_{13}=0.1^\circ$,
while on the right we fix $A_0=\{-1,0,1\}$ TeV, and take several
choices for $\theta_{13}=0.1^\circ,1^\circ,5^\circ$ with $\tan
\beta=10$ and $\theta_{13}=0.1^\circ$ for $\tan \beta=3$.
We vary $m_0$ and $M_{1/2}$ in such a way that we satisfy the 
requirement of a viable $\Omega h^2$ in the co-annihilation region.
The seesaw parameters have been taken as $R=1$,
with hierarchical right-handed neutrinos, 
$M_{N_1}=10^{10}$ GeV, $M_{N_2}=10^{11}$ GeV, with 
$M_{N_3}$ varied as to satisfy BR($\mu \to e \gamma$) $\leq 1.2
\times 10^{-11}$ and BR($\tau \to \mu \gamma$)$\leq 4.5
\times 10^{-8}$. 
Dotted lines denote points where the kinematical constraints are
outside the ``standard window'' and dashed lines are for $m_h \leq
114.4$ GeV, while satisfying the ``standard window'' requirement.
We have displayed three points $\alpha,\beta,\gamma$
used for the subsequent discussion in the text.
}\label{fig:deltaMS:chi2:A0}
\end{center}
\end{figure}

A similar study is conducted on the right panel of 
Fig.~\ref{fig:deltaMS:chi2:A0}, taking discrete values
of $A_0$, and studying different combinations of $\tan \beta$ and
$\theta_{13}$. We notice that of all the SUSY seesaw parameters 
likely to be measurable, $A_0$, $\tan \beta$, 
and $\theta_{13}$, are those expected to be 
measured/reconstructed at a later stage. Just like in the 
previous figure, the two  regimes for the slopes  again denote
the bounds for BR($\ell_i \to \ell_j \gamma$) (ascending), and 
$Y^\nu \sim 1$ (descending).
The curves corresponding to $\tan \beta=10$ and maximal
$\theta_{13}$ present very low $\Delta m_{\tilde \ell}/m_{\tilde
  \ell}$: this is a direct consequence of having to take comparatively
low values of the heaviest right-handed neutrino mass in order to
comply with the BR($\mu \to e \gamma$) bound (easily saturated for
$\theta_{13}=5^\circ$). 

From Figs.~\ref{fig:deltaMS:chi2:A0}, it is clear that, even in the
very conservative case of $R=1$, without 
the reconstruction of the mSUGRA parameters and measurement of
$\theta_{13}$, very little can be said regarding the expected values
of $\Delta m_{\tilde \ell}/m_{\tilde \ell}$, apart from some remarkable
exceptions, which we proceed to discuss. 
Let us address  the hypothetical measurements of
$m_{\chi_2^0}$ and $\Delta m_{\tilde \ell}/m_{\tilde \ell}$
corresponding to three points highlighted in the right panel of 
Figs.~\ref{fig:deltaMS:chi2:A0} ($\alpha,\ \beta ,\ \gamma$). 
A measurement close to point $\gamma$ would provide very little information
regarding the underlying 
source of LFV: different choices of either $M_{N_3}$, $\tan \beta$ or
$\theta_{13}$ could easily account for such an observation. The case
denoted by $\beta$ is already more interesting: although large values
of  $\Delta m_{\tilde \ell}/m_{\tilde \ell}$ (in association with a
heavy $\chi_2^0$) can be obtained for very large $|A_0|$, 
complying with the bound on the LSP relic density becomes increasingly
more complicated in these regimes, so that a correct $\Omega h^2$ might 
eventually preclude compatibility of a type-I SUSY seesaw with
$\beta$. A set of measurements $\sim \alpha$ (and $\beta$ to a certain extent)
would certainly provide the most challenging scenario: 
such a mass splitting, in agreement with current bounds on
low-energy LFV observables, and for such a light gaugino spectrum,
cannot be accounted for by a type-I SUSY seesaw (in the $R=1$ limit). 
Another mechanism of flavour violation (or at least flavour
non-universality) should be at work in this case, e.g. non-universal
soft-breaking slepton masses. This will also be true for $\theta_i
\neq 0$, as in this case the low-energy LFV observables
would be enhanced making it even more difficult to
account for a set of measurements $\sim \alpha$ and $\beta$, while respecting
the current bounds on low-energy LFV observables.

\bigskip
We will now depart from the conservative (albeit singular) $R=1$ case,
allowing for additional sources of flavour violation through the
$\theta_i$ angles. Given that the right-handed neutrino sector (both
spectrum and mixings) is experimentally unreachable,\footnote{If the
  SUSY seesaw is indeed responsible for LFV observables within
  experimental sensitivity, as well as for the BAU via leptogenesis, then
  the seesaw scale lies in general well above the TeV range ($\sim
  10^{10} \text{ GeV} - 10^{15}$ GeV).} this 
translates into having parameters about which one has no direct information. 
As mentioned in Section~\ref{susy:seesaw:lfv}, one can impose indirect
constraints on the $\theta_i - M_{N_j}$ parameter space, choosing 
$R$-matrix angles and heavy neutrino hierarchies suggested by
phenomenological arguments, such as generating the observed BAU from
thermal leptogenesis and complying with lepton EDMs. For simplicity,
and motivated by the analysis of the $R=1$ limit, we have selected
additional scenarios, that will play 
the r\^ole of seesaw ``benchmark'' points: 
three configurations of the heavy neutrino spectrum and
reactor angle $\theta_{13}$ are summarised in
Table~\ref{table:seesaw:benchmark} and  can 
be applied to the different mSUGRA points
(P1$^\prime$, etc.). P$^{\prime (\prime \prime)}$ denotes a case of
  nearly degenerate $N_1$ and $N_2$ ($N_2$ and $N_3$), while 
P$^{\prime\prime \prime}$ is the limit of a strongly hierarchical
right-handed spectrum, with $M_{N_3}$ close to its maximal value (as
allowed by the perturbativity bound on $Y^\nu$).
We do not consider the case of degenerate RH neutrinos 
as the associated phenomenology 
will not differ significantly from the $R=1$  case already discussed.

\begin{table}[ht]
\begin{center}
\begin{tabular}{|c|c|c|c|c|}	
\hline
Point & 
$M_{N_1}$ (GeV) & $M_{N_2}$ (GeV) & $M_{N_3}$ (GeV) & 
$\theta_{13}$ \\ \hline
P$^{\prime}$& $10^{10}$ & $5 \times 10^{10}$ & $5 \times 10^{13}$ &
$0.1^\circ$\\ \hline 
P$^{\prime\prime}$&$10^{10}$ & $10^{12}$ & $5 \times 10^{12}$ & $1^\circ$\\ \hline
P$^{\prime\prime\prime}$&$10^{10}$ & $10^{12}$ & $10^{15}$ & $0.1^\circ$\\ \hline
\end{tabular}
\end{center}
\caption{Seesaw benchmark points. For the remaining parameters we have
taken $R=1$, and $\varphi_1 = \varphi_2 = 
\delta = 0$.}\label{table:seesaw:benchmark}
\end{table}

In order to summarise the results, 
we now display a comprehensive scan over  the seesaw parameters, in
particular over the complex angles of the $R$ matrix. 
We consider mSUGRA benchmark point P1 with $M_{N_i} = \{ 10^{10},
10^{11}, 10^{13} \}$ GeV, and randomly scan over $| \theta_i | \in [0,
\pi]$ and $\arg \theta_i \in [-\pi,\pi]$. We also select four values
of $\theta_1$ and vary $\theta_{2,3}$ as favoured by
leptogenesis~\cite{Antusch:2006gy} (see end of
Section~\ref{susy:seesaw:lfv}), and highlight these regions via a
different colour code, for illustrative purposes only.

\begin{figure}[ht!]
\begin{center}
\begin{tabular}{cc}
\hspace*{-10mm}
\epsfig{file=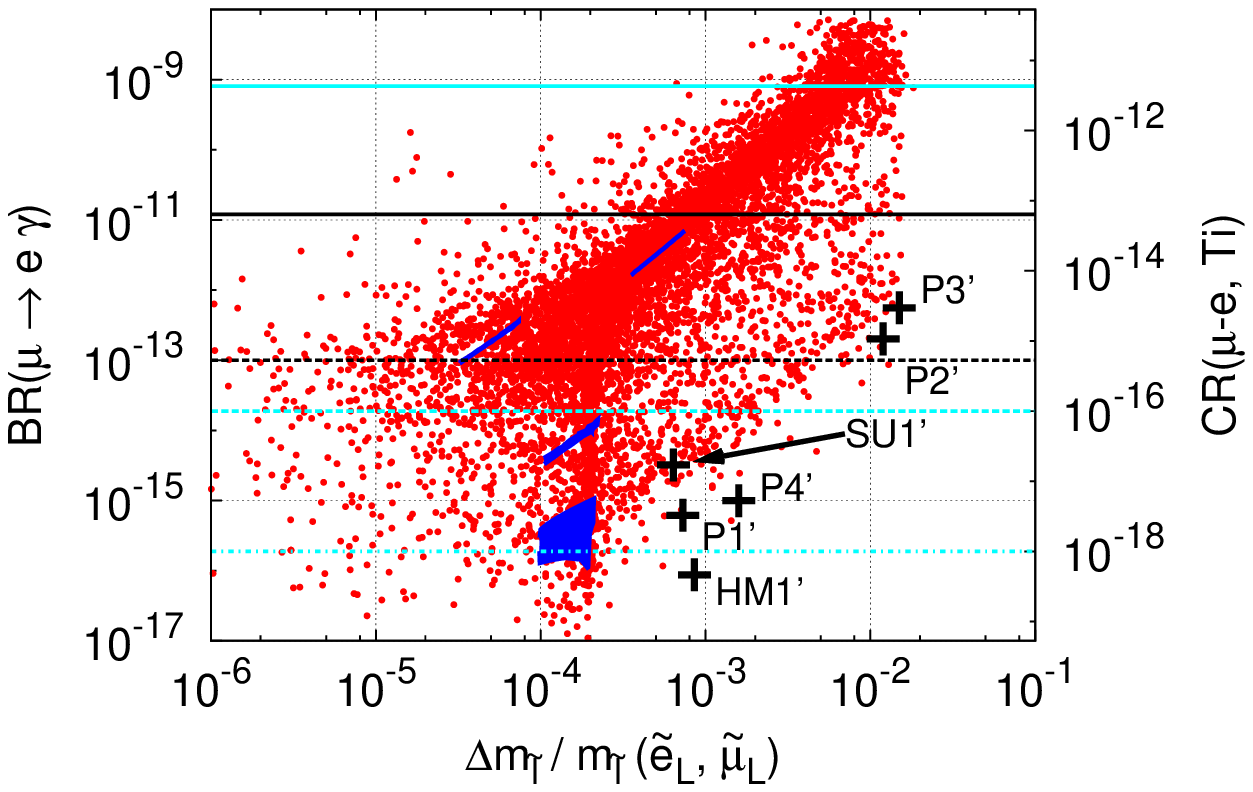, clip=, angle=0, width=90mm}&
\epsfig{file=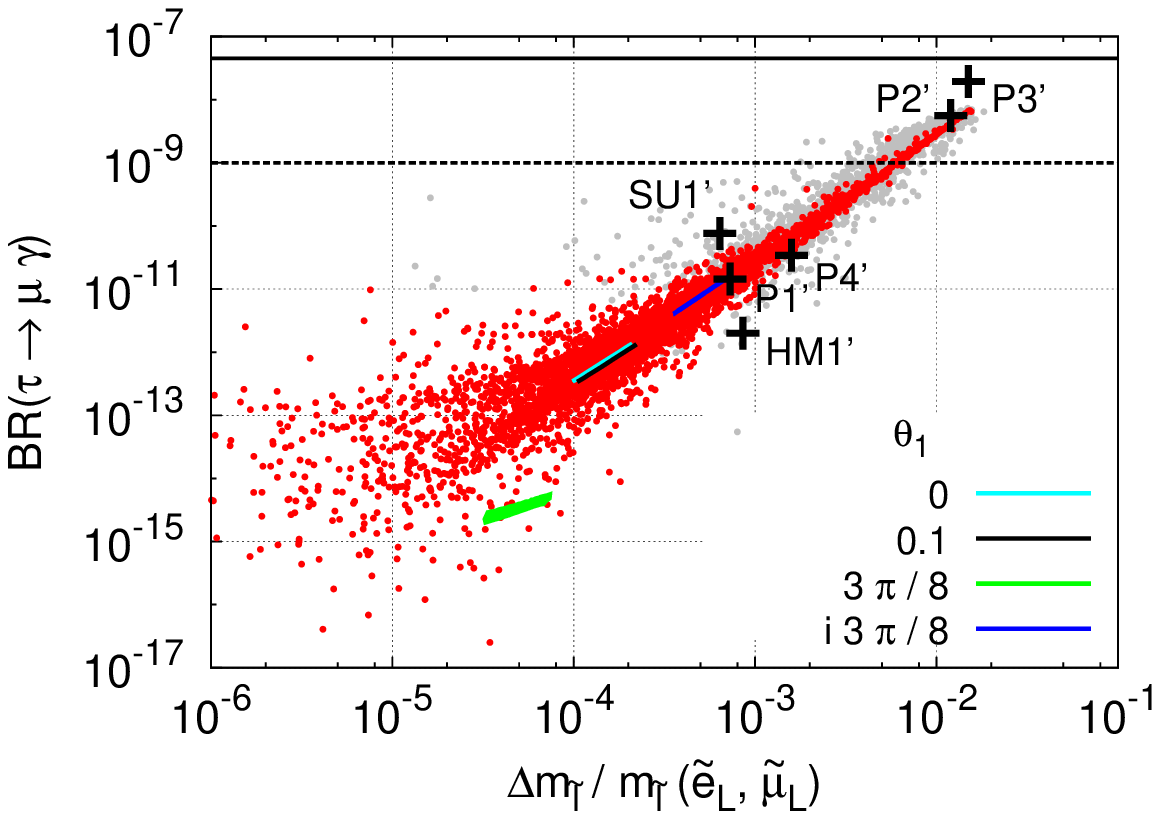, clip=, angle=0, width=80mm}
\end{tabular}
\caption{Left (right) panel BR($\mu \to e \gamma$) (BR($\tau \to \mu
  \gamma$))  as a function of the
mass difference $\tilde e_L - \tilde \mu_L$, normalised to the average
$\tilde e_L,\tilde \mu_L $ mass, for
seesaw variations of point P1. We display the corresponding
predictions of CR($\mu-e$, Ti) on the right secondary y-axis.
Horizontal lines denote the corresponding current bounds/future
sensitivities. We have taken $\theta_{13}=0.1^\circ$, $M_{N_1}=10^{10}$ 
GeV, $M_{N_2}=10^{11}$ GeV, and $M_{N_3}=10^{13}$ GeV,
and the complex $R$ matrix angles have been randomly varied as
$|\theta_i| \in [0, \pi]$ and $\arg(\theta_i) \in [-\pi,
\pi]$. The
crosses correspond to the $R=1$ case of the benchmark points
 P$^\prime$ (see Table~\ref{table:points1} and
 \ref{table:seesaw:benchmark}).
In each panel the four highlighted regions correspond to $\theta_1=0,
0.1, 3 \pi/8$ and $i\, 3 \pi/8$, with $\theta_{2,3}$ scanned as favoured
by leptogenesis (see text for discussion). 
In the left panel we show
in solid blue the leptogenesis favoured regions  for
different values of $\theta_1=0, 0.1, 3 \pi / 8$ and $i\, 3 \pi / 8$,
from lower  
to higher BR($\mu \to e \gamma$). On the right panel these regions are
identified in the inset.
}\label{fig:BRCR.MS:Rcomplex}
\end{center}
\end{figure}

The full realisation of a type-I seesaw leads to very rich scenarios
(albeit less predictive) for flavour violation, as can be seen
from Fig~\ref{fig:BRCR.MS:Rcomplex}. Recall that for an mSUGRA
configuration similar to P1 (see e.g.~Fig.~\ref{fig:MS:theta13}), the
associated BR was $\mathcal{O}(10^{-16})$, with a mass splitting
around $2 \times 10^{-4}$. Under a generic choice of $\theta_i$, the
associated amount of FV is extremely enhanced (even already excluded
by current bounds in some cases). This confirms that, 
barring cancellations, the case $R=1$ clearly constitutes a
case of minimal flavour violation, inducing low values for the BRs and
CR. Regarding the highlighted regions (corresponding to $\theta_{2,3}$ in
the ranges given at the end of Section~\ref{susy:seesaw:lfv}), the
distinct disconnected regions correspond, for increasing values of
BR($\mu \to e \gamma$), to $\theta_1=0, 0.1, 3 \pi/8$ and $i\, 3 \pi/8$. 
For a SUSY spectrum similar to P1, a type-I seesaw could easily
account for slepton mass differences within the sensitivity of both
the LHC and of low-energy flavour dedicated experiments (possibly
associated to  viable leptogenesis scenario). 

For the other seesaw benchmark points, an identical scan would
translate in scatter regions of comparable ranges, similarly
positioned with respect to the different benchmark point. 

In Fig.~\ref{fig:BRCR.MS:Rcomplex:M3:theta13}, we conduct a general
scan over the $\theta_i$ parameter space, again displaying different
low-energy LFV observables as a function of the $\tilde e_L - \tilde
\mu_L$ mass difference. Given the amount of collider simulations
conducted for the LHC benchmark points~\cite{Ball:2007zza,ATLAS}, we
choose for this overview of the SUSY seesaw the LHC points P5-HM1 and
P6-SU1. We randomly scan over $|\theta_i| \lesssim \pi$, and
$\arg(\theta_i) \in [-\pi, \pi]$, taking $\theta_{13}=0.1^\circ$, and
choosing three representative values for $M_{N_3}$.

As can be seen from the first panel of
Fig.~\ref{fig:BRCR.MS:Rcomplex:M3:theta13}, if a SUSY type-I seesaw is indeed
at work, and $\theta_{13}$ has been constrained to be extremely small,
a measurement of $\Delta m_{\tilde \ell}/m_{\tilde \ell} (\tilde e_L ,
\tilde \mu_L)$ between $0.1\%$ and 1\%, in association with a
reconstructed sparticle spectrum similar to P5-HM1, would be
accompanied (with a significant probability) by the observation of
BR($\mu \to e \gamma$) at MEG.  On the other hand, even for very large
values of $M_{N_3}$, the constraints on the parameter space from
BR($\mu \to e \gamma$) preclude the observation of a $\tau \to \mu
\gamma$ transition for an HM1-like SUSY spectrum.  From the comparison
of both left and right panels, it is also manifest that the slepton
mass splittings are predominantly generated from mixings involving the
$\tau-\mu$ sector: this can be seen from the strongly correlated
behaviour of $\Delta m_{\tilde \ell}/m_{\tilde \ell} (\tilde e_L ,
\tilde \mu_L)$ and BR($\tau \to \mu \gamma$), implying that both are
governed by the term proportional to $({Y^\nu}^\dagger L Y^\nu)_{23}$
(see Section~\ref{lfv:lhc}). In this case, the three seesaw benchmark
points appear almost superimposed on the $R=1$ (i.e. $\theta_i=0$)
central diagonal region, and their corresponding $\Delta m_{\tilde
\ell}/m_{\tilde \ell}$ and BRs follow the LLog dependency
(i.e. BR $\propto M_{N_3}^2 \log^2 M_{N_3} $).

Although LHC production prospects have to be taken into account, 
when compared to P5-HM1, P6-SU1 offers  a less promising framework  for
the observation of sizable mass splittings at the LHC (unless a
precision of around $10^{-3}$ for $\Delta m_{\tilde \ell}/m_{\tilde
\ell} (\tilde e_L , \tilde \mu_L)$ can indeed be achieved). In the
latter case, it is expected that a determination of 
$\Delta m_{\tilde \ell}/m_{\tilde \ell} (\tilde e_L , \tilde \mu_L)$
be accompanied by 
evidence of LFV in muon decays. However the most interesting lepton
flavour signature of P6-SU1 is related to its potential to induce
large BR($\tau \to \mu \gamma$), within the future sensitivity of
SuperB~\cite{Bona:2007qt}: a measurement of 
$\Delta m_{\tilde \ell}/m_{\tilde \ell} (\tilde e_L , \tilde \mu_L)$ 
$\sim 0.1\% - 1\%$ at the LHC would imply
BR($\tau \to \mu \gamma$) $\gtrsim 10^{-9}$, and would hint towards a
heavy seesaw scale, $M_{N_3} \gtrsim 10^{13}$ GeV. 
For shortness, we do not present the analog of
Fig.~\ref{fig:BRCR.MS:Rcomplex:M3:theta13} for the mass difference
$\tilde \mu_L - \tilde \tau_2$, as little new information is conveyed 
by these figures. Moreover, we have also verified that larger values
of $\theta_{13}$ would only have the small effect of slightly
augmenting the concentration of the points  around the central region.

\begin{figure}[ht!]
\begin{center}
\begin{tabular}{cc}
\epsfig{file=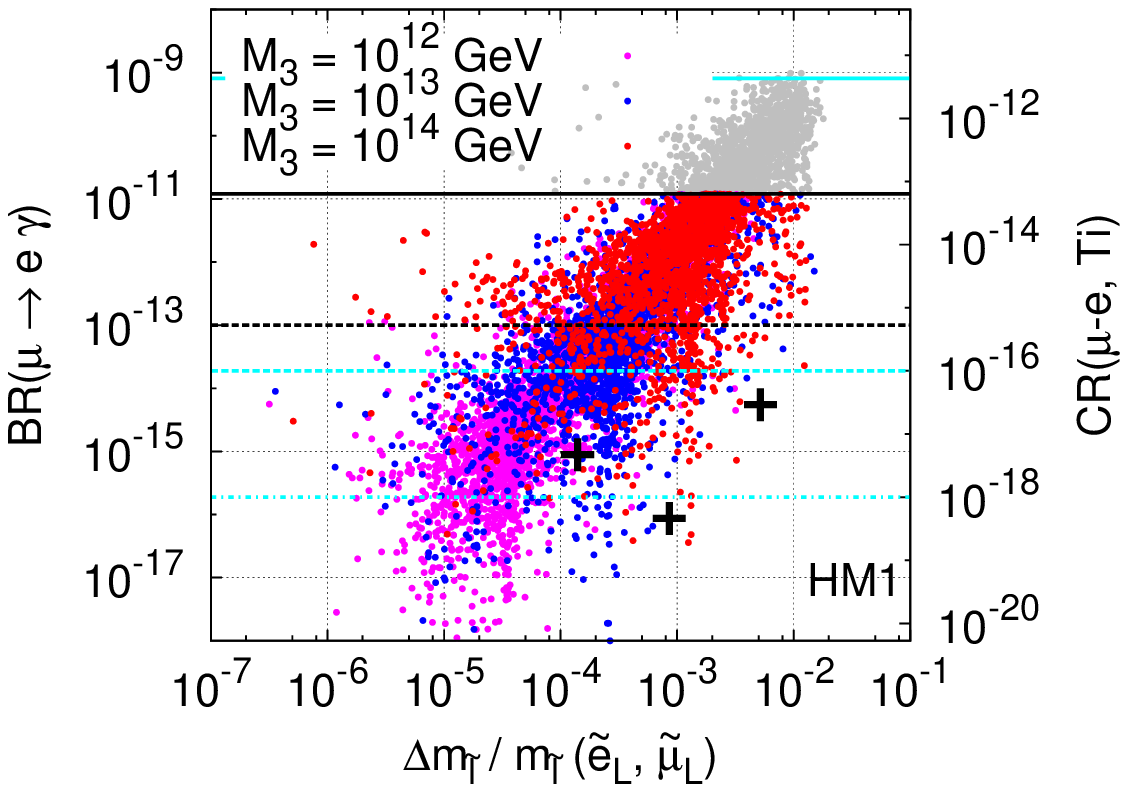, clip=, angle=0, width=80mm}&
\epsfig{file=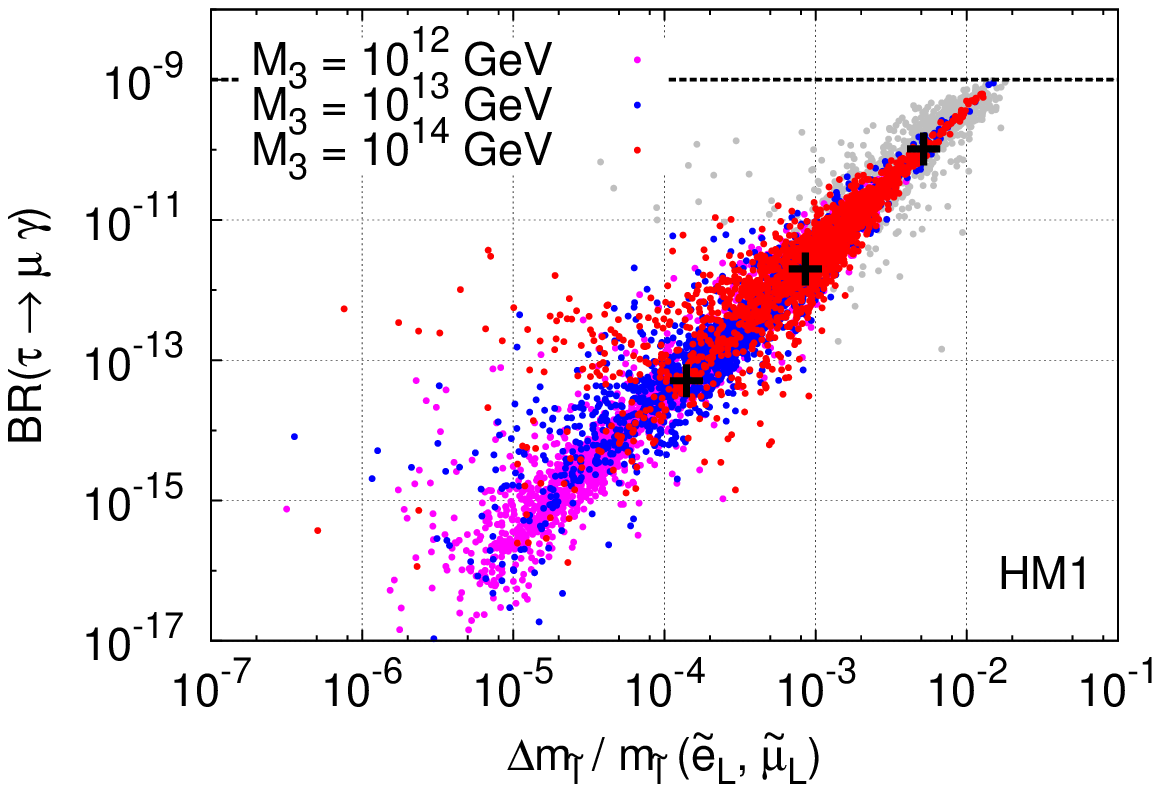, clip=, angle=0, width=80mm}\\
\epsfig{file=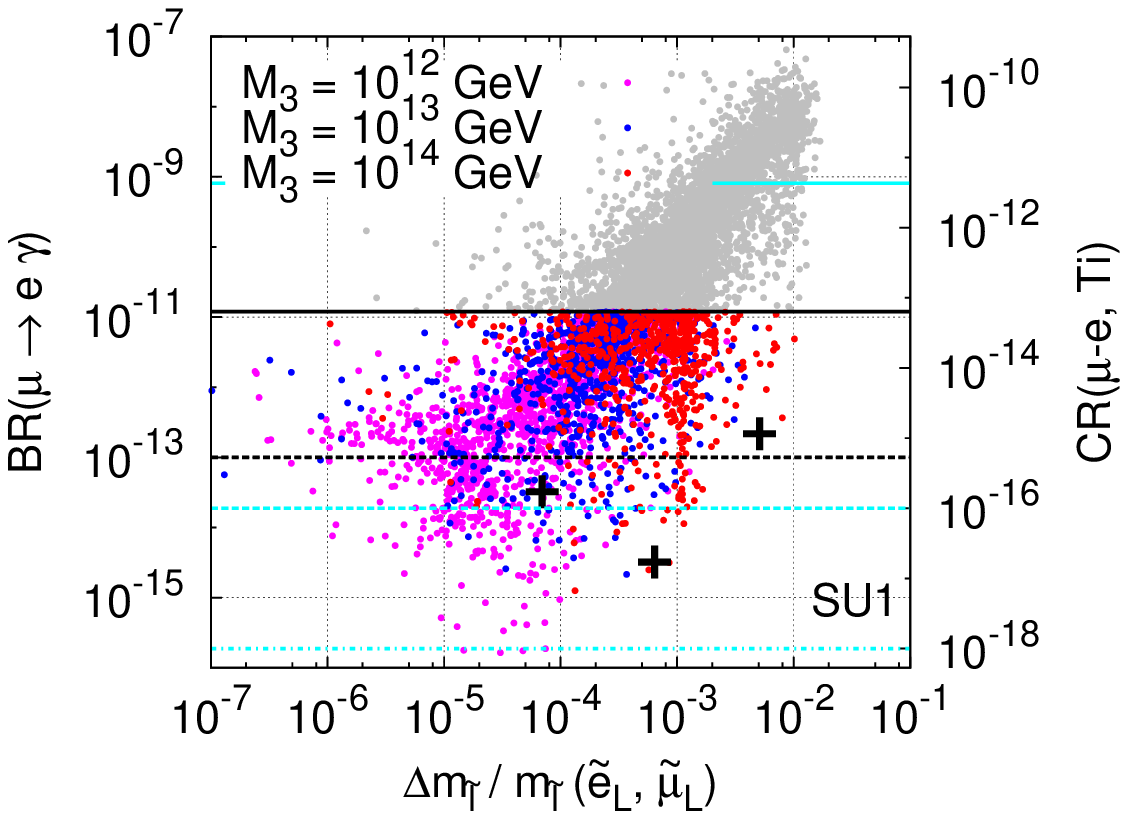, clip=, angle=0, width=80mm}&
\epsfig{file=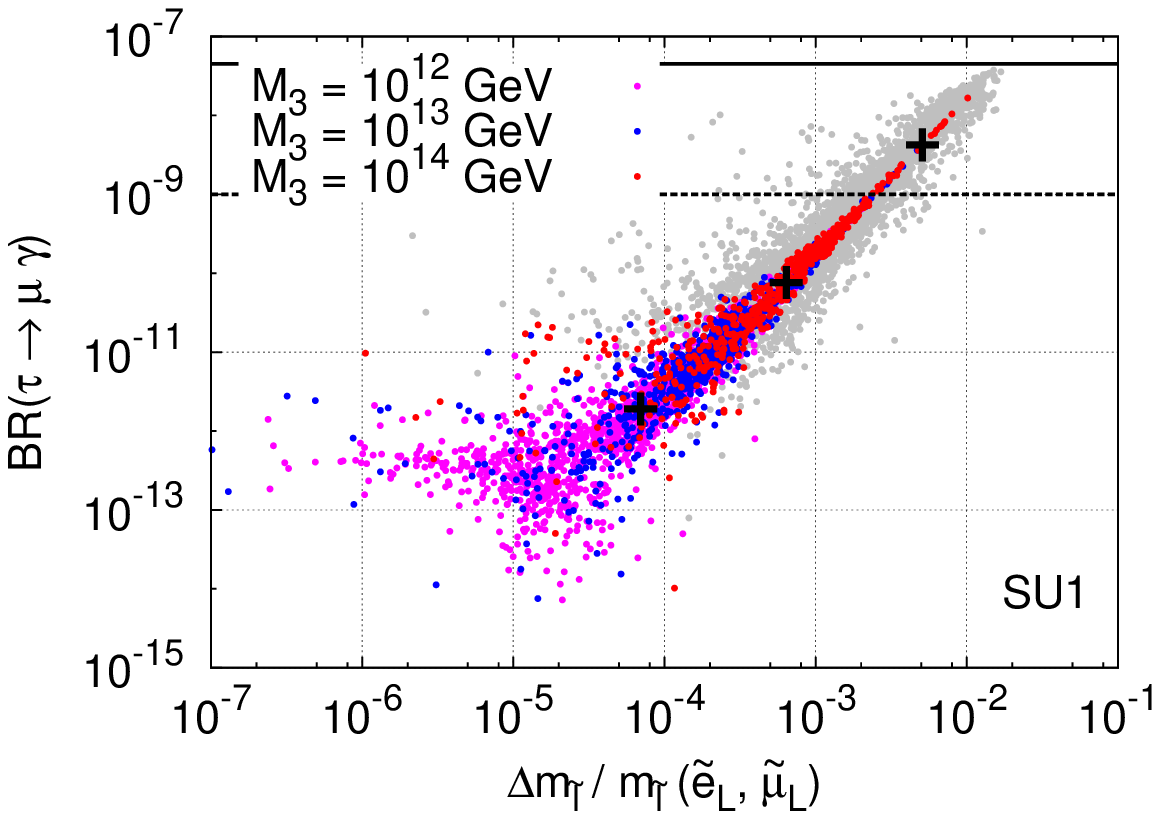, clip=, angle=0, width=80mm}
\end{tabular}
\caption{
Upper left (right) panel: BR($\mu \to e \gamma$) (BR($\tau \to \mu
\gamma$)) on the left y-axis as a function of the
mass difference $\tilde e_L - \tilde \mu_L$, normalised to the average
$\tilde e_L,\tilde \mu_L $ mass, for seesaw variations of
point P5-HM1. We display the corresponding
predictions of CR($\mu-e$, Ti) on the secondary right y-axis.
 Lower panels: same as above, but for point P6-SU1.
Horizontal lines denote the corresponding current bounds/future
sensitivities.  The distinct coloured regions correspond to three
different values of $M_{N_3}=\{ 10^{12},\ 10^{13}, \ 10^{14}\}$ GeV.
The remaining parameters were set as $M_{N_1}=10^{10}$ GeV,
$M_{N_2}=10^{11}$ GeV, $\theta_{13}=0.1^\circ$ and the complex $R$
matrix angles have been randomly varied as $|\theta_i| \in [0, \pi]$,
and $\arg(\theta_i) \in [-\pi, \pi]$. The crosses correspond to the
different seesaw benchmark points: from smaller to larger mass
splittings one has HM1$^{\prime \prime}$ (SU1$^{\prime \prime}$),
HM1$^{\prime}$ (SU1$^{\prime}$), HM1$^{\prime \prime \prime}$
(SU1$^{\prime \prime \prime}$), for the upper (lower) panels.
}\label{fig:BRCR.MS:Rcomplex:M3:theta13}
\end{center}
\end{figure}

The analysis we have done for a few illustrative SUSY benchmark points
can be reproduced for any other cMSSM realisation. In a hopefully not
too distant future, when fundamental mSUGRA parameters will have been
reconstructed, and a measurement of LFV observables (BR($\tau \to \mu
\gamma$), and CR($\mu-e$) in nuclei, for instance) will have also been
reported,  
one will then be able to predict the mass splittings associated 
to this (these) region(s) of the SUSY seesaw parameter space.  
Should an additional
measurement of the slepton mass splittings correspond to the above
prediction, one can say that the present seesaw realisation is in
striking agreement with the data we will so far have collected.  
On the other hand, if the measurement of the mass splittings lies outside
the predictions (as obtained by the SUSY seesaw, possibly in a
leptogenesis motivated region), we will be led to the conclusion that
one of the underlying hypothesis 
(either this  seesaw realisation or a type-I seesaw as the dominant or even
unique LFV source) has to be reconsidered, or even strongly
disfavoured.

\subsection{Flavour violating neutralino decays: di-lepton
  distributions in the SUSY seesaw}

To conclude our study of LFV at the LHC, we reconduct the analysis of
Section~\ref{subsec:di-lepton-cmssm}, but now in the framework of the
SUSY seesaw. As mentioned before, models of
supersymmetric LFV may be manifest in di-lepton distributions, either
through the relative separation of the kinematical edges corresponding
to $\tilde e_L$ in $m_{ee}$ and those of $\tilde \mu_L$ in $m_{\mu \mu}$
(implying that $m_{\tilde e_L} \neq m_{\tilde \mu_L}$), or via the
appearance of new edges in a given di-lepton mass distribution.

In Fig.~\ref{fig:edges:seesaw:cMSSM}, we display the BR($\chi_2^0 \to
\mu \,\mu \,\chi_1^0$) as a function of the di-muon invariant mass $m_{\mu
\mu}$ for different SUSY seesaw points, comparing the
distributions with those of the cMSSM (formerly shown in
Fig.~\ref{fig:msugra.di-leptonx}). For simplicity, we do not present here 
the peaks corresponding to the  $Z$ and $h$ intermediate states in $\chi_2^0 \to \ell  \,\ell \,
\chi_1^0$ decays. The values of the edges are presented in
Table~\ref{table:seesaw:dilepton:values}, and should be compared to
those listed in Table~\ref{table:cmssm:di-lepton:values} for the pure
cMSSM case.

\begin{table}[ht]
\begin{center}
\begin{tabular}{|c|r|r|r|r|c|c|}
\hline
\multirow{2}{*}{$\tilde{l}_X$}  &
\multicolumn{6}{c|}{$m_{ll}(\tilde{l}_X)$ (GeV) for type-I SUSY seesaw} 
\\ \cline{2-7}
& \multicolumn{1}{c|}{P1$^{\prime \prime \prime}$}      &    
\multicolumn{1}{c|}{P2$^{\prime}$}     
& \multicolumn{1}{c|}{P3$^{\prime}$}    &    
\multicolumn{1}{c|}{P4$^{\prime \prime \prime}$}     
& \multicolumn{1}{c|}{P5-HM1$^{\prime \prime \prime}$}   &
\multicolumn{1}{c|}{P6-SU1$^{\prime \prime \prime}$}
\\ \hline     
$\tilde{e}_R$   & 115.8 & 125.9 
& 150.8 & 434.5 & 128.4 & 92.2  
\\ \hline
$\tilde{e}_L$   & 136.4 &  93.1 
& 83.7 & 188.2 & 256.0 & 62.3 
\\ \hline
$\tilde{\mu}_R$ & 115.7 & 125.8 
& 150.7 & 434.3 & 128.2 & 92.1 
\\ \hline
$\tilde{\mu}_L$ & 141.6 & 95.5 
& 85.6 & 212.8 & 256.3 & 66.6 
\\ \hline
$\tilde{\tau}_1$ & 81.8 & 77.1 
& 76.6 & 40.9 & 53.6 & 67.1 
\\ \hline
$\tilde{\tau}_2$ & 135.4 & 111.8 
& 105.2 & 300.4 & 263.3 & 56.7 
\\ \hline 
\end{tabular}
\end{center}
\caption{$m_{ll}(\tilde{l}_X)$ (GeV) for type-I SUSY seesaw points
  (see Tables~ \ref{table:points1} and \ref{table:seesaw:benchmark}),
where $l$ is any of the charged leptons and
$X$ stands for left- and right-handed sleptons (all families).
}\label{table:seesaw:dilepton:values} 
\end{table}

As is manifest from Fig.~\ref{fig:edges:seesaw:cMSSM}, and readily
confirmed from Table~\ref{table:seesaw:dilepton:values}, the impact of
the seesaw at the level of the di-muon mass distributions is quite
spectacular, particularly in the appearance of a third edge in most
of the benchmarks considered. With the exception of P1$^{\prime \prime
  \prime}$, all other distributions exhibit now the edge corresponding
to the presence of an intermediate $\tilde \tau_2$, implying that the decay
occurs via $\chi_2^0 \to \tilde \tau_2 \mu \to  \mu\, \mu\, \chi_1^0$. 
For instance, for point P2$^{\prime}$, the BR($\chi_2^0 \to \mu \mu
\chi_1^0$) via intermediate $\tilde \mu_L$, $\tilde \mu_R$ and $\tilde
\tau_2$  are  2.6\%, 1.1\% and 1.6\%, respectively. 

In Fig.~\ref{fig:edges:seesaw:cMSSM2} we compare the di-muon with the
di-electron distribution, for the previous seesaw benchmark points.  
We point here that unlike the smuon case the di-electron distribution does
not change with respect to the cMSSM case. From this figure one further
observes that selectron and smuon edges exhibit a clear separation. 
For a c.o.m. energy $\sim 7$ TeV at the LHC, and an integrated
luminosity of $1\ \text{fb}^{-1}$, only a few events would be
observable. But for $\sqrt{s} = 14$ TeV (and an integrated luminosity
of $100\ \text{fb}^{-1}$), the expected number of events (without
background analysis nor detector simulation) is $\mathcal{O}(10^3)$
for P1$^{\prime\prime\prime}$, P2$^{\prime}$ and P3$^{\prime}$, 
while considerable poorer prospects are expected for P4$^{\prime\prime\prime}$ 
and HM1$^{\prime\prime\prime}$. SU1$^{\prime\prime\prime}$
offers the most promising scenario, with more that $10^4$ expected
events.

Comparing Table~\ref{table:seesaw:dilepton:values} and 
Table~\ref{table:cmssm:di-lepton:values}, one further verifies that 
in the type-I SUSY seesaw the
mass splittings are indeed a $LL$ sector phenomenon (notice that for
both right-handed smuons and selectrons the edges remain identical up to
$\sim$ 0.1 GeV) and are essentially restricted to the $\tilde \mu -
\tilde \tau$ sectors, since the edges corresponding to $\tilde e_L$ also
remain unaffected. 

Interestingly, the fact that the SUSY seesaw leads to increased mass
splittings only for the left-handed sleptons
might provide another potential fingerprint for this mechanism of
LFV. Compiling all the data collected throughout our numerical analysis, we
have found that the maximal splitting between right-handed smuons and
selectrons, in all the cases studied, is 
\begin{equation}
\left.
\frac{\Delta m_{\tilde \ell}}{m_{\tilde \ell}} (\tilde \mu_R, \tilde
e_R)\right|_\text{max}\, \approx \, 0.09 \%\,.
\end{equation}
Recall that throughout the preceding subsections we have verified that 
within the SUSY seesaw $\Delta m_{\tilde \ell}/m_{\tilde \ell} (
\tilde \mu_L, \tilde e_L) $ could easily reach values of a few \%. 
Should the LHC measure mass splittings between right-handed sleptons
of the first two families that are significantly above the 0.1\%
level, this could provide important indication to the fact that
another mechanism of FV is at work (for instance, an effective
parametrization of flavour violating effects in the lepton sector, as
done in~\cite{Buras:2009sg}, induces similar mass splittings for both
right- and left-handed sleptons). Among the many possibilities, 
a likely hypothesis would be the 
non-universality of the slepton  soft-breaking terms. 

\begin{figure}[ht!]
\begin{center}
\begin{tabular}{c}
\epsfig{file=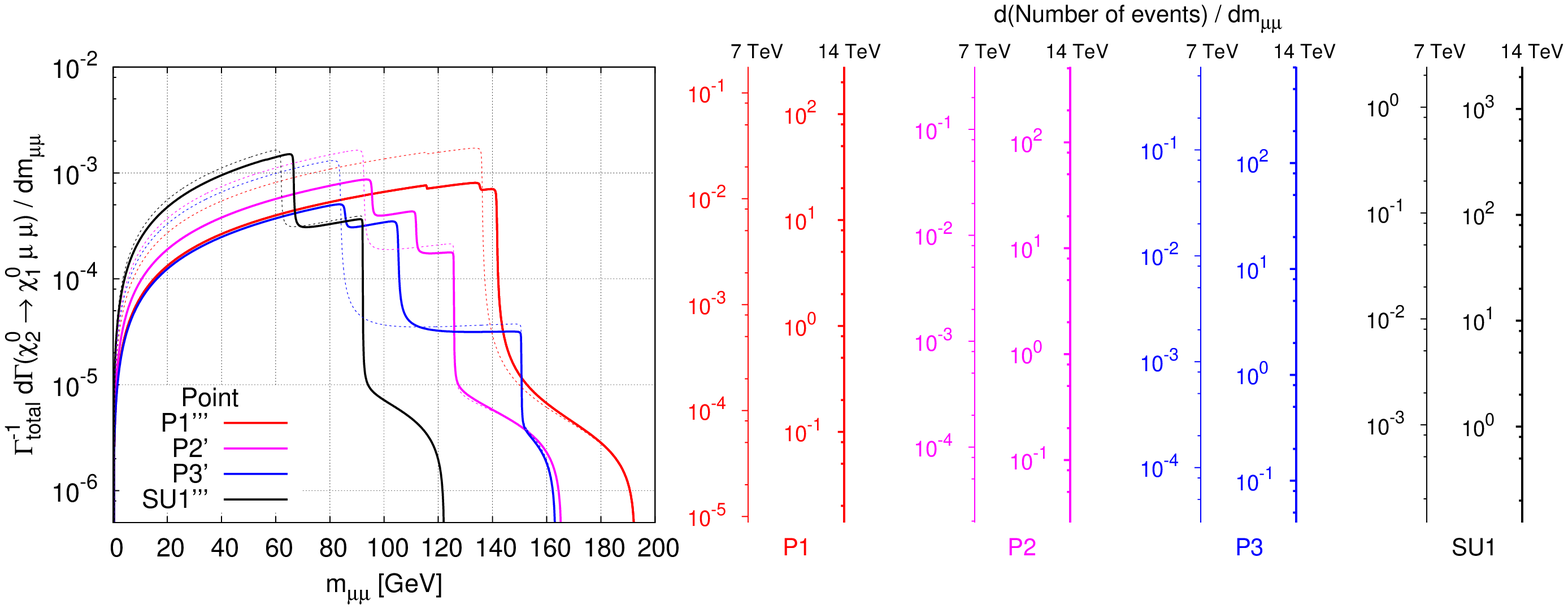, clip=, angle=0, width=170mm}\\
\epsfig{file=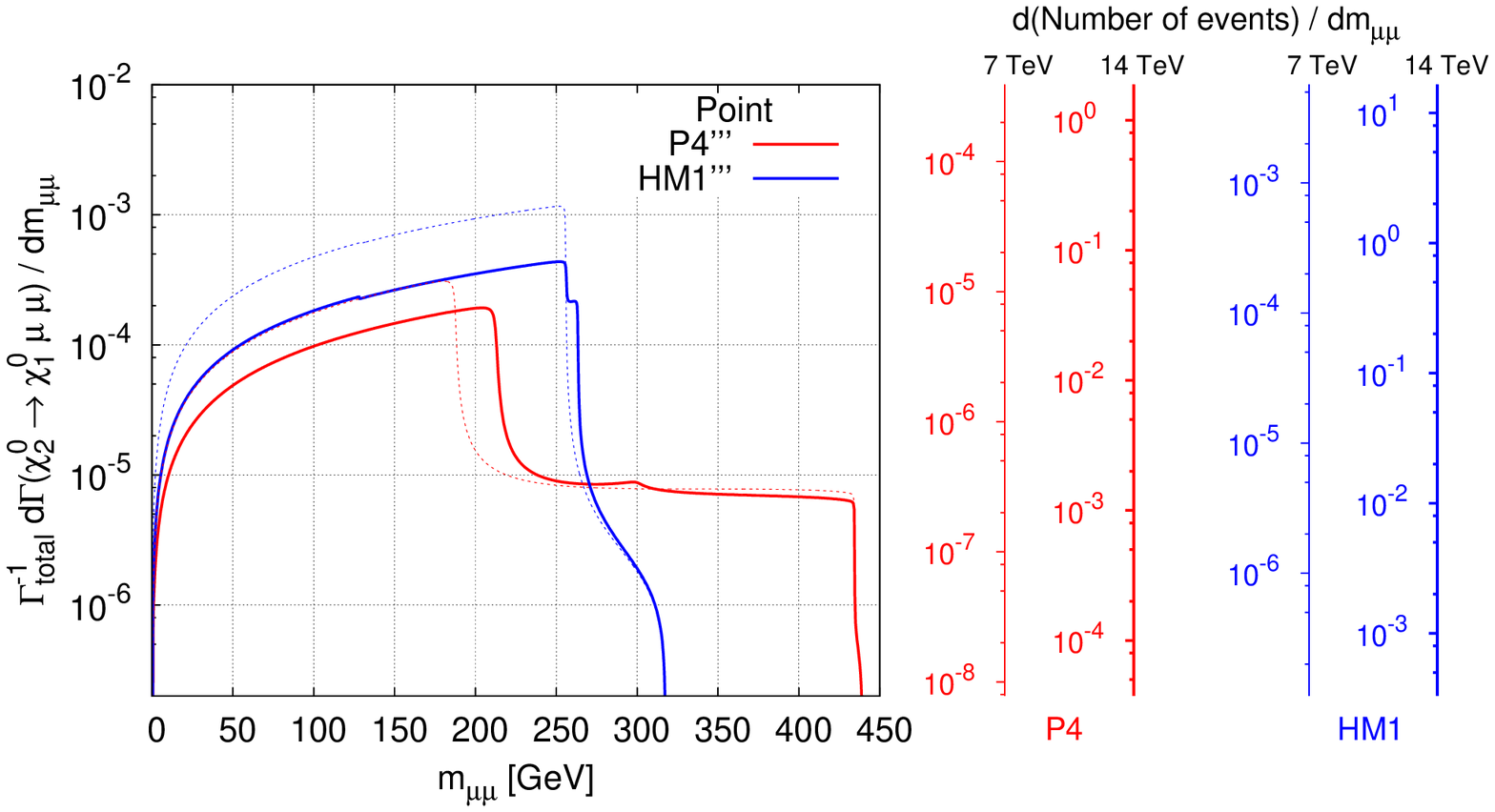, clip=, angle=0, width=120mm}\\
\end{tabular}
\caption{BR($\chi_2^0 \to \mu \mu \chi_1^0$) as a function of the
di-muon invariant mass $m_{\mu \mu}$ (in GeV) for different SUSY seesaw
points (see Tables~\ref{table:points1} and \ref{table:seesaw:benchmark}).
Upper panel: 
P1$^{\prime \prime \prime}$ (red), 
P2$^{\prime}$ (pink), P3$^{\prime}$ (blue) and 
P6-SU1$^{\prime \prime \prime}$ (black). 
Lower panel:
P4$^{\prime \prime \prime}$ (red) and P5-HM1$^{\prime \prime \prime}$
(blue).
Dotted (coloured) lines denote in both panels the curves for the
corresponding cMSSM case. 
Secondary (right) y-axes denote the corresponding expected number of events for 
$\sqrt s = 7$  TeV and 14 TeV, respectively with 
$\mathcal{L}=1\ \text{fb}^{-1}$ and $\mathcal{L}=100\ \text{fb}^{-1}$.}
\label{fig:edges:seesaw:cMSSM}
\end{center}
\end{figure}

\begin{figure}[h!]
\begin{center}
\begin{tabular}{c}
\epsfig{file=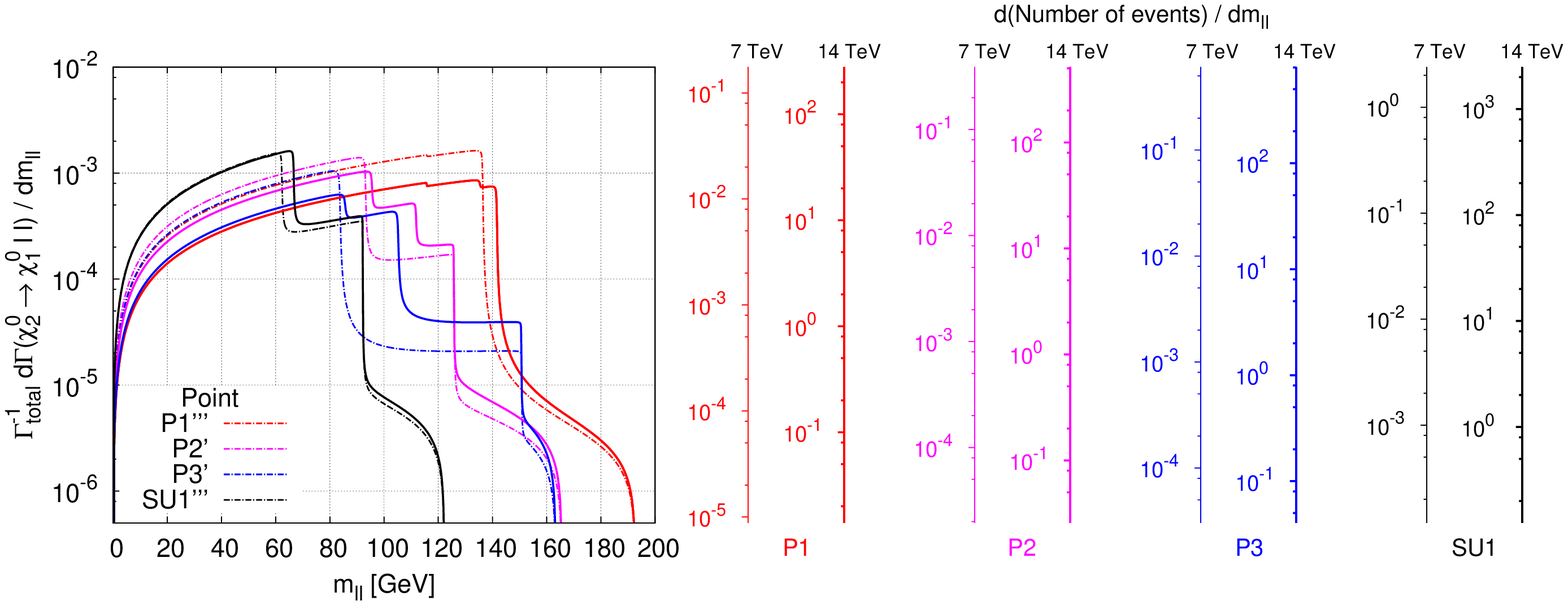, clip=, angle=0, width=170mm}\\
\epsfig{file=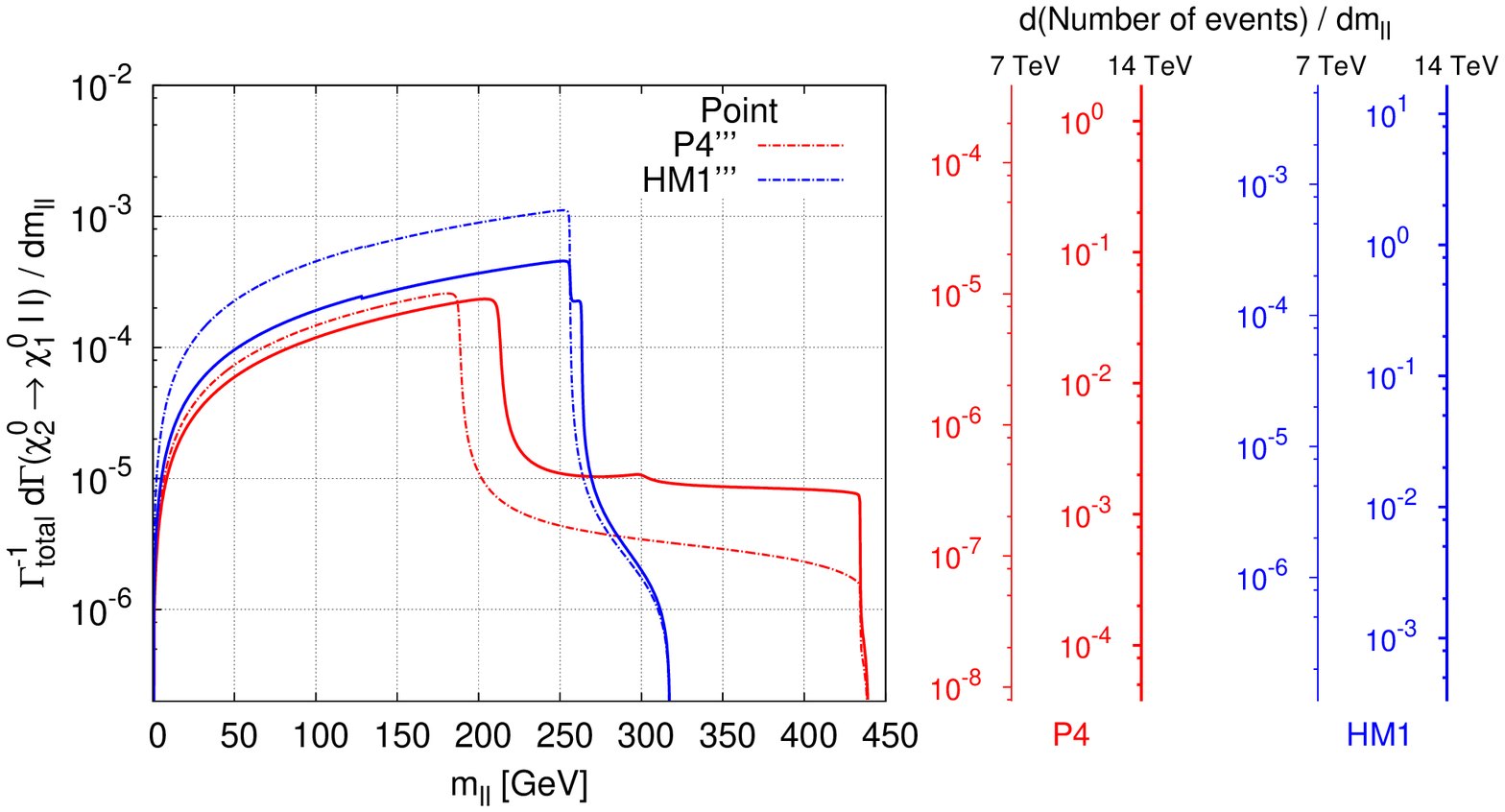, clip=, angle=0, width=120mm}\\
\end{tabular}
\caption{BR($\chi_2^0 \to \ell \ell \chi_1^0$) as a function of the
di-lepton invariant mass $m_{\ell \ell}$ ($\ell=e,\mu $) (in GeV) for
P1$^{\prime \prime \prime}$ (red), P2$^{\prime}$ (pink), P3$^{\prime}$ (blue) and 
P6-SU1$^{\prime \prime \prime}$ (black) (see Tables~\ref{table:points1} and
  \ref{table:seesaw:benchmark}).
Lower panel:
P4$^{\prime \prime \prime}$ (red) and P5-HM1$^{\prime \prime \prime}$
(blue).
Full (dashed) lines denote in both panels the curves for di-muon (di-electron)
distributions.
Secondary-right y-axes denote the corresponding expected number of events for 
$\sqrt s = 7$  TeV and 14 TeV, respectively with 
$\mathcal{L}=1\ \text{fb}^{-1}$ and $\mathcal{L}=100\ \text{fb}^{-1}$.}
\label{fig:edges:seesaw:cMSSM2}
\end{center}
\end{figure}

Finally, we display the prospects for direct flavour violation in
$\chi_2^0$ decays: in addition to the possibility of having staus in
the intermediate states, one can also have opposite-sign, different
flavour final state di-leptons.
In particular, one can have $\chi_2^0 \to \mu \tau \chi_1^0$, with a
non-negligible associated branching ratio. For $\sqrt{s} = 14$ TeV and
$\mathcal{L}=100\ \text{fb}^{-1}$, the expected number of events
(again without background analysis nor detector simulation) is
$\mathcal{O} (10^3)$ for P1$^{\prime\prime\prime}$, P2$^{\prime}$,
P3$^{\prime}$ and SU1$^{\prime\prime\prime}$. This is shown in
Fig.~\ref{fig:edges:seesaw:cMSSM:FV}.
For P4$^{\prime\prime\prime}$ and HM1$^{\prime\prime\prime}$ a
significantly smaller
number is expected. 

\begin{figure}[h!]
\begin{center}
\epsfig{file=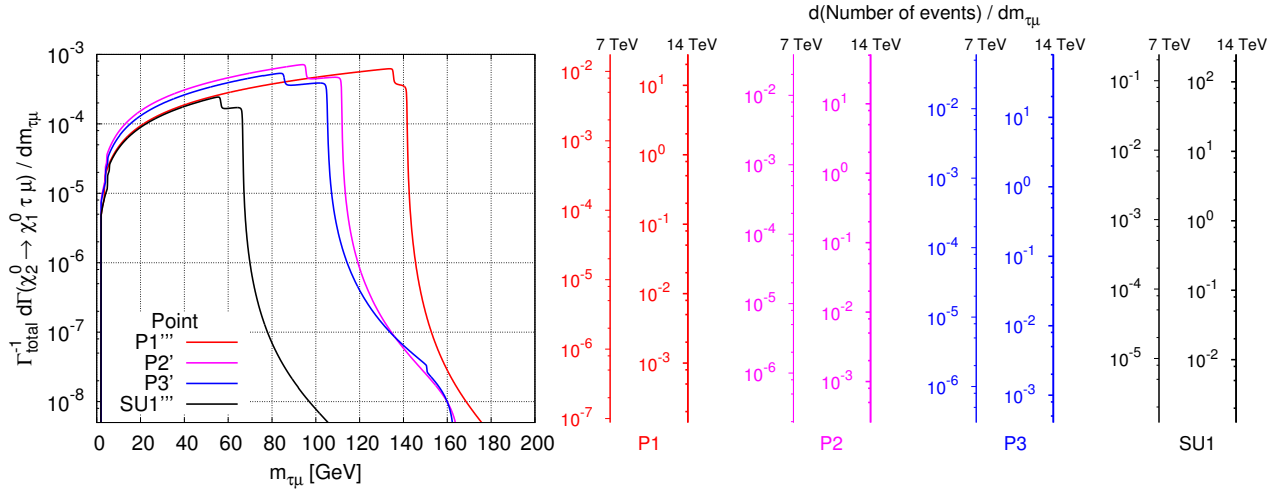, clip=, angle=0, width=170mm}
\caption{Flavour violating 
BR($\chi_2^0 \to \mu \tau \chi_1^0$) as a function of the
di-lepton invariant mass $m_{\tau \mu}$ (in GeV) for the seesaw benchmark points
P1$^{\prime \prime \prime}$ (red), P2$^{\prime}$ (pink), P3$^{\prime}$ (blue) and 
P6-SU1$^{\prime \prime \prime}$ (black) (see Tables~\ref{table:points1} and
\ref{table:seesaw:benchmark}).
On the right y-axis, we also display the expected number of events
for $\sqrt s = 7$ TeV (with $\mathcal{L}=1\ \text{fb}^{-1}$) and 
$\sqrt s = 14$ TeV (for $\mathcal{L}=100\ \text{fb}^{-1}$).
}\label{fig:edges:seesaw:cMSSM:FV}
\end{center}
\end{figure}

\section{Conclusions}\label{sec:conclusions}

In this work we have studied lepton flavour violation in high- and
low-energy observables in the framework of a type-I SUSY seesaw.  If
the seesaw is indeed responsible for both neutrino masses and leptonic
mixings, and accounts for low-energy LFV observables within future
sensitivity reach, interesting slepton phenomena are expected to be
observed at the LHC.  Under the assumption of a unique source - the
neutrino Yukawa couplings -, the interplay between these high- and
low-energy LFV observables allows to derive some information about the
seesaw parameters and, for specific configurations of the model,
disfavour the type-I SUSY seesaw as being the unique source of LFV.

We began our analysis by considering the mSUGRA parameter space,
looking for regions where the conditions for a successful
reconstruction of the slepton masses can be met: the observation of a
significant number of events of the type $\chi_2^0\to
\ell^\pm\ell^\mp\chi_1^0$ with sufficiently hard leptons in the final
state implies that the mSUGRA parameters should follow specific
relations.  In general, the most promising scenarios are encountered
for low to intermediate $\tan\beta$ and a light sparticle spectrum.
 In these regions the correct relic density is always obtained via
 $\tilde \tau_1-\chi_1^0 $ co-annihilation.
We have illustrated these features by considering different points in
mSUGRA parameter space. Among them, and in order to address LHC
prospects, we included in our analysis two LHC benchmark points: SU1
(ATLAS) and HM1 (CMS). 
 
The analysis of the slepton mass splittings in the cMSSM (in the
absence of flavour violation in the lepton sector) reveals that the
expected values are very small due to the tiny $LR$ mixing and RGE
effects. Although marginally observable for the smuon-stau sector, the
mass differences for the first two generations of mostly LH sleptons
is at best of order of $0.03$ \%. At the LHC, the cMSSM smuon and
selectron masses could be reconstructed from the (double) triangular
di-lepton invariant mass distributions (in the best case scenario)
with identical kinematical edges for both di-muon and di-electron mass
distributions.

In the minimal implementation of a type-I SUSY seesaw ($R=1$, i.e. not
taking into account possible mixings in the right-handed neutrino
sector), the slepton spectrum reflects the mixing introduced at low
energies due to the non-trivial structure of the Yukawa couplings
(given by the $U^{\text{MNS}}$ leptonic mixing matrix).  These effects
are only manifest for the left-handed sector, potentially leading to
maximal mixing between left-handed smuons and staus.  This has
motivated us to introduce the concept of quasi-degenerate flavour
content sleptons and ``effective" mass splittings.  Especially for
larger values of $|A_0|$, the mass splittings between the first two
generations are significantly enhanced with respect to the pure cMSSM
case. Even in this limit of $R=1$, mass splittings of a few percent
can be easily found, and have associated lepton flavour violating
low-energy observables within reach of the future LFV experiments.

Regarding the seesaw parameters that are likely to be measured, the
Chooz angle only has an indirect effect on the slepton mass
differences, independently of the entries of the $R$ matrix and the
seesaw scale $M_{N_3}$.  For larger values of $\theta_{13}$ slepton
mass differences within LHC sensitivity have associated
BR($\mu\to e \gamma$) already excluded by current experimental
bounds.

 Given the dimensionality of the full SUSY seesaw parameter space, we
 have selected a set of benchmark points, in particular the LHC points
 SU1 and HM1,
  to carry the analysis of the general seesaw case (that is $R\neq
 1$), considering different values of $\theta_{13}$ and distinct
 right-handed neutrino spectra. A measurement of a $\tilde e_L -
 \tilde \mu_L$ mass splitting 
 between 0.1\% and 1\% for P5-HM1 implies that MEG should observe a
 $\mu\to e \gamma$ signal. Provided the seesaw is the unique source of
 LFV, no signal is expected to be observed by SuperB.  The most
 interesting LFV signature of P6-SU1 is that, contrary to P5-HM1, a
 measurement of a mass splitting between left-handed selectrons and
 smuons would imply a BR($\tau \to\mu \gamma$) within SuperB
 reach. Furthermore, such observation would strongly hint towards a
 heavy seesaw scale $M_{N_3}\gtrsim 10^{13}$ GeV.

Despite the richness of the SUSY seesaw regarding the interplay of
slepton mass splittings and low-energy flavour violation, the most
spectacular result would be definitely manifest in the di-lepton mass
distributions obtained at the LHC.  In addition to the clear
separation between the edges of di-muon and di-electron distributions
(or equivalently, the observation of slepton mass splittings) one
expects the appearance of an additional third kinematical edge for
most of the benchmark points considered (which exhibited only two
edges for the pure cMSSM case). The latter would signal flavour
violation in $\chi_2^0$ and/or $\tilde \ell$ decays.

Interestingly, irrespective of the specific seesaw configuration, the
mass differences of right-handed sleptons are hardly sensitive to
$Y^\nu$-induced radiative effects (at leading order).  Should the LHC
observe mass splittings between right-handed sleptons of the first two
families significantly above the per mille level, this would strongly
hint towards the presence of another source of flavour violation
(other than the seesaw).

It is important to stress that although a joint set of LFV observables
might contribute to disfavour a type-I seesaw as the underlying 
mechanism of neutrino mass generation (and lepton flavour violation), 
one cannot  exclude the possibility (however unlikely) that
effects such as slepton mass splittings or flavour
violating decays originate from a non-trivial structure of the SUSY soft
breaking Lagrangian (the sleptonic part) at the GUT scale.

\section{Acknowledgements}
We are grateful to J. Orloff for many enlightening discussions, and to 
F. R. Joaquim for his valuable remarks. 
This work has been done partly under the ANR project CPV-LFV-LHC {NT09-508531}. 
The work of  A. J. R. F.  has been supported by {\it Funda\c c\~ao
  para a Ci\^encia e a 
Tecnologia} through the fellowship SFRH/BD/64666/2009. 
A. J. R. F. and J. C. R. also acknowledge the financial support from
the EU Network grant UNILHC PITN-GA-2009-237920 and from {\it
Funda\c{c}\~ao para a Ci\^encia e a Tecnologia} grants CFTP-FCT UNIT
777, CERN/FP/83503/2008 and PTDC/FIS/102120/2008. 
Finally, A. J. R. F., J. C. R. and A. M. T.  are indebted to the
LPT-Orsay for the hospitality, and for the funding from ANR.

\end{document}